\documentclass[eqsecnum,preprint]{revtex4}

\makeatletter       

\renewcommand{\p@subsection}{}

\renewcommand{\p@subsubsection}{}

\makeatother

\newcommand{\lbfig}[1]{\refstepcounter{fig} \label{#1} }
\newcounter{fig}

\usepackage{amssymb}
\usepackage{amsmath}
\usepackage{bbm}
\usepackage{epsf}
\usepackage{epsfig}
\usepackage{psfrag}
\usepackage{citesort}

\newcommand{\beq}{\begin{equation}}
\newcommand{\eeq}{\end{equation}}
\newcommand{\beqa}{\begin{eqnarray}}
\newcommand{\eeqa}{\end{eqnarray}}
\newcommand{\Tr}{\text{Tr}}

\newcommand{\del}{\partial}

\newcommand{\tk}{\tilde{k}_0}

\newcommand{\deldag}{\mathbin{\partial\mkern-10.5mu\big/}}
\newcommand{\mDdag}{\mathbin{{\mathbf D}\mkern-14mu\big/}}
\newcommand{\kdag}{\mathbin{k\mkern-9.8mu\big/}}
\newcommand{\ndag}{\mathbin{n\mkern-9mu\big/}}
\newcommand{\hdag}{\mathbin{h\mkern-10mu\big/}}
\newcommand{\npdag}{\mathbin{n'\mkern-13.5mu\big/}}

\newcommand{\bm}{\mathbf{m}}
\newcommand{\bX}{\mathbf{X}}
\newcommand{\bY}{\mathbf{Y}}

\def\DBR#1#2{\bigl\{#1 \bigr\}\bigl\{#2 \bigr\}} 
\def\Slash#1{#1\kern-0.55em\raise.05ex\hbox{/}}
\def\slash#1{#1\kern-0.5em\raise.05ex\hbox{{$\scriptstyle /$}}}

\begin{document}

\pagenumbering{Roman}


\begin{titlepage}

\vskip 0.3in

\preprint{BNL-72343-2004-JA, HD-THEP-03-62}

\vskip 0.6in

\centerline{\LARGE \bf Transport equations for chiral fermions to order $\hbar$}
\vskip 0.2in
\centerline{\LARGE \bf and electroweak baryogenesis: Part~I}

\vskip 0.4in

\begin{center}

  {\Large \bf
   Tomislav Prokopec$^{\diamond}$,
   Michael G. Schmidt$^\diamond$ \\
   \vskip 0.2in
   and Steffen Weinstock$^\bullet$}

\vskip 0.3in
  {\it $^\diamond$ Institut f\"ur Theoretische Physik, Universit\"at Heidelberg\\
           Philosophenweg 16, D-69120 Heidelberg, Germany} \\

  \vskip 0.1in
  {\it $^\bullet$ Nuclear Theory Group, Brookhaven National Laboratory, Upton, NY 11973-5000, USA}

\end{center}

\vskip 0.4in

\centerline{\bf Abstract}
\vskip 0.2in

This is the first in a series of two papers. In this first part,
we use the Schwinger-Keldysh formalism to derive semiclassical Boltzmann
transport equations, accurate to order $\hbar$,
for massive chiral fermions, scalar particles,
and for the corresponding CP-conjugate states. 
Our considerations include complex mass terms and 
mixing fermion and scalar fields, such that 
CP-violation is naturally included, rendering the equations
particularly suitable for studies of baryogenesis
at a first order electroweak phase transition.
We provide a quantitative criterion for when the reduction to
the diagonal kinetic equations in the mass eigenbasis is justified,
leading to a quasiparticle picture even in the case of mixing scalar
or fermionic particles.
Within the approximations we make, it is possible
to first study the Boltzmann equations without the collision term.
In a second paper~\cite{PSW_2} we discuss the collision terms
and reduce the Boltzmann equations to fluid equations.

\vskip 0.4in

\footnoterule
\vskip 3truemm
{\small\tt
\noindent
$^\diamond$T.Prokopec@, M.G.Schmidt@thphys.uni-heidelberg.de \\
$^\bullet$weinstock@bnl.gov
}

\end{titlepage}

\pagenumbering{arabic}


\tableofcontents


\cleardoublepage
\section{Introduction}

\vskip 0.05in

 The question of how Boltzmann transport equations emerge from 
quantum field theory is an old 
problem~\cite{Schwinger:1961,KadanoffBaym:1962,Keldysh:1964}. It seems that 
a consensus has been reached, according to which a suitable
(loop) truncation of the self-energies in the Kadanoff-Baym equations,
supplied by an on-shell reduction, leads to Boltzmann transport equations
for the dynamics of quasiparticles in phase space. 
While procedures leading to classical
Boltzmann equations -- which formally correspond to taking the
limit $\hbar\rightarrow 0$ of the Kadanoff-Baym equations --
seem to be quite clear and well 
established~\cite{ChouSuHaoYu:1985,CalzettaHu:1986}, 
no systematic treatment has been pursued beyond the classical approximation, 
which would lead to a kinetic description valid to linear order in an $\hbar$ 
expansion
(a notable exception is a study of interference between the tree-level
and one-loop decay rates of relevance for grand unified 
baryogenesis~\cite{KolbWolfram:1980}.)
The inclusion of $\hbar$ corrections is absolutely
necessary for any treatment of baryogenesis, which represents the primary
impetus for the work presented here,
since CP-violating effects vanish in the limit
when $\hbar\rightarrow 0$. More concretely,
one deals with particle transport in high temperature relativistic plasmas
with Minimal Standard Model particle content,
often extended to include exotic, as-of-yet undiscovered new particles,
{\it e.g.}  various supersymmetric partners. 
While $\hbar$ expansions are often
identified with loop expansions in field theory, here we associate $\hbar$
primarily to a gradient expansion
(which is a field-theoretical generalization of the WKB expansion) 
with respect to space-time variations of background fields.
The relevant examples of background fields for baryogenesis are
the mass terms generated by the (Higgs) order parameter 
at a first order electroweak phase transition; other examples include
classical (possibly stochastic) background gauge fields. 
One important hurdle, which must be overcome in the development of any
formalism that aspires to accurately describe transport in plasmas beyond
the classical level, is the establishment of a quasiparticle picture.
This entails a demonstration of the existence and the
identification of the relevant quasiparticle excitations in the plasma, 
which can be used to encode the relevant dynamical information.
This is the problem we solve here.
Apart from baryogenesis, potential applications of the formalism developed 
in this work include propagation of neutrinos
in dissipative media, transport of electrons and holes in quantum wires and 
quantum semiconductor devices, dynamics of Bose-Einstein condensates, 
{\it etc.}

\vskip 0.05in

The standard approach to electroweak 
baryogenesis~\cite{KuzminRubakovShaposhnikov:1985}
requires a study of the generation and transport of CP-violating 
flows~\cite{CohenKaplanNelson:1990+1991} arising from interactions of fermions
with expanding phase transition fronts of a first order electroweak
phase transition. The most prominent theoretical problem in such
scenarios over the past few years has been to
find a systematic derivation of the appropriate transport equations, including
the CP-violating sources. In particular, there has been controversy in the 
literature regarding what are the dominant 
sources appearing in the transport equations and how to compute them. 
This work is an attempt to resolve this daunting question. 

\vskip 0.05in

It is known that a strong first order transition is a necessary requirement
for electroweak baryogenesis~\cite{Shaposhnikov:1986,Shaposhnikov:1987}.
While absent in the
Minimal Standard Model~\cite{KajantieLaineRummukainenShaposhnikov,
RummukainenTsypinKajantieLaineShaposhnikov:1998,CsikorFodorHeitger:1998,
AokiCsikorFodorUkawa:1999,
BuchmullerPhilipsen:1994,BergerhoffWetterich:1994}, 
a strong transition can be realized in its extensions, which include 
two Higgs doublet models~\cite{LaineRummukainen:2001},
the Minimal Supersymmetric Standard Model 
(MSSM)~\cite{CarenaQuirosWagner:1996,CarenaQuirosWagner:1998,
ClineKainulainen:1996,Laine:1996,
Losada97,BodeckerJohnLaineSchmidt:1997,LaineRummukainen:1998,
PilaftsisWagner:1999,CsikorFodorHegedusJakovacKatzPiroth:2000,
Espinosa:1996,deCarlosEspinosa:1997},
and its nonminimal extensions (NMSSM)~\cite{Pietroni:1993,
DaviesFroggattMoorhouse:1996,HuberSchmidt:1999}.
In the NMSSM,  not only the possibility of electroweak baryogenesis
is given in wide regions of the parameter space, but it can also have stronger
CP-violation, increasing the amount of the actually produced baryon 
asymmetry~\cite{HuberJohnLaineSchmidt}.

\vskip 0.05in

Considering the derivation of the relevant transport equations, 
there are essentially two approaches in literature. The first approach, 
which we shall refer to as the {\it semiclassical force
 mechanism}~\cite{JoyceProkopecTurok:1995,JoyceProkopecTurok:1996}, is based 
on the observation that, when treated in the WKB approximation,
a CP-violating quadratic part of the fermionic Lagrangean
exhibits a CP-violating shift at first order in gradients in the dispersion
relation, which manifests itself as a CP-violating semiclassical
force in the Boltzmann equation. This force then appears 
in the fluid transport equations, or equivalently diffusion equations 
for the relevant particle densities. In the original 
work~\cite{JoyceProkopecTurok:1995,JoyceProkopecTurok:1996} baryogenesis 
from scattering of top quarks and tau leptons in two Higgs doublet models
was considered. In the subsequent
work~\cite{ClineJoyceKainulainen:1998}
the semiclassical force formalism was extended to include the case of fermion
mixing, which is relevant for example for the {\it chargino} induced 
baryogenesis in the Minimal Supersymmetric Standard Model 
(MSSM)~\cite{ClineJoyceKainulainen:1998,Kainulainen:1999,ClineKainulainen:2000,
ClineJoyceKainulainen:2000+2001,HuberJohnSchmidt:2001} and in its nonminimal
extensions (NMSSM)~\cite{HuberSchmidt:2000,HuberSchmidt:2001}.

\vskip 0.05in

 The second approach, commonly referred to as 
{\it spontaneous baryogenesis}~\cite{CohenKaplanNelson:1991}, is based
on the observation that in the presence of CP-violation the (fermion)
hypercharge currents are not conserved. As a consequence, the energy levels for
CP-conjugate fermionic states are mutually shifted. 
In the presence of scatterings and transport, 
this shift then leads to different populations
for chiral fermions and the corresponding antifermions,
thus sourcing baryogenesis. It was then pointed out in 
Refs.~\cite{DineThomas:1994}, \cite{JoyceProkopecTurok:1996}
and \cite{ComelliPietroniRiotto:1995}
that the hypercharge current, which sources spontaneous baryogenesis,
is suppressed by the square of the mass.
(This must be so simply because the hypercharge current is conserved in the
limit when the mass of particles vanishes, 
so that the spontaneous source must vanish, too.)
Mostly in order to incorporate the possibility of fermionic mixing, 
the spontaneous baryogenesis mechanism was 
subsequently developed by several groups, 
which adopted the idea to baryogenesis studies mediated by the stops and 
charginos of the MSSM~\cite{HuetNelson:1995+1996,Riotto:1995,Riotto:1998,
CarenaQuirosRiottoViljaWagner:1997,
CarenaMorenoQuirosSecoWagner:2000,
CarenaQuirosSecoWagner:2002} and NMSSM~\cite{DaviesFroggattMoorhouse:1996}.

\vskip 0.05in

 From the current literature, it is in fact not clear whether the semiclassical
force and spontaneous sources represent identical or different sources,
or if there perhaps has been a conceptual problem in calculating (one or both of) the
sources. The situation was made even more controversial by a recent
work~\cite{CarenaQuirosSecoWagner:2002}, in which it was claimed that
the spontaneous baryogenesis and semiclassical force sources
correspond formally to the same source in fluid transport equations,
arguing that the method used for computing the semiclassical force source was
incorrect. 

\vskip 0.05in

This is to be compared with the methodology 
advocated in the work based on the WKB 
method~\cite{ClineJoyceKainulainen:2000+2001},
and more recently in~\cite{KainulainenProkopecSchmidtWeinstock:2001,
KainulainenProkopecSchmidtWeinstock:2002,
KainulainenProkopecSchmidtWeinstock:2002b,
KainulainenProkopecSchmidtWeinstock:2002c}
based on the Schwinger-Keldysh formalism~\cite{Schwinger:1961},
or more precisely on a gradient expansion of the Kadanoff-Baym 
equations~\cite{KadanoffBaym:1962} in a weak coupling regime. In order
to clarify the point of disagreement, we first extend the approach developed
in~\cite{KainulainenProkopecSchmidtWeinstock:2001,
KainulainenProkopecSchmidtWeinstock:2002}.
We consider the dynamics of non-equilibrium Wigner functions for massive chiral
fermions which couple to a CP-violating Higgs condensate of advancing
phase fronts to first order in a gradient expansion. 
Assuming that the relevant mass parameter depends only on one
spatial coordinate, which models reasonably well the limit of
large almost planar bubble walls of a strong first order phase transition,
implies a conserved spin, and the constraint equation for the Wigner function
is solved by a spectral on-shell {\it ansatz}, thus proving 
the validity of the {\it quasiparticle picture} of the plasma. 
This result holds both in the case of one fermion
and in the case of mixing fermions. 

\vskip 0.05in

 An important difference, regarding the source calculation
in spontaneous baryogenesis and semiclassical force baryogenesis, 
is the choice of the basis in flavor space, which leads to qualitative 
differences in the resulting baryon production.
In studies of chargino-mediated baryogenesis flavor mixing is of  
crucial importance. While the proponents of spontaneous baryogenesis
argue that the weak interaction (flavor) basis (of charginos) 
is the right basis to work in, since it is favored by the 
interactions~\cite{CarenaQuirosSecoWagner:2002}, the semiclassical force camp 
contends that the mass eigenbasis has apparent advantages. 
Of course, any physical quantity must be basis independent,
but only when one includes flavor mixing in transport
equations~\cite{KonstandinProkopecSchmidtWeinstock:2004},
which -- as of yet --  has not been done.
We strongly favor the mass eigenbasis:
as we prove in section~\ref{Constraint and kinetic equations},
when working in the mass eigenbasis and to order $\hbar$ accuracy, 
the diagonal densities decouple from the off-diagonals.
The decoupling works uniquely in the mass eigenbasis.
The proof is valid provided the mass eigenvalues are not nearly degenerate,
that is when $\hbar k\cdot \partial \ll \delta(m_d^2)$ is satisfied, where 
$\delta(m_d^2)$ signifies the split in the mass eigenvalues. While this
strongly indicates that the mass eigenbasis is singled out, the definite
resolution of the controversy awaits a basis independent treatment.

\vskip 0.05in

 The paper is organized as follows. 

\vskip 0.05in

 In section~\ref{From the 2PI effective action to kinetic theory}
we review the Schwinger-Keldysh formalism, suitable for out-of-equilibrium
dynamics of field theoretical correlators, and present a derivation
of the relevant Kadanoff-Baym equations in Wigner space
for our model Lagrangean of chiral fermions coupled to scalars.
We include the case of flavor mixing in both the fermionic and
the scalar sector.

\vskip 0.05in

In section~\ref{Reduction to the on-shell limit} we study 
an on-shell reduction of the Kadanoff-Baym equations for a single scalar field.
We show that it is consistent to include the (real part of the)
self-energy and the collision term to both the constraint and kinetic 
equation, provided one truncates the self-energy
and the collision term at the same order in the coupling constant. 
This resolves one of the important -- and up-to-now unanswered -- questions 
of Boltzmann-like kinetics of relativistic systems. 
Even though we do not attempt to generalize our result
to the kinetics of fermions, it is quite plausible that the same conclusion can
be reached. In the subsequent sections we assume that this is the case.

\vskip 0.05in

A proof that no source appears in the flow term of the kinetic equations
for mixing scalars at order $\hbar$ for planar walls is presented in 
section~\ref{Kinetics of scalars: tree-level analysis}. 
We complete the proof originally presented 
in~\cite{KainulainenProkopecSchmidtWeinstock:2001}. Next, we discuss scalar
kinetic equations for the CP-conjugate scalar densities, which 
are then used in Paper~II~\cite{PSW_2}
to identify the CP-violating source in the scalar collision term.

\vskip 0.05in

Section~\ref{Kinetics of fermions: tree-level analysis} 
is devoted to the tree-level kinetics of mixing fermions in the presence 
of a CP-violating, complex mass matrix. An important application includes
chargino and neutralino mediated baryogenesis in supersymmetric extensions 
of the Minimal Standard Model. We begin by identifying a conserved quantity,
which for planar bubble walls and in the wall frame corresponds to 
the spin in the direction 
$\vec n \propto (k_xk_z,k_yk_z,k_0^2 - k_x^2 - k_y^2)$ , which {\it does not}
correspond to helicity, which has been erroneously identified as the conserved
quantity in most of the literature on semiclassical force baryogenesis.
Then we construct Wigner functions which are block diagonal in the
conserved spin. Just like in the scalar case, 
we prove that to order $\hbar$ the diagonal and off-diagonal densities 
decouple in the mass eigenbasis (we fill in a gap in the original derivation
in Ref.~\cite{KainulainenProkopecSchmidtWeinstock:2001}),
provided the mass eigenvalues
are not nearly degenerate, $\hbar k\cdot \partial \ll \delta(|m_d|^2)$.
The final part of the section contains the derivation of Boltzmann-like
transport equations for the fermion distribution functions, with
particular emphasis on the kinetics of CP-violating densities.
The relevant CP-violating sources are identified, some of which
are unaccounted for in the existing literature.
For example, we show that a CP-even deviation from equilibrium in 
the distribution function can contribute as a CP-violating source at order 
$\hbar$.

\vskip 0.05in

In Paper~II~\cite{PSW_2} we address the collision terms of the Boltzmann
equations, which contain further CP-violating sources. We furthermore
simplify the Boltzmann equations to a set of fluid equations,
based on which one can study the dynamics of CP-violating currents.


\cleardoublepage
\section{From the 2PI effective action to kinetic theory}
\label{From the 2PI effective action to kinetic theory}

\subsection{Green Functions}

The {\it in-out-formalism} of quantum field theory is well suited for
the description of particle scattering experiments, in which the system is
prepared to be in a definite {\it in-state} at $t\rightarrow-\infty$ and 
the question is what is the probability amplitude for the system to be found
in a definite {\it out-state} at $t\rightarrow+\infty$. 
In statistical physics, however, it is often of interest to follow the temporal
evolution of a system. Starting with definite initial conditions at some
time $t=t_0$, we ask for the expectation values of physical
quantities at finite times.
A theoretical framework for such problems was first suggested by Schwinger
in 1961~\cite{Schwinger:1961}
and then developed further by Keldysh~\cite{Keldysh:1964} and others.
An extension of field theory capable of dealing with non-equilibrium problems
is obtained by defining the time arguments of all quantities on a path 
$\cal C$ that leads from the initial time $t_0$ to $t$ and then back
to $t_0$, as illustrated in figure~\ref{figure-CTP}. 
All integrals and derivatives have then to be performed along
that path, and the usual time ordering becomes time ordering $T_{\cal C}$ 
along ${\cal C}$. This formalism is called Schwinger-Keldysh formalism
or Closed Time Path (CTP) formalism~\cite{ChouSuHaoYu:1985}.
\begin{figure}[tbp]
\centerline{\hspace{.in} 
\epsfig{file=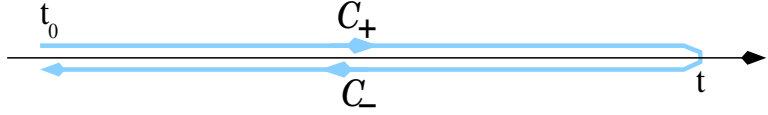, width=4.in,height=0.6in}
}
\vskip -0.1in
\caption{\small
The complex time contour for the Schwinger-Keldysh non-equilibrium formalism.
}
\lbfig{figure-CTP}
\end{figure}
The definitions of the scalar and the fermionic Green functions are
\begin{eqnarray}
  i\Delta(u,v)
  &\equiv& \langle \Omega| T_C \, \phi(u) \phi^\dagger(v) |\Omega\rangle
\label{Green_scalar_CTP}
\\
  iS_{\alpha\beta}(u,v)
  &\equiv& \langle \Omega| T_{\cal C} \, \psi_\alpha(u) \bar{\psi}_\beta(v) 
      |\Omega\rangle 
\,,
\label{Green_fermionic_CTP}
\end{eqnarray}
where $|\Omega\rangle $ denotes the physical state of the system, and
$\phi$ and $\psi$ are the scalar and fermionic fields, respectively.
The contour $\cal C$ can now be split into a ${\cal C}_+$ branch from $t_0$
to $t$ and a ${\cal C}_-$ branch from $t$ to $t_0$, as shown 
in figure~\ref{figure-CTP}. An important simplification, which will 
help us to perform the Wigner transform, instrumental for
gradient expansion, is an extension of the integration path
to $t_0\rightarrow -\infty$ and $t\rightarrow \infty$. 
This simplification is suitable for problems of our interest, which 
comprise plasmas close to chemical and thermal equilibrium with efficient
equilibration processes, as it is indeed the case with the electroweak plasma. 
In this case the influence of initial conditions can be neglected. 
Denoting the branch on which the time argument lies by an index
$a=\pm$, we can rewrite the formalism using ordinary time arguments.
We then have
\begin{eqnarray}
    \int_{\cal C} d^4u      &\rightarrow& \sum_a a \int_{-\infty}^\infty d^4u
\nonumber \\
     \delta_{\cal C}(u-v)   &\rightarrow& a \delta_{ab} \delta(u-v)     
\nonumber\\
     i\Delta(u,v)           &\rightarrow& i\Delta^{ab}(u,v)
\nonumber\\
     iS(u,v)           &\rightarrow& iS^{ab}(u,v)
\,.
\label{translation}
\end{eqnarray}
The additional factors of $a$ in the integral and in the $\delta$-function
appear since the ${\cal C}_-$ branch runs backward in time.

\vskip 0.1in

In this Keldysh formulation one usually defines the following 
four scalar Green functions:
\begin{eqnarray}
  i\Delta^{++}(u,v) &\equiv& i\Delta^t(u,v) \;
   = \langle\Omega|
        T [\phi(u) \phi^\dagger(v) ]
     |\Omega\rangle   
\nonumber\\
  i\Delta^{+-}(u,v) &\equiv& i\Delta^<(u,v)
   = \langle\Omega|
        \phi^\dagger(v) \phi(u) 
     |\Omega\rangle   
\nonumber\\
  i\Delta^{-+}(u,v) &\equiv& i\Delta^>(u,v)
   = \langle\Omega|
        \phi(u) \phi^\dagger(v) 
     |\Omega\rangle   
\nonumber\\
  i\Delta^{--}(u,v) &\equiv& i\Delta^{\bar t}(u,v) \;
   = \langle\Omega|
        \overline T [ \phi(u) \phi^\dagger(v) ] 
     |\Omega\rangle   
\,,
\label{Green_scalar_index}
\end{eqnarray}
and similarly the fermionic Green functions 
\begin{eqnarray}
  iS_{\alpha\beta}^{++}(u,v) &\equiv& iS_{\alpha\beta}^t(u,v)
   = \;\;\,  \langle\Omega|
                T [ \psi_\alpha(u) \bar{\psi}_\beta(v)]
             |\Omega\rangle   
\nonumber\\
  iS_{\alpha\beta}^{+-}(u,v) &\equiv& iS_{\alpha\beta}^<(u,v)
   = -       \langle\Omega|
                \bar{\psi}_\beta(v) \psi_\alpha(u)
             |\Omega\rangle   
\nonumber\\
  iS_{\alpha\beta}^{-+}(u,v) &\equiv& iS_{\alpha\beta}^>(u,v)
   = \;\;\,  \langle\Omega|
              \psi_\alpha(u) \bar{\psi}_\beta(v) 
             |\Omega\rangle   
\nonumber\\
  iS_{\alpha\beta}^{--}(u,v) &\equiv& iS_{\alpha\beta}^{\bar{t}}(u,v)
   = \;\;\,   \langle\Omega|
                \overline{T} [\psi_\alpha(u) \bar{\psi}_\beta(v)]
              |\Omega\rangle   
\,,
\label{Green_fermionic_index}
\end{eqnarray}
where ${T}$ ($\overline{T}$) denotes time (`anti-time') ordering
and the additional minus sign in the second fermion line is due to
the anticommutation property of the fermionic fields.

\vskip 0.1in

These definitions
imply immediately the following hermiticity properties for the Wightman
functions
\begin{eqnarray}
\label{S_herm_config}
    \left( i \Delta^{<,>}(u,v) \right)^\dagger &=&  i \Delta^{<,>}(v,u) 
\\
    \left( i\gamma^0 S^{<,>}(u,v) \right)^\dagger &=&  i\gamma^0 S^{<,>}(v,u)
\,.
\label{Delta_herm_config}
\end{eqnarray}
The four two-point functions~(\ref{Green_scalar_index})
and~(\ref{Green_fermionic_index}) are not
completely independent. Indeed, with the definition of time and anti-time
ordering one finds for the chronological (Feynman) and anti-chronological
Green functions
\begin{eqnarray}
  G^t(u,v)       &=& \theta(u_0-v_0) G^>(u,v) + \theta(v_0-u_0)G^<(u,v)
\nonumber\\
  G^{\bar{t}}(u,v) &=& \theta(u_0-v_0) G^<(u,v) + \theta(v_0-u_0)G^>(u,v)
\,,
\label{Gtt_by_G<>}
\end{eqnarray}
where we have used the generic notation $G = \{\Delta,S\}$. 
From now on we use this notation in relations which are identical
for bosonic and fermionic Green functions.

\vskip 0.1in

From the definitions of the retarded and advanced propagators
\begin{eqnarray}
  G^r &\equiv& G^t - G^< = G^> - G^{\bar t} 
\nonumber\\
  G^a &\equiv& G^t - G^>  =  G^< - G^{\bar t}
\label{G-ra}
\end{eqnarray}
and the definitions of their hermitean and antihermitean parts
\begin{eqnarray}
G_h &\equiv& \frac 12 (G^r+G^a)
\nonumber\\
G_a &\equiv& \frac {1}{2i} (G^a-G^r)
    =        \frac {i}{2} (G^>-G^<)
    \equiv   {\cal A}
\,,
\label{G:ha}
\end{eqnarray}
one obtains the spectral relation
\beq
  G_h(u,v) = -i \, {\rm sign}(u^0-v^0)\, {\cal A}(u,v)
\,,
\label{G-h}
\eeq
where ${\rm sign}(u^0-v^0)=\Theta(u^0-v^0)-\Theta(v^0-u^0)$, 
$u^\mu = (u^0,\vec u)$.
${\cal A}$ is called the spectral function.

\vskip 0.1in

\subsection{Lagrange density}
\label{Lagrange density}

We shall now consider the dynamics of fermionic and scalar particles
in the presence of a scalar condensate which may give space-time dependent
masses to both types of particles, and we shall assume 
that scalars and fermions couple {\it via} a Yukawa interaction term. 
The tree-level action is then
\begin{eqnarray}
         I[\phi,\psi] &=& \int_{\cal C} d^4 u \, {\cal L}
\label{action:tree-level}
\\
  {\cal L} &=&
              i \bar{\psi} \deldag \psi
            - \bar{\psi}_L m \psi_R - \bar{\psi}_R m^* \psi_L
            + \left(\del_\mu\phi \right)^\dagger\left( \del^\mu\phi\right)
            - \phi^\dagger M^2\phi
            + {\cal L}_{\rm int}
\,,
\label{lagrangean}
\end{eqnarray}
where the Yukawa interaction Lagrangean reads
\begin{eqnarray}
     {\cal L}_{\rm int}
 &=& -  \bar{\psi}_L  y\phi          \psi_R
     -  \bar{\psi}_R (y\phi)^\dagger \psi_L    
\nonumber\\
 &=& - \bar{\psi}\left( P_R \otimes y\phi + P_L \otimes (y\phi)^\dagger
                 \right) \psi 
\,,
\label{Yukawa}
\end{eqnarray}
and we can rewrite the fermionic mass term as
\beq
    - \bar{\psi}_L m \psi_R - \bar{\psi}_R m^* \psi_L
  = - \bar{\psi} (\mathbbm{1}\otimes m_h + i\gamma^5 \otimes m_a) \psi
\,.
\label{lag_ferm_mass}
\eeq
Fermionic fields with definite chirality $\psi_{R,L} = P_{R,L}\psi$
are defined with the help of the chirality projection operator
$P_{R,L}=(\mathbbm{1}\pm\gamma^5)/2$, and 
$\gamma^5 = i \gamma^0\gamma^1\gamma^2\gamma^3$.
Our analysis includes the possibility of particle flavor mixing, but for
notational simplicity we have suppressed scalar and fermionic flavor labels
(we shall explicitly state so when we consider the special case of the single
field dynamics without mixing.)
Thus, $\phi$ and $\psi$ are vectors in the scalar and fermionic flavor
space, respectively, and $y$ is a vector in the scalar
and a matrix in the fermionic flavor space. 
Then $\otimes$ denotes a direct product of the spinor and fermionic flavor
spaces. $M$ and $m$ are complex matrices in the scalar and fermionic flavor
space, respectively, whose non-diagonal elements couple fields of
different flavors to each other.
Even though the Lagrangean~(\ref{lagrangean}) cannot fully describe the
physics of the Standard Model, since it does not contain gauge fields,
it contains many essential elements of the physics of CP-violation
and interactions of importance for baryogenesis. 
Since the main purpose of this work
is to establish controlled calculational methods, the
Lagrangean~(\ref{lagrangean}) should be a good toy model for
studying transport, in particular for baryogenesis.
 
\vskip 0.1in

An important example of the scalar field condensate is the Higgs field 
at a first order electroweak phase transition in variants of the standard
model. This phase transition proceeds via nucleation and growth of 
the broken phase bubbles of a nonzero Higgs condensate,
which varies at the interface of the symmetric and the broken phases,
the so called bubble wall.
The phase transition bubbles quickly grow large in comparison with the
thickness of the phase transition front, so for studies of local transport
phenomena at the scale of the bubble walls one can to a good approximation
assume planar symmetry. We will explicitly make use of this symmetry in
later sections.
To allow for CP-violation we choose the fermion mass to be a complex matrix
\begin{equation}
     m(u) = m_h(u) + i m_a(u) = |m(u)|\mbox{e}^{i\theta(u)} 
\,,
\label{fermion_varying mass}
\end{equation}
where $m_h$ and $m_a$ denote the hermitean and antihermitean part of $m$,
respectively. 
The last part of this equation is to be understood as an equation
for the components.
CP-violation can be mediated either through $m_a$ or 
complex off-diagonal entries of $m_h$. 
The scalar mass matrix is by construction hermitean, 
${M^2}^\dagger = M^2$.
Nevertheless, in the case of flavor mixing CP-violation can be
mediated through the off-diagonal elements of $M^2$, provided they
are complex.

\subsection{Two-particle-irreducible (2PI) effective action}
\label{Two-particle-irreducible (2PI) effective action}

We shall now review how to derive the equation of motion
for the Green functions in the CTP-formalism. Note first that
the tree-level equation can be obtained in a straightforward manner
as follows. By varying the action~(\ref{action:tree-level})
with respect to $\phi^\dagger$, one obtains the familiar Klein-Gordon
equation
\begin{equation}
  \big( - \del^2_u  - M^2(u) \big) \phi(u) = 0 
\,.
\label{eom_scalar_field}
\end{equation}
After multiplying this from the left by $\phi^\dagger(v)$ and taking
the expectation value one gets the tree-level equation of 
motion for the Wightman function~(\ref{Green_scalar_index}):
\begin{equation}
  \big( - \del^2_u - M^2(u) \big) i\Delta^<(u,v) = 0 
\,.
\label{scalar-eom:tree-level}
\end{equation}
By an analogous procedure one finds that the same equation holds
for $i\Delta^>$. The equations of motion for 
$i\Delta^{t}$ and $i\Delta^{\bar t}$
are obtained by imposing the appropriate time ordering, such that  
they acquire on the {\it r.h.s.} the $\delta$-function sources
$i\delta^4(u-v)$ and $-i\delta^4(u-v)$, respectively. 
Similarly one finds that at tree-level the fermionic Wightman
functions~(\ref{Green_fermionic_index}) obey the Dirac equation
\begin{equation}
   \left( i\deldag_u - m_h(u) -i\gamma^5m_a(u) \right) iS^{<,>}(u,v)  = 0
,
\label{fermionic-eom:tree-level}
\end{equation}
and the equations of motion for $iS^{t}$ and $iS^{\bar t}$   
again acquire $i\delta^4(u-v)$ and $-i\delta^4(u-v)$
on the {\it r.h.s.}, respectively.

\vskip 0.1in

In order to include interactions into the equations of motion,
we use the 2PI effective action approach, which was 
originally developed for equilibrium problems in condensed matter physics
(where it is better known as $\Phi$-derivable actions)
by Luttinger and Ward~\cite{LuttingerWard:1960}
and in relativistic field theory by  
Cornwall, Jackiw and Tomboulis~\cite{CornwallJackiwTomboulis:1974} and 
then adapted to non-equilibrium situations by Chou, Su, Hao and
Yu~\cite{ChouSuHaoYu:1985}, and by Calzetta and Hu~\cite{CalzettaHu:1986}.
In this approach one uses the two-particle irreducible (2PI) out-of-equilibrium
formulation for the CTP effective action $\Gamma$, which
is a functional not only of the expectation
value of the field $\varphi(u)\equiv \langle\Omega|\phi(u)|\Omega\rangle$,
but also of the two-point functions $i\Delta(u,v)$ and $iS(u,v)$
defined in~(\ref{Green_scalar_CTP}-\ref{Green_fermionic_CTP}).
The main advantage of this formalism compared to the more standard 
1PI approach is that, at a given order in the loop expansion, varying
the 2PI action probes much more accurately the field configuration space.
Technically speaking,  at a given order in the loop expansion
the 2PI effective action typically resums a much larger set of
diagrams than the corresponding 1PI effective action. 
In absence of sources, the equations of motion are obtained simply by
extremising the effective action with respect to 
$\varphi(u)$, $\Delta(u,v)$ and $S(u,v)$:
\begin{eqnarray}
   \frac{\delta\Gamma[\varphi,\Delta,S]}{\delta\varphi} &=& 0
\nonumber\\
    \frac{\delta\Gamma[\varphi,\Delta,S]}{\delta\Delta} &=& 0
\nonumber\\
        \frac{\delta\Gamma[\varphi,\Delta,S]}{\delta S} &=& 0
\,.
\label{eom:2PI}
\end{eqnarray}
These equations are obtained by the variation of the 2PI effective
action, and hence they correspond to the Schwinger-Dyson equations,
with the self-energy approximated by the 1PI diagrams. 
Since we are here not interested in modeling the dynamics of 
the scalar field condensate, we shall not study the equation of motion 
for $\varphi$.
For our purposes we consider $\varphi$ as given
and simply absorb the effects of $\varphi$ into space-time dependent
mass terms $m$ and $M^2$. The quantum dynamics of a self-interacting
scalar field in the presence of a condensate describing the inflaton decay 
is considered in the framework of the 2PI effective action approach
for example in~\cite{BergesSerreau:2002,
AartsAhrensmeierBaierBergesSerreau:2002,CalzettaHu:2002},
while the equivalent problem for classical scalar fields is studied
in~\cite{KhlebnikovTkachev:1996,ProkopecRoos:1996}.
Various aspects of the out-of-equilibrium dynamics and thermalization
of quantum fields are considered 
in~\cite{Berges:2002,BergesBorsanyiSerreau:2002,AartsBerges:2001,BergesCox:2000}.
Here we develop a gradient approximation for the two-point function,
starting with the 2PI effective action, which is appropriate for studying
the dynamics of fields in the presence of slowly varying backgrounds.

\vskip 0.1in

\begin{figure}[tbp]
\centerline{\hspace{.in} 
\epsfig{file=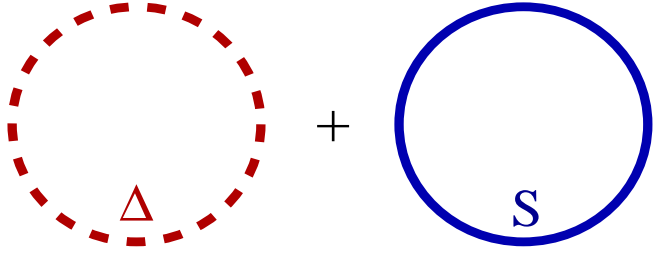, width=3.4in,height=1.3in}
}
\vskip -0.1in
\caption{\small
The one-loop diagrams contributing to the 2PI effective
action~(\ref{2PI_effective_action}).
}
\lbfig{figure:2pi-1loop}
\end{figure}
The 2PI effective action corresponding to the classical
action~(\ref{action:tree-level})
can be written in the form~\cite{CornwallJackiwTomboulis:1974}
\begin{equation}
  \Gamma[\Delta,S] =  i \Tr ({\Delta^{(0)}}^{-1}\!\Delta)
                   -  i \Tr ({S^{(0)}}^{-1}\!S)
                   +  i \Tr \ln \Delta^{-1}
                   -  i \Tr \ln S^{-1}
                   +  \Gamma_2[\Delta,S]
\,,
\label{2PI_effective_action}
\end{equation}
where the minus signs in the fermionic terms are related to Pauli statistics.
We use a condensed notation, with $\Tr$ denoting both integration
over space-time and summation over spinor and flavor indices.
The first two terms in~(\ref{2PI_effective_action}) are the classical
(tree-level) actions,
the next two terms correspond to the one-loop vacuum diagrams
shown in figure~\ref{figure:2pi-1loop}, while the last term
$\Gamma_2$ stands for the sum of all two-particle irreducible vacuum graphs.
In Paper~II we illustrate how to compute $\Gamma_2$ in practice.
The inverse free propagators ${\Delta^{(0)}}^{-1}$ and ${S^{(0)}}^{-1}$
can be read off from the classical
action~(\ref{action:tree-level}-\ref{lagrangean}) rewritten as
\beq
    I[\phi,\psi]
  =     \int_{\cal C} d^4u \, d^4v \,
              \phi^\dagger(u) {\Delta^{(0)}}^{-1}(u,v) \phi(v)
     +  \int_{\cal C} d^4u \, d^4v \,
              \bar\psi(u) {S^{(0)}}^{-1}(u,v) \psi(v)
     +  \int_{\cal C} d^4 u {\cal L}_{\rm int}
\,.
\label{scalar_class_eff_action}
\eeq
They satisfy
equations~(\ref{scalar-eom:tree-level}-\ref{fermionic-eom:tree-level})
and are given by
\begin{eqnarray}
{\Delta^{(0)}}^{-1}(u,v) &=& \big(\!-\del^2_u-M^2(u)\big)
                              \delta^4_{\cal C}(u-v)
\nonumber\\
  {S^{(0)}}^{-1}(u,v)    &=& \left( i\deldag_u - m_h - i\gamma^5 m_a \right)
                                \delta_{\cal C}^4(u-v)
\,.
\label{inverse_free_props}
\end{eqnarray}
We now take the functional derivatives of the effective
action~(\ref{2PI_effective_action}) with respect to $\Delta$ and $S$
to obtain
\begin{eqnarray}
    \frac{\delta\Gamma[\Delta,S]}{\delta \Delta(v,u)}
        &=& i {\Delta^{(0)}}^{-1}(u,v)
         -  i \Delta^{-1}(u,v)
         +    \frac{\delta\Gamma_2[\Delta,S]}{\delta \Delta(v,u)}
         = 0
\nonumber\\
    \frac{\delta\Gamma[\Delta,S]}{\delta S(v,u)}
        &=& - i {S^{(0)}}^{-1}(u,v)
            + i S^{-1}(u,v)
            +   \frac{\delta\Gamma_2[\Delta,S]}{\delta S(v,u)}
         = 0
\,.
\label{functional-derivative_of_Gamma}
\end{eqnarray}
By making use of the definitions for the scalar and fermionic self-energies
\begin{eqnarray}
  \Pi(u,v)    &\equiv& \;\;\, i \frac{\delta\Gamma_2[\Delta,S]}{\delta \Delta(v,u)}
\label{Pi}
\\
  \Sigma(u,v) &\equiv&    -   i \frac{\delta\Gamma_2[\Delta,S]}{\delta S(v,u)}
\label{Sigma}
\end{eqnarray}
and multiplying from the right by $\Delta$ and $-S$, respectively,
we can recast~(\ref{functional-derivative_of_Gamma}) as the
Schwinger-Dyson equations
\begin{eqnarray}
    \big(-\del^2_u - M^2(u) \big) i\Delta(u,v)
         &=& i\delta_{\cal C}^4(u-v) 
          +  \int_{\cal C} d^4w \, \Pi(u,w) i\Delta(w,v)
\label{eom_contour:scalar}
\\
 \big( i\deldag_u - m_h(u) - i\gamma^5 m_a(u) \big) iS(u,v)
         &=& i\delta_{\cal C}^4(u-v) 
          +  \int_{\cal C} d^4w \, \Sigma(u,w) iS(w,v)
\,.
\label{eom_contour:fermionic}
\end{eqnarray}

\vskip 0.1in

So far we have written everything in the complex contour notation. 
In the index notation the self-energies are
\begin{eqnarray}
  \Pi^{ab}(u,v)
     &=& \;\;\, iab \frac{\delta\Gamma_2[\Delta,S]}{\delta \Delta^{ba}(v,u)} 
\label{Pi:ab}
\\
  \Sigma^{ab}(u,v)
     &=& - iab \frac{\delta\Gamma_2[\Delta,S]}{\delta S^{ba}(v,u)} 
\,,
\label{Sigma:ab}
\end{eqnarray}
and the equations of 
motion~(\ref{eom_contour:scalar}-\ref{eom_contour:fermionic}) read
\begin{eqnarray}
    \big(-\del^2_u - M^2(u) \big) i\Delta^{ab}(u,v)
  &=& ai\delta_{ab}\delta^4(u-v) 
   +    \sum_c c \int d^4w \, \Pi^{ac}(u,w) i\Delta^{cb}(w,v)
\label{eom_index:scalar}
\\
    \big(i\deldag_u - m_h(u) - i\gamma^5 m_a(u) \big) iS^{ab}(u,v)
  &=& ai\delta_{ab}\delta^4(u-v) 
    + \sum_c c \int d^4w \, \Sigma^{ac}(u,w) iS^{cb}(w,v)
\,.
\label{eom_index:fermionic}
\end{eqnarray}
These are the fundamental quantum dynamical equations corresponding 
to the action~(\ref{action:tree-level}-\ref{lagrangean}).
They look deceptively simple, since the full complexity of the problem
is hidden in the self-energies
, which are 
very complicated functionals of the Green functions and in general not known
completely.

\vskip 0.1in

From Eqs.~(\ref{Green_scalar_index}-\ref{Green_fermionic_index})
and~(\ref{eom_index:scalar}-\ref{eom_index:fermionic}) we infer the
bosonic equations
\begin{eqnarray}
(-\partial^2-M^2) \Delta^{r,a}
  -  \Pi^{r,a} \odot \Delta^{r,a}   &=& \delta 
\label{Deltaeom-ra}
\\
(-\partial^2-M^2) \Delta^{<,>}
  - \Pi^r \odot \Delta^{<,>}   &=& \Pi^{<,>} \odot \Delta^a
\,,
\label{Deltaeom-<>}
\end{eqnarray}
and the fermionic equations 
\begin{eqnarray}
(i\partial\!\!\!/-m_h - i\gamma^5 m_a) S^{r,a}
  -  \Sigma^{r,a}\odot S^{r,a}   &=& \delta
\label{Seom-ra}
\\
(i\partial\!\!\!/-m_h - i\gamma^5 m_a) S^{<,>}
  - \Sigma^r \odot S^{<,>}   &=& \Sigma^{<,>}\odot S^a
\,,
\label{Seom-<>}
\end{eqnarray}
where $\odot$ stands for integration over the intermediate variable.
We use the notation 
$\Pi^t \equiv \Pi^{++}$, {\it etc.} 
({\it cf.} Eqs.~(\ref{Green_scalar_index}-\ref{Green_fermionic_index}))
and have defined retarded and advanced self-energies
in analogy with~(\ref{G-ra}).
As we have already pointed out, not all Green functions are independent:
$G^r$ and $G^a$ can be expressed in terms of $G^<$ and $G^>$
({\it cf.} Eqs.~(\ref{Gtt_by_G<>}-\ref{G-ra})), implying that 
the equations for the retarded and advanced propagators are redundant.
It is not difficult to show that Eqs.~(\ref{Deltaeom-ra}-\ref{Seom-<>})
are consistent provided the relations
\begin{eqnarray}
  \Pi^t(u,v)         &=& \delta^4(u-v)\Pi^{\tt sg}(u) 
                      +  \theta(u_0-v_0) \Pi^>(u,v) 
                      +  \theta(v_0-u_0) \Pi^<(u,v)
\nonumber
\\
  \Pi^{\bar{t}}(u,v) &=& \delta^4(u-v)\Pi^{\tt sg}(u) 
                      +  \theta(u_0-v_0) \Pi^<(u,v) 
                      +  \theta(v_0-u_0) \Pi^>(u,v)
\,,
\label{Pit-Pi<>}
\end{eqnarray}
as well as corresponding relations for the fermionic self-energies,
are satisfied.
This should be the case for any reasonable approximation of the
self-energies. The additional singular terms $\Pi^{\tt sg}(u)$ and
$\Sigma^{\tt sg}(u)$ we have allowed for appear naturally in some
theories and can be absorbed by the mass terms (mass renormalization). 
An example of a singular self-energy contribution is the one-loop
tadpole of a scalar theory with quartic self-interaction. The one-loop
expressions for the self-energies that we will derive in
Paper~II
indeed satisfy~(\ref{Pit-Pi<>}) and the corresponding fermionic equations.

\vskip 0.1in

\begin{figure}[tbp]
\begin{center}
\epsfig{file=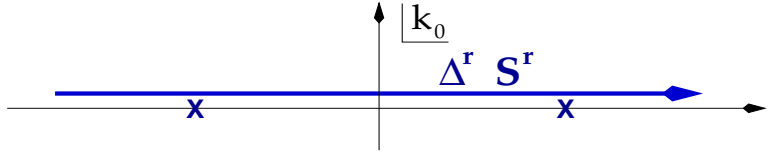, height=0.8in,width=4.in}
\end{center}
\vskip -0.3in
\lbfig{figure:retarded-contour}
\caption{%
\small
Integration contour for the retarded 
propagators $\Delta^r$ and $S^r$ in~(\ref{Delta-ra:free}-\ref{S-ra:free}).
}
\end{figure}
\begin{figure}[tbp]
\begin{center}
\epsfig{file=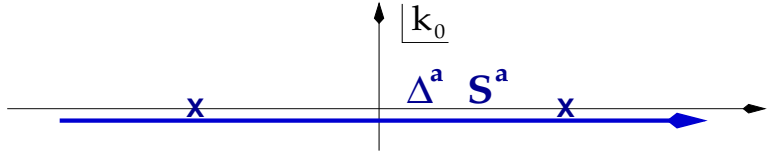, height=0.8in,width=4.in}
\end{center}
\vskip -0.3in
\lbfig{figure:advanced-contour}
\caption{%
\small
Integration contour for the advanced propagators
$\Delta^a$ and $S^a$ in~(\ref{Delta-ra:free}-\ref{S-ra:free}).
}
\end{figure}
To clarify the physical meaning of 
Eqs.~(\ref{Deltaeom-ra}-\ref{Seom-<>}), we observe that 
the equations for the retarded and advanced 
propagators~(\ref{Deltaeom-ra}) and~(\ref{Seom-ra}) describe mostly 
the spectral properties of the system. Indeed, this can be seen for example
from the tree-level solutions, which in the limit of constant mass terms and 
no flavor mixing read
\begin{eqnarray}
  \Delta_{\tt 0}^{r,a}(u,v) 
          &=& \int \frac{d^4k}{(2\pi)^4}{\rm e}^{-ik\cdot(u-v)}
                   \frac{1}{k^2 - M^2 \pm i{\rm sign}(k_0)\epsilon}
\label{Delta-ra:free}
\\
  S^{r,a}_{\tt 0}(u,v) &=& \int \frac{d^4k}{(2\pi)^4}{\rm e}^{-ik\cdot(u-v)}
                            \frac{k\!\!\!/ + m_h - i\gamma^5 m_a}
                             {k^2 - m_h^2 - m_a^2 \pm i{\rm sign}(k_0)\epsilon}
\,,
\label{S-ra:free}
\end{eqnarray}
where we use the infinitesimal pole shifts $\mp i\epsilon$ 
($\epsilon \rightarrow 0+$)
to indicate the standard integration prescription for the retarded
and advanced propagators, respectively, shown in
figures~\ref{figure:retarded-contour} 
and~\ref{figure:advanced-contour}.
The spectral functions~(\ref{G:ha}) are then
\begin{eqnarray}
    {\cal A}_{\phi\tt 0} &=& \int \frac{d^4k}{(2\pi)^4}{\rm e}^{-ik\cdot(u-v)}
                              \pi {\rm sign}(k_0)\delta(k^2 - M^2)
\label{Aphi:free}
\\
    {\cal A}_{\psi\tt 0} &=& \int \frac{d^4k}{(2\pi)^4}{\rm e}^{-ik\cdot(u-v)}
                              (k\!\!\!/ + m_h - i\gamma^5 m_a)
                              \pi {\rm sign}(k_0)\delta(k^2 - m_h^2 - m_a^2)
\,.
\label{Apsi:free}
\end{eqnarray}
The $\delta$-functions
indicate 
that the frequencies of the plasma excitations are constrained to lie on the 
mass shell, which for scalars is given by
\begin{equation}  
    k_0 = \pm \omega_\phi
\,,\qquad 
   \omega_\phi = \sqrt{\vec k^2 + M^2}
\label{mass-shell:free-scalars}
\end{equation}
and for fermions by
\begin{equation}  
\quad
      k_0 = \pm \omega_0 \,,
\qquad
    \omega_0 = \sqrt{\vec k^2 + m_h^2 + m_a^2}
\,.
\label{mass-shell:free-fermions}
\end{equation}

\vskip 0.1in

On the other hand, the equations of motion for the Wightman
functions~(\ref{Deltaeom-<>}) and~(\ref{Seom-<>})
describe mostly statistical (kinetic) properties of the system. 
This can be seen for example from considering the thermal equilibrium 
solutions to these equations, which are given in
section~\ref{Equilibrium Green functions} below.
The kinetic equations~(\ref{Deltaeom-<>}) and~(\ref{Seom-<>}) 
can be rewritten in the form of the Kadanoff-Baym (KB) equations
\begin{eqnarray}
(-\partial^2-M^2) \Delta^{<,>} 
    -     \Pi_h\odot \Delta^{<,>}
    -     \Pi^{<,>}\odot \Delta_h 
   &=&    {\cal C}_\phi 
  \equiv  \frac 12 (\Pi^> \odot \Delta^< - \Pi^< \odot \Delta^>)
\label{Deltaeom-<>b}
\\
(i\partial\!\!\!/ - m_h - i\gamma^5 m_a)S^{<,>} 
    -     \Sigma_h\odot S^{<,>}
    -     \Sigma^{<,>}\odot S_h &=& {\cal C}_\psi 
  \equiv  \frac 12 ( \Sigma^> \odot S^< - \Sigma^< \odot S^>)
\,,
\label{Seom-<>b}
\end{eqnarray}
where ${\cal C}_\phi$ and ${\cal C}_\psi$ denote the scalar and fermionic
collision terms, respectively, and the hermitean parts of the self-energies
are defined analogously to those for the Wightman functions (\ref{G:ha}).
With reasonable approximations for the self-energies, the Kadanoff-Baym 
equations
are suited to study kinetics of out-of-equilibrium quantum systems.
The terms $\Pi_h\Delta^<$ and $\Sigma_hS^<$ on the {\it l.h.s.} 
of~(\ref{Deltaeom-<>b}-\ref{Seom-<>b}) represent the self-energy contributions,
which are sometimes considered as a (nonlocal) contribution to the mass terms.
We postpone the discussion of how inclusion of the self-energy may change
the plasma dynamics and in particular the quasiparticle picture
to a future work.
The collision terms ${\cal C}_\phi$ and ${\cal C}_\psi$,
as we will see in Paper~II,
contain the standard gain and loss terms responsible for equilibration,
but they may also contain additional CP-violating sources.
Finally, the terms $\Pi^<\Delta_h$ and $\Sigma^<S_h$ induce broadening of
the on-shell dispersion relations and may cause breakdown of the quasiparticle
picture. We shall consider the role of these terms in some detail
in section~\ref{Reduction to the on-shell limit}.
There we will need the width
\beq
\Gamma_\phi = \frac{1}{2i} (\Pi^a-\Pi^r) = \frac i2 (\Pi^>-\Pi^<)
\label{Gammapsiphi}
\eeq
and the hermitean part of the scalar self-energy
\beq
\Pi_h = \frac 12 (\Pi^r+\Pi^a) 
          = \Pi^t - \frac 12 (\Pi^>+\Pi^<)
\,,
\label{DeltaS_h}
\eeq
which satisfy the spectral relation ({\it cf.} Eq.~(\ref{G-h}))
\beq
\Pi_h(u,v) =  
       \frac 12 {\rm sign}(u^0-v^0)\,[\Pi^{>}(u,v)-\Pi^{<}(u,v)]
       = -i {\rm sign}(u^0-v^0)\, \Gamma_\phi(u,v)
\,.
\label{Sigma-h}
\eeq
Analogous relations hold for the fermionic self-energies.

\subsection{Wigner representation and gradient expansion}
\label{Wigner representation and gradient expansion}

In equilibrium the Green functions $\Delta(u,v)$ and $S(u,v)$
depend only on the relative (microscopic) coordinate $r = u-v$. 
This dependence corresponds to the internal fluctuations
that typically take place on microscopic scales.
In non-equilibrium situations, however, $\Delta$ and $S$
depend also on the average (macroscopic) coordinate $x = (u+v)/2$.
This dependence describes the system's behavior on large, macroscopic scales.
For example, if the system couples to an external field which varies in space 
and time, the Green functions will show a corresponding dependence
on the average coordinate.
Thick bubble walls at a first order electroweak phase transition
represent precisely an example of such an external field. 

\vskip 0.1in

In order to separate the microscopic scale fluctuations from the behavior
on macroscopical scales one typically performs a Wigner transform, which 
is a Fourier transform with respect to the relative coordinate $r$.
The Green function in the Wigner representation, called Wigner function, is
\begin{equation}
  G(k,x) = \int d^4r \, {\rm e}^{ik\cdot r} \, G( x+r/2 , x-r/2 ) 
\,.
\label{G_Wigner}
\end{equation}
In the Wigner representation the hermiticity properties (\ref{S_herm_config}) 
and~(\ref{Delta_herm_config}) become simply
\beqa
  \big( i\Delta^{<,>}(k,x) \big)^\dagger       &=&  i\Delta^{<,>}(k,x)
\label{herm_scal_Wigner}
\\
  \big( i\gamma^0S^{<,>}(k,x) \big)^\dagger &=&  i\gamma^0S^{<,>}(k,x)
\,.
\label{herm_ferm_Wigner}  
\eeqa
In order to transform the equations of motion, we make use of
the general relation
\begin{equation}
  \int d^4(u-v) \, {\rm e}^{ik\cdot(u-v)}
  \int d^4w A(u,w) B(w,v) 
  = {\rm e}^{-i\diamond}\{A(k,x)\}\{B(k,x)\} \,,
\label{convolution_Wigner}
\end{equation}
where $x$ is the average coordinate and the diamond operator
is defined by
\begin{equation}
     \diamond \left\{\cdot\right\} \left\{\cdot\right\}
   = \frac{1}{2} \big(   \del^{(1)}\cdot\del^{(2)}_k
                        - \del^{(1)}_k\cdot\del^{(2)} 
                  \big)\{\cdot\}\{\cdot\} 
\,.
\label{diamond}
\end{equation}
The superscripts $(1)$ and $(2)$ refer to the first and second argument,
respectively, and $\del \equiv \del_x$.
Since the nonlocal terms have the form of~(\ref{convolution_Wigner}), 
it is a simple exercise to show that 
the Kadanoff-Baym equations~(\ref{Deltaeom-<>b}-\ref{Seom-<>b}),
when written in the Wigner representation, become
\begin{eqnarray}
  \Big(  k^2 - \frac{1}{4} \del^2 + ik\cdot\del 
       - M^2{\rm e}^{-\frac{i}{2}\stackrel{\leftarrow}{\del}\cdot\del_k}
  \Big)\Delta^{<,>}
       - {\rm e}^{-i\diamond}\{\Pi_h\}\{ \Delta^{<,>}\}
       - {\rm e}^{-i\diamond}\{\Pi^{<,>}\}\{\Delta_h\}
   =     {\cal C}_\phi
\label{Wigner-space:scalar_eom}
\\
  \Big(     \kdag
          + \frac i2 \deldag 
          - m_h
            {\rm e}^{-\frac{i}{2}\stackrel{\leftarrow}{\del}\cdot\del_k}
          - i\gamma^5m_a
            {\rm e}^{-\frac{i}{2}\stackrel{\leftarrow}{\del}\cdot\del_k}
  \Big)S^{<,>}
       -   {\rm e}^{-i\diamond}\{\Sigma_h\}\{ S^{<,>}\}
       -   {\rm e}^{-i\diamond}\{\Sigma^{<,>}\}\{ S_h\}
   =       {\cal C}_\psi
\,,
\label{Wigner-space:fermionic_eom}
\end{eqnarray}
where the collision terms are
\begin{eqnarray}
  {\cal C}_\phi &=&  \frac 12 {\rm e}^{-i\diamond}
                     \big( \{\Pi^>\} \{\Delta^<\} 
                         - \{\Pi^<\} \{\Delta^>\} 
                     \big)
\label{Cphi}
\\
  {\cal C}_\psi &=&  \frac 12 {\rm e}^{-i\diamond}
                     \big( \{\Sigma^>\}\{S^<\} 
                         - \{\Sigma^<\}\{S^>\}
                     \big)
\,.
\label{Cpsi}
\end{eqnarray}
We shall analyze these equations in gradient expansion:
we assume that the variation of the background field,
and therefore also the variation of the Green functions and self-energies,
with respect to the macroscopic coordinate $x$ is small when compared with
the momenta of plasma excitations. Then we can perform an expansion
in derivatives with respect to $x$.
Formally, this criterion reads
\begin{equation}
     \del \ll k
. 
\label{gradient-expansion:criterion}
\end{equation}
That is, the background field is assumed to have a characteristic length scale
that is large in comparison to the de~Broglie wavelength of the particles
in the plasma. 

\vskip 0.1in

This assumption of a slowly varying
background field seems to be justified in our case~\cite{ThickWalls}:
in the MSSM, for instance,
the width of the bubble wall $L_w$ is roughly $10/T$, where $T$ is the 
temperature of the plasma. The typical momentum of a particle in the plasma
is of the order $T$, so that the de~Broglie wavelength $l_{dB} \sim 1/T$ 
is indeed small when compared to $L_w$.
Since the expansion in powers of derivatives can be viewed as an expansion in
powers of the Planck constant $\hbar$, the gradient expansion is equivalent to
a semiclassical expansion. We expect that the leading order terms in the
gradient expansion correspond to classical kinetic equations, 
while higher order derivatives represent quantum corrections.

\subsection{Sum rules and Equilibrium Green functions}
\label{Equilibrium Green functions}

It is easily shown that, as a consequence of the equal-time
commutation relation for scalars, the Wigner transform of their
spectral function ${\cal A}_\phi$ (\ref{G:ha})
obeys the spectral sum rule
\begin{equation}
 \int_{-\infty}^\infty \frac{dk_0}{\pi}k_0 {\cal A}_\phi(k,x) = 1
\,,
\label{sum_rule_scalars}
\end{equation}
while the anticommutation rule for fermions leads to
\begin{equation}
 \int_{-\infty}^\infty \frac{dk_0}{\pi} \gamma^0 {\cal A}_\psi(k,x)
            = \mathbbm{1}
\label{sum_rule_fermions}
\end{equation}
for the fermionic spectral function.
These sum rules are important, since they have to be imposed as 
consistency conditions on the solutions for the Wigner functions
$G^<$ and $G^>$, or for $G^a$ and $G^r$, as indicated in~(\ref{G:ha}). 
Furthermore, making use of these spectral sum rules
and the Kubo-Martin-Schwinger (KMS) relations
\beqa
  \Delta^>_{\rm eq}(k) &=& \;\;\, {\rm e}^{\beta k_0} \Delta^<_{\rm eq}(k) 
\nonumber\\
       S^>_{\rm eq}(k) &=& -{\rm e}^{\beta k_0} S^<_{\rm eq}(k)
\,,
\label{KMS}
\eeqa
where $\beta=1/T$ denotes the inverse temperature,
one can unambiguously obtain the thermal equilibrium
propagators~\cite{LeBellac:1996}. 
They are translationally invariant in space and time
and hence, when written in the Fourier space, defined by the Fourier
transform with respect to the relative coordinate $r = u-v$, 
they display a dependence on the momentum only: 
\begin{equation}
  G_{\rm eq}(k) = \int d^4r \, {\rm e}^{ik\cdot r}G_{\rm eq}(r)
\,.
\label{Green_fermionic_Fourier}
\end{equation}
The bosonic and fermionic equilibrium propagators are the solutions of 
the Klein-Gordon~(\ref{scalar-eom:tree-level}) and the Dirac
equation~(\ref{fermionic-eom:tree-level}), respectively, which in the 
Fourier space read
\begin{eqnarray}
                \big( k^2 - M^2 \big) i\Delta_{\rm eq}^{<,>}(k) &=& 0 
\,,\qquad\qquad\quad\;\,
                \big( k^2 - M^2 \big) i\Delta_{\rm eq}^{r,a}(k)  = i
\nonumber
\\
   \left( \kdag - m_h -i\gamma^5m_a \right) iS_{\rm eq}^{<,>}(k)  &=& 0
\,,\qquad
   \left( \kdag - m_h -i\gamma^5m_a \right) iS_{\rm eq}^{r,a}(k)   = i
\,,
\label{kspace:fermionic-bosonic:tree-level}
\end{eqnarray}
with the normalization~(\ref{sum_rule_scalars}-\ref{sum_rule_fermions})
and the boundary conditions~(\ref{KMS}). 
The scalar equilibrium Green functions are
\begin{eqnarray}
 i\Delta_{\rm eq}^t(k) &=& \;\;\, 
                           \frac{i}{k^2-M^2+i{\rm sign}(k_0)\epsilon}
                        +  2\pi \delta(k^2-M^2) {\rm sign}(k_0) 
                             n^\phi_{\rm eq}(k_0)
\label{Green_scalar_eq_t}
\\
 i\Delta_{\rm eq}^{\bar{t}}(k) &=& -\frac{i}{k^2-M^2+i{\rm sign}(k_0)\epsilon}
                                +   2\pi \delta(k^2-M^2) {\rm sign}(k_0)
                                       (1 + n^\phi_{\rm eq}(k_0))
\label{Green_scalaw_eq_tbar}\\
 i\Delta_{\rm eq}^<(k) &=& \;\; 2\pi \delta(k^2-M^2) {\rm sign}(k_0)
                                 n^\phi_{\rm eq}(k_0)
\label{Green_scalar_eq_<}
\\
 i\Delta_{\rm eq}^>(k) &=& \;\; 2\pi \delta(k^2-M^2) {\rm sign}(k_0)
                                 (1+ n^\phi_{\rm eq}(k_0))
\,,
\label{Green_scalar_eq_>} 
\end{eqnarray}
where the Bose-Einstein distribution appears:
\begin{equation}
  n^\phi_{\rm eq}(k_0) = \frac{1}{{\rm e}^{\beta k_0} - 1}
\,.
\label{BoseEinstein}
\end{equation}
Similarly, for the fermions we have 
\begin{eqnarray}
  iS_{\rm eq}^t(k) &=& \;\;\,\frac{i(\kdag + m_h - i\gamma^5m_a)}
                            {k^2 - |m|^2 + i{\rm sign}(k_0)\epsilon}
                    -  2\pi (\kdag +  m_h - i\gamma^5m_a) 
                             \delta(k^2-|m|^2)
                             {\rm sign}(k_0) n_{\rm eq}(k_0)
\label{Green_fermionic_eq_t}\\
  iS_{\rm eq}^{\bar{t}}(k)
                   &=&  -\frac{i(\kdag + m_h - i\gamma^5m_a)}
                             {k^2 - |m|^2 + i{\rm sign}(k_0)\epsilon}
                    + 2\pi  (\kdag + m_h - i\gamma^5m_a) 
                             \delta(k^2 - |m|^2) {\rm sign}(k_0)
                             (1- n_{\rm eq}(k_0)) 
\qquad\,
\label{Green_fermionic_eq_tbar}
\\
 iS_{\rm eq}^<(k)  &=&  -2\pi (\kdag+ m_h - i\gamma^5m_a)
                                \delta(k^2 - |m|^2) {\rm sign}(k_0) 
                                n_{\rm eq}(k_0)
\label{Green_fermionic_eq_<}
\\
 iS_{\rm eq}^>(k)  &=&  \;\;\, 2\pi (\kdag + m_h - i\gamma^5m_a) 
                                \delta(k^2 - |m|^2) {\rm sign}(k_0)
                                (1 - n_{\rm eq}(k_0))
\,,
\label{Green_fermionic_eq_>} 
\end{eqnarray}
with the Fermi-Dirac distribution function
\begin{equation}
  n_{\rm eq}(k_0) = \frac{1}{{\rm e}^{\beta k_0} + 1}
\label{FermiDirac}
\end{equation}
and $ |m|^2 \equiv  m_h^2 + m_a^2$.
For the case of several mixing particle species, the equilibrium Green
functions are diagonal matrices in flavor space, and the given relations hold
for the diagonal elements.


\cleardoublepage
\section{Reduction to the on-shell limit}
\label{Reduction to the on-shell limit}

In this section~\footnote{Based on work with Kimmo Kainulainen.}
 we pay particular attention to the self-consistency
of the on-shell limit as a controlled expansion in the coupling constants of
the theory. To this aim we consider the role of the dissipative terms
on the {\it l.h.s.}
of the propagator equations~(\ref{Deltaeom-ra})
and the Kadanoff-Baym
equation~(\ref{Wigner-space:scalar_eom}).
We shall see that including these terms into the retarded and
advanced propagators results in a Breit-Wigner form for the spectral function.
By making use of the spectral decomposition ansatz for the Wigner function
with the Breit-Wigner form for the spectral function, we shall study 
under which conditions the kinetic equation reduces to the on-shell form.

\vskip 0.1in

For simplicity, we consider here the scalar case.
It is quite plausible that similar techniques can be used to extend
the analysis presented in this section to the case of fermions.
Let us begin by rewriting  equations~(\ref{Deltaeom-ra})
and~(\ref{Wigner-space:scalar_eom}) more compactly,
\begin{eqnarray}
 {\rm e}^{-i\diamond}\{\Omega_\phi^2 \pm i\Gamma_\phi\}\{\Delta^{r,a}\} &=& 1
\label{Deltaeom-ra2}
\\
 {\rm e}^{-i\diamond}\{\Omega_\phi^2\}\{\Delta^{<,>}\} 
       -   {\rm e}^{-i\diamond}\{\Pi^{<,>}\}\{\Delta_h\}
       &=&   {\cal C}_\phi  
\,,
\label{Wigner-space:scalar:compact}
\end{eqnarray}
where ${\cal C}_\phi$ is defined in Eq.~(\ref{Cphi}), and we defined
\begin{eqnarray}
  \Omega_\phi^2 &=&  k^2  -  M^2 - \Pi_h
\,.
\label{Omegas}
\end{eqnarray}
Upon adding and subtracting the complex conjugate
of~(\ref{Deltaeom-ra2}), we arrive at the following 
propagator equations for the case of a single scalar
field up  to second order in gradients
\begin{eqnarray}
  \cos\diamond \DBR{\Omega_\phi^2\pm i\Gamma_\phi}{\Delta^{r,a}} &=& 1 
\label{spec1}
\\
  \sin\diamond \DBR{\Omega_\phi^2\pm i\Gamma_\phi}{\Delta^{r,a}} &=& 0
,
\label{spec2}
\end{eqnarray}
where we used the fact that $\Delta^{r,a}$ commute with  
$\Omega_\phi^2$ and $\Gamma_\phi$. This commutation property
is special to the single field case however. Indeed, in the case of mixing 
scalars, $\Delta^{r,a}$ are matrices which in general
do not commute with $\Omega_\phi^2$ and $\Gamma_\phi$. 

\vskip 0.1in

Now observe that Eq.\ (\ref{spec2}) can be obtained from Eq.\
(\ref{spec1}) by
an application of the differential operator $\tan\diamond$, and hence it gives
no new information.  Because $\cos\diamond$ is an even function, we then see
that the corrections to the propagator equation appear first time
at the second order in gradients~\cite{Henning:1995}. 
Therefore, to the first order in gradients, we have simply
\begin{equation}
\Delta^{r,a} = (\Omega_\phi^2 \pm i\Gamma_\phi)^{-1}
,
\label{Delta:ra1}
\end{equation}
or equivalently ({\it cf.}~(\ref{G:ha}))
\begin{eqnarray}
\Delta_h &=& \frac{\Omega_\phi^2}
             {\Omega_\phi^4 + \Gamma_\phi^2}
\label{Delta_h}
\\
{\cal A}_\phi &=& \frac{\Gamma_\phi}
             {\Omega_\phi^4 + \Gamma_\phi^2}.
\label{A_phi}
\end{eqnarray}
The kinetic and constraint equations can be obtained from the
Kadanoff-Baym equations~(\ref{Wigner-space:scalar:compact})
as the hermitean and antihermitean parts,  
\begin{eqnarray}
 - \diamond\DBR{\Omega^2_\phi}{i\Delta^{<,>}} 
 +  \diamond\DBR{i\Pi^{<,>}}{\Delta_h}
    &=&  \frac{1}{2} \big(\Pi^> \Delta^<- \Pi^< \Delta^>\big) .
\label{ke:scalar}
\\
 \Omega_\phi^2 \Delta^{<,>}
   + \frac{i}{2} \diamond\big(\DBR{\Pi^>}{\Delta^<}
                             -\DBR{\Pi^<}{\Delta^>}\big)
   &=& \Pi^< \Delta_h
,
\label{ce:scalar}
\end{eqnarray}
where we truncated to second and first order in gradients, respectively.
Because of the hermiticity property~(\ref{herm_scal_Wigner}) 
of the scalar Wigner function, both kinetic and constraint equations must 
be satisfied simultaneously. The constraint equation selects the physical
set of solutions among all solutions of the kinetic equation.

\vskip 0.1in

Note that the equations (\ref{Delta_h}-\ref{A_phi}) are only formal solutions
in the sense that $\Pi_h$ and $\Gamma_\phi$ are functionals
of $i\Delta^{<,>}$ and $iS^{<,>}$.
Consequently, equations~(\ref{ke:scalar}-\ref{ce:scalar}) are 
complicated integro-differential equations. 
However, in many cases deviations from equilibrium are small, 
so that one can linearize them
in deviation from equilibrium, such that $\Pi_h$ and $\Gamma_\phi$ 
can be computed by using the equilibrium distribution 
functions~(\ref{Green_scalar_eq_t}-\ref{FermiDirac}).

\vskip 0.1in

Inspired by the spectral form of the equilibrium solutions,
we shall assume that close to equilibrium
the following spectral decomposition holds
\begin{equation}
     i\Delta^< = 2{\cal A}_\phi n_\phi
,\qquad\quad
     i\Delta^> = 2{\cal A}_\phi (1+ n_\phi)
\,,
\label{spectral-decomposition:scalar}
\end{equation}
where $n_\phi=n_\phi(k,x)$ is some unknown function representing 
a generalized distribution function on phase space. 
Similarly, for the self-energies, we can formally write
\begin{equation}
i\Pi^< \equiv 2\Gamma_\phi n_{\Pi}
,\qquad
i\Pi^> \equiv 2\Gamma_\phi (1+n_{\Pi}).
\label{spectral-decomposition:Pi}
\end{equation}
We have adopted a somewhat nonstandard definition for $\Gamma_\phi$;
the more standard definition is obtained by the simple replacement
$\Gamma_\phi\rightarrow 2 k_0\Gamma_\phi$.
At the moment $n_{\Pi}$ is just an arbitrary variable replacing
$\Pi^{<,>}$, but it will become closely related to the particle
distribution function in the end. It is important to note that unlike
$n_\phi$, it is {\em not} to be considered a free variable in the equations.

Then using the definitions
(\ref{spectral-decomposition:scalar}-\ref{spectral-decomposition:Pi}) 
and the identity
\begin{equation}
\diamond \DBR{f}{gh} = g \diamond\DBR{f}{h} + h \diamond\DBR{f}{g}
,
\label{identity}
\end{equation}
we rewrite Eqs.\ (\ref{ke:scalar}-\ref{ce:scalar}) as
\begin{eqnarray}
  {\cal A}_\phi\diamond\DBR{\Omega_\phi^2}{n_\phi}
    + \Gamma_\phi\diamond\DBR{\Delta_h}{n_\phi}
  &=& \Gamma_\phi{\cal A}_\phi \big(n_\phi-n_{\Pi}\big)
\label{ke:scalar2}
\\
  \Omega^2_\phi{\cal A}_\phi n_\phi
  + \diamond\DBR{\Gamma_\phi{\cal A}_\phi}{n_\phi}
  &=&  \Gamma_\phi n_{\Pi} \Delta_h.
\label{ce:scalar2}
\end{eqnarray}
We can now see that taking $n_{\Pi}\rightarrow n_\phi$ solves 
the kinetic equation~(\ref{ke:scalar2}) to the zeroth order in gradients,
establishing the promised connection between $n_{\Pi}$ and $n_\phi$. 
This also shows that, to the first order in gradients, it is consistent
to replace $n_{\Pi}$ by the zeroth order solution $n_\phi$, 
whenever the $\diamond$ operator acts on $n_{\Pi}$. In fact, we
already took this into account by replacing $n_{\Pi}$ by $n_\phi$ in the
$\Gamma_\phi$-terms on the {\it l.h.s.} 
of Eqs.~(\ref{ke:scalar2}-\ref{ce:scalar2}). 

\vskip 0.1in

We shall now see how to reduce these equations
to an on-shell approximation, and in particular ask to what extent
the on-shell limit is a good approximation to the plasma dynamics
close to equilibrium.
The equations
can be simplified further when the width $\Gamma_\phi$ is small. 
While this is in general true in the weak coupling limit, due care
must be exercised in making this approximation, because both $\Pi_h$ and
$\Gamma_\phi$ are often controlled by the same coupling constants.

\subsection{Propagator equation}
\label{sec: Propagator equation}

In the weak coupling limit $\Gamma_\phi \rightarrow 0$ and
to first order in gradients, the spectral function~(\ref{A_phi}) 
reduces to the singular spectral form,
\begin{equation}
{\cal A}_\phi { \;\; \stackrel{\Gamma_\phi\rightarrow 0}{\longrightarrow} \;\;}
       {\cal A}_s \; = \; \pi\; {\rm sign}(k_0) \;
                          \delta\big(\Omega_\phi^2\big)
,
\label{As}
\end{equation}
where the correction is of order $\Gamma_\phi$. 
This singular hypersurface defines the usual quasiparticle dispersion 
relation
\begin{equation}
  \Omega^2_\phi \equiv k_0^2 - \vec k^2 - M^2 - \Pi_h(k_0,\vec k,x) = 0
,
\label{quasiDR2}
\end{equation}
which defines the spectrum of the quasiparticle excitations of the system. 
This equation
has in general two distinct solutions 
$k_0 = \omega_k$ and $k_0 = -\bar \omega_k$, corresponding to particles and
antiparticles, respectively. Thus the spectral function (\ref{As}) breaks
up into two clearly separate contributions:
\begin{equation}
  {\cal A}_s =
          \frac{\pi }{2 \omega_k} Z_k\; \delta(k_0 - \omega_k)
        - \frac{\pi }{2 \bar \omega_k} \bar Z_k \;
                            \delta(k_0 + \bar\omega_k),
\label{As2}
\end{equation}
where the wave function renormalization factors $Z_k$ and $\bar Z_k$ are
\begin{equation}
  2\omega_k Z^{-1}_k =
      \left | {\partial \Omega_\phi^2}/{\partial k_0}
      \right|_{k_0 = \omega_k}
,\qquad
  2\bar \omega_k \bar Z^{-1}_k =
     \left | {\partial \Omega_\phi^2}/{\partial k_0}
     \right|_{k_0=-\bar\omega_k}
.
\label{Zpm}
\end{equation}

\vskip 0.1in

The definition of $\Delta_h$ in~(\ref{Delta_h})
becomes more problematic in the limit $\Gamma_\phi \rightarrow 0$ because
of the appearance of the so called Landau ghost poles, when perturbative
expressions are used for the self-energy $\Pi_h$~\cite{Henning:1995}.
Therefore, when needed, we shall relate $\Delta_h$ to ${\cal A}_\phi$ 
{\it via} equations~(\ref{spec1}-\ref{spec2}) before taking 
the zero width limit. Let us note however that the spectral representation
\begin{equation}
  \Delta_h(k,x) { \;\; \stackrel{\Gamma\rightarrow 0}{\longrightarrow} \;\;}
                       {\rm Re} \int d k_0'
                       \frac{{\cal A}_\phi(k_0',\vec k,x)}
                            {k_0 - k_0' + i\epsilon}
\label{PGR}
\end{equation}
remains valid for $\Delta_h$ even in the on-shell limit \cite{Henning:1995}.

\vskip 0.1in

The full spectral function ${\cal A}_\phi$ must obey the
sum-rule~(\ref{sum_rule_scalars}).
The fact that the singular function ${\cal A}_s$~(\ref{As2}) fails this rule
when interactions are included, tells us something important about the physical
nature of the quasiparticle approximation. Indeed, one is here describing the
spectrum of inherently collective plasma excitations by a simple single
particle picture. The amount by which ${\cal A}_s$ fails the sum-rule can then
be attributed to the effect of collective plasma excitations. This gives a
quantitative estimate for the goodness, but also indicates the limits of
applicability of the quasiparticle approximation.

\vskip 0.1in

Since both $\Pi_h$ and $\Gamma_\phi$ are controlled
by the same coupling constants, they typically acquire contributions
at the same order in the coupling constant, so that one cannot take the limit
$\Gamma_\phi\rightarrow 0$ as an independent approximation. 
The theories containing tadpoles, such as the $\lambda \phi^4$-theory,
seem at a first sight to be an exception. However, the tadpoles give rise
to a singular non-dissipative contribution to the self-energy, which
can be absorbed by the mass term, implying that they cannot be responsible
for thermalization. This means that in order to study thermalization properly,
one has to include the higher order nonlocal dissipative contributions, 
which typically contribute to both the self-energy and width at the same order
in the coupling constant. Consequently, there is a danger that one might lose
the consistency of the single particle picture. Nevertheless, as we shall now see,
this is not the case. Namely, the two quantities appear in the spectral function
in quite distinct ways such that the kinetic equation allows a consistent on-shell
limit assuming equal accuracy in the computation of $\Pi_h$ and $\Gamma_\phi$.

\subsection{Kinetic equation in the on-shell limit}
\label{Kinetic equation in the on-shell limit}

Our aim is now to study the conditions under which the kinetic and constraint
equations~(\ref{ke:scalar2}-\ref{ce:scalar2}) reduce to the on-shell
approximation for the effective plasma degrees of freedom found in the previous
section. Bearing in mind the discussion concerning consistency of the
quasiparticle limit, we start by using the
expressions~(\ref{Delta_h}-\ref{A_phi}) with 
a finite width $\Gamma_\phi$. The term linear in $\Gamma_\phi$ on the
{\it l.h.s.} of~(\ref{ke:scalar2}) is often simply dropped without 
a careful consideration~\cite{GreinerLeupold:1998}.  However, making use
of the identity $\Gamma_\phi\Delta_h = \Omega_\phi^2{\cal A}_\phi$,
which is accurate to the leading order in gradients 
({\it cf.} Eq.~(\ref{ce:scalar2})), and
expressions~(\ref{Delta_h}-\ref{A_phi}), we can in fact combine it with the
first term in the kinetic equation to obtain
\begin{equation}
       \frac{-2\Gamma_\phi^3}{(\Omega_\phi^4+\Gamma_\phi^2)^2} \;
             \diamond \DBR{\Omega_\phi^2}{n_\phi}
     - \frac{2\Gamma_\phi^2\Omega_\phi^2}{(\Omega_\phi^4+\Gamma_\phi^2)^2} \;
           \diamond \DBR{\Gamma_\phi}{n_\phi}  
     = \Gamma_\phi{\cal A}_\phi \big(n_\phi-n_{\Pi}\big)
\,.
\label{ke:scalar3}
\end{equation}
Note that in the weak coupling limit and to leading order in gradients
\begin{equation}
       \frac{-2\Gamma_\phi^3}{(\Omega_\phi^4+\Gamma_\phi^2)^2} 
             \stackrel{\Gamma_\phi\rightarrow 0}{\longrightarrow}
              {\cal A}_s,
\label{ke:scalar3b}
\end{equation}
such that Eq.~(\ref{ke:scalar3}) can be rewritten as  
\begin{equation}
       {\cal A}_s\; \diamond \DBR{\Omega_\phi^2}{n_\phi}
     - 2\Gamma_\phi{\cal A}_s\Delta_h \;
           \diamond \DBR{\Gamma_\phi}{n_\phi}  
     = \Gamma_\phi{\cal A}_\phi \big(n_\phi-n_{\Pi}\big)
\,.
\label{ke:scalar4}
\end{equation}
Under the reasonable assumption that $\Gamma_\phi$ is a smooth function of
the energy, we get from~(\ref{ke:scalar4}) a consistent expansion of
the final result in powers of $\Gamma_\phi$ evaluated at the pole.
In this expansion the contributions from the off-shellness cancel so that up to
order ${\cal O}(\Gamma_\phi)$ the integrated equation involves only the leading
on-shell excitations, defined by the singular spectral function~(\ref{As2}).
In other words, the integrated kinetic equation exhibits the important property
of {\em closure} onto the on-shell excitations in the weak coupling limit. We
can then replace (\ref{ke:scalar4}) by the singular 
{\em on-shell kinetic equation}
\begin{equation}
 {\cal A}_s \diamond\DBR{\Omega_\phi^2}{n_\phi}
     = \Gamma_\phi{\cal A}_s \big(n_\phi-n_{\Pi}\big)
.
\label{ke:singular-onshell}
\end{equation}
We again emphasize that it is actually consistent to include non-local
loop contributions to $\Pi_h$ on the {\it l.h.s.} despite the fact that
the singular single particle limit does not, strictly speaking, exist for
the excitations. This is simply because the corrections due to off-shellness
cancel up to ${\cal O}(\Gamma_\phi)$. This is fortunate, because applying
the weak coupling limit in the strict sense to~(\ref{ke:singular-onshell}) 
indeed requires either computing $\Pi_h$ and $\Gamma_\phi$ to the same order,
or simply neglecting $\Gamma_\phi$, leading to a collisionless Boltzmann equation.
A somewhat orthogonal approach has been taken by Leupold~\cite{Leupold:2000}
in which, based on  particle number conservation, the author advocates a
modification in the relation between the Wigner function and the on-shell
distribution function. 

\vskip 0.1in

The final step is to show that the constraint equation~(\ref{ce:scalar2}) can
be written as
\begin{equation}
  \Omega^2_\phi{\cal A}_s\big(n_\phi-n_{\Pi}\big)
           =   -\diamond\DBR{\Gamma_\phi{\cal A}_s}{n_\phi}.
\label{ce:scalar:singular}
\end{equation}
In the limit $\Gamma_\phi \rightarrow 0$ the constraint equation becomes
identically solved when the quasiparticle on-shell condition~(\ref{quasiDR2}) 
is satisfied.  In other
words, the on-shell condition and constraint equations become degenerate.
Note that we can here refer to the on-shell condition, not necessarily restricted
to the singular approximation, but in the broader sense we derived the on-shell
kinetic equation~(\ref{ke:singular-onshell}).
Indeed, the constraint equation can always be
satisfied by the on-shell solutions as given by Eq.~(\ref{ke:singular-onshell}), 
whereby it only gives rise to an integral constraint for the off-shell
excitations.
These however, do not belong to the set of dynamical degrees of freedom
for the kinetic equation~(\ref{ke:singular-onshell}), which only contains 
the on-shell excitations as defined by the singular overall projection
operator ${\cal A}_s$. In this sense the kinetic 
equation~(\ref{ke:singular-onshell}) and the constraint equation decouple.
Of course, the off-shell excitations in the constraint 
equation~(\ref{ce:scalar:singular}) become relevant when the off-shell effects
are included into the kinetic equation.

We emphasise that the derivation of the on shell kinetic
equation~(\ref{ke:singular-onshell}) applies to the special case
of one scalar field in weak coupling limit and close to equilibrium.
It would be of interest to extend the analysis to the case of mixing 
scalar and fermionic fields.


\cleardoublepage
\section{Kinetics of scalars: tree-level analysis}
\label{Kinetics of scalars: tree-level analysis}

In this section we analyze the tree-level dynamics of the scalar sector
of our theory with mixing and emphasise its ramifications.
The hermiticity property (\ref{herm_scal_Wigner}) of the scalar Wigner
function implies that the hermitean and the antihermitean part of the
equation of motion
\begin{equation}
    \left( k^2 - \frac14 \partial^2 + i  k\cdot\partial 
 - M^2(x)e^{-\frac i2\stackrel{\leftarrow}{\partial}\,\cdot\,\partial_k}
  \right) i\Delta^< = 0
\label{scalars-eom}
\end{equation}
ought to be satisfied simultaneously. 
A simple analysis reveals that the hermitean part corresponds to the 
{\it kinetic} equation, while the antihermitean part corresponds to the 
{\it constraint} equation. Roughly speaking, the kinetic equation describes 
the dynamics of quantum fields, while the constraint equation 
constrains the space of solutions of the kinetic
equation~\cite{KainulainenProkopecSchmidtWeinstock:2001}. 
In the simple case when $N=1$ it is immediately clear that the first quantum
correction to the constraint equation, the real part of (\ref{scalars-eom}),
is of second order and to the kinetic equation (imaginary part) of third order
in gradients or second order in $\hbar$.
To extract the spectral information to second order in
$\hbar$ is quite delicate since the constraint equation
in~(\ref{scalars-eom}) contains derivatives.
The situation is more involved in the case of more than one mixing field.
In this case it is convenient to rotate into the mass eigenbasis
\begin{equation}
     M_d^2 = U M^2U^\dagger ,
\label{rotu}
\end{equation}
where $U$ is the unitary matrix that diagonalizes $M^2$. 
In this propagating basis equation~(\ref{scalars-eom}) becomes
\begin{equation}
    \left( k^2 - \frac14 {\cal D}^2 
      + i  k\cdot{\cal D}
      - M_d^2e^{-\frac i2\stackrel{\leftarrow}{{\cal D}}\,\cdot\,\partial_k}
     \right) \Delta_d^< = 0,
\label{scalars-eom-d}
\end{equation}
where $\Delta^<_d \equiv U \Delta^< U^\dagger$ and
the `covariant' derivative is defined as
\def\calD{{\cal{D}}}
\begin{equation}
\calD_\mu = \partial_\mu -i \left[{\Xi}_\mu,\;\cdot\;\right],
\qquad
{\Xi}_\mu = iU\partial_\mu U^\dagger
\,.
\label{calD}
\end{equation}
Making use of $(i\Delta_d^<)^\dagger= i\Delta_d^<$,
$\calD_\mu^\dagger = \calD_\mu$, which implies 
$({\calD_\mu}i\Delta_d^<)^\dagger = i\Delta_d^<\overleftarrow{\cal D}_\mu 
  = {\calD_\mu}i\Delta_d^<$,
and $\hat M_{c,s}^2 i \Delta_d^< 
   = \frac 12 \{\hat M_{c,s}^2, i \Delta_d^<\} 
   + \frac 12 [\hat M_{c,s}^2, i \Delta_d^< ] $,
we can identify the antihermitean part of~(\ref{scalars-eom-d})
as the constraint equation
\begin{equation}
  \Big(k^2  - \frac14 {\cal D}^2\Big) i\Delta_d^< 
 - \frac 12\left\{\hat M^2_c, i\Delta_d^< \right\}
              + \frac i2 \left[\hat M^2_s, i\Delta_d^< \right]
              = 0,
\label{scalars-ce-d}
\end{equation}
and the hermitean part is the kinetic equation
\begin{equation}
k\cdot{\cal D}i\Delta_d^<
     + \frac 12 \left\{  \hat M^2_s, i\Delta_d^< \right\}
     + \frac i2 \left[ \hat M^2_c, i\Delta_d^< \right]
     = 0
\,.
\label{scalars-ke-d}
\end{equation}
We defined
\begin{eqnarray}
\hat M^2_c &=& 
  M^2_d\cos\frac 12\stackrel{\leftarrow}{{\cal D}}\cdot\,\partial_k
\nonumber\\
\hat M^2_s &=& 
   M^2_d\sin\frac 12\stackrel{\leftarrow}{{\cal D}}\cdot\,\partial_k
\,,
\label{McMs}
\end{eqnarray}
and $[\,\cdot\,,\,\cdot] $, $\{\,\cdot\,,\,\cdot\}$ denote a commutator
and an anticommutator, respectively.
The constraint~(\ref{scalars-ce-d}) and kinetic equation~(\ref{scalars-ke-d})
are formally still exact. Since we are interested in order $\hbar$ 
correction to the classical approximation, it is good to keep in mind that one
can always restore $\hbar$ dependencies by the simple replacements
$\partial \rightarrow \hbar \partial$ and $i\Delta_d^< \rightarrow 
\hbar^{-1} i \Delta_d^<$. We work to order $\hbar$ accuracy, 
so it suffices to truncate the constraint equation~(\ref{scalars-ce-d}) 
to first order in gradients, and the kinetic equation~(\ref{scalars-ke-d})
to second order in gradients:
\begin{eqnarray}
  k^2 i\Delta_d^< 
 - \frac 12\left\{M^2_d, i\Delta_d^< \right\}
              + \frac i4 \left[{\cal D}M^2_d, \partial_k i\Delta_d^< \right]
              &=& 0,
\label{scalars-ce-d2}
\\
k\cdot{\cal D}\, i\Delta_d^<
     + \frac 14 \left\{ {\cal D} M^2_d, \partial_k i\Delta_d^< \right\}
   + \frac i2 \left[ M^2_d \Big(1 
   - \frac 18(\stackrel{\leftarrow}{{\cal D}}\cdot\,\partial_k)^2 \Big),
                        i\Delta_d^< \right]
     &=& 0
\,.
\label{scalars-ke-d2}
\end{eqnarray}
We shall now consider these equations in more detail.
The off-diagonal elements of $i\Delta^<_d$ 
in both the constraint~(\ref{scalars-ce-d2}) and the kinetic 
equation~(\ref{scalars-ke-d2}) are sourced by the diagonal elements
through the terms involving commutators, which are suppressed by at least
$\hbar$ with respect to the diagonal 
elements~\footnote{Since $M_d^2$ is a diagonal matrix, the commutator 
$\frac i2 \left[ M^2_d, i\Delta_d^< \right]$ in~(\ref{scalars-ke-d2}),
which is formally of order $\hbar^{-1}$, 
contributes to the off-diagonal equations {\it via} terms that contain 
only the off-diagonal elements.}.
On the other hand, 
the off-diagonals source the diagonal equations through terms that 
are of the same order as the diagonals. 
For the CP-violating diagonal densities that are suppressed at least by
one power of $\hbar$, this immediately implies that the CP-violating 
off-diagonals are at least of the order $\hbar^2$, and thus cannot source
the diagonal densities at order $\hbar$. By a similar argument, 
the second order term in the commutator in~(\ref{scalars-ke-d2}) can be 
dropped, since it can induce effects only at order $\hbar^2$. 

This analysis is however incomplete for the following reason. 
When the rotation matrices
${\Xi}_\mu = iU\partial_\mu U^\dagger$ in~(\ref{calD}) 
contain CP-violation, the off-diagonal elements can in principle
be CP-violating already at order $\hbar$, and hence source the diagonals
at the same order. In order to get more insight into the role of
the off-diagonals, we now analyze the case of two mixing scalars.
For notational simplicity we omit the index $d$ for diagonal in the following.
All quantities have to be taken in the rotated basis,
where the mass is diagonal.
The constraint equations~(\ref{scalars-ce-d2}), when written in components,
are
\beqa
      \Big(k^2 - M_{\tt ii}^2 \Big)i\Delta_{\tt ii}^< 
  \mp \frac 14 \delta(M^2)
               \Big(  \Xi_{\tt 12}\cdot \del_{k}i\Delta_{\tt 21}^<
                    + \Xi_{\tt 21}\cdot \del_{k}i\Delta_{\tt 12}^<
               \Big)
  &=& 0,
\label{scalars-ce-11-22}
\\
      \Big(  k^2
           - \bar M^2 
           + \frac i4 (\del\delta(M^2)) \cdot \del_k 
      \Big)
           i\Delta_{\tt 12}^< 
    + \frac 14 \delta(M^2) \Xi_{\tt 12} \cdot \del_k \delta(i\Delta^<)
  &=& 0
\,,
\label{scalars-ce-12}
\eeqa
where
$\bar M^2 \equiv {\rm Tr}(M^2)/2 = (M_{\tt 11}^2+M_{\tt 22}^2)/2$,
$\delta(M^2) \equiv M_{\tt 11}^2-M_{\tt 22}^2$, 
$\delta(i\Delta^<) \equiv i\Delta^<_{\tt 11}-i\Delta^<_{\tt 22}$, 
and the equation for $i\Delta_{\tt 21}^< = (i\Delta_{\tt 12}^<)^*$ is obtained 
by taking the complex conjugate of~(\ref{scalars-ce-12}). 
To order $\hbar$ the diagonal equations~(\ref{scalars-ce-11-22}) are solved
by the spectral on-shell solution
\begin{eqnarray}
 i\Delta_{\tt ii}^<(k,x) &=& 
  2\pi \delta\Big(k^2 - M_{\tt ii}^2\Big) {\rm sign}(k_0)
            n^\phi_i(k,x)
\nonumber\\
             &=& \frac{\pi}{\omega_{\phi i}}
                 \left[ 
                    \delta(k_0-\omega_{\phi i})
                   -\delta(k_0+\omega_{\phi i})
                 \right]n^\phi_i(k,x)
\,,
\label{ce-diag-solution}
\end{eqnarray}
where $   \omega_{\phi i}
        = [\vec k^{\,2}+M_{\tt ii}^2]^{1/2}$,
and $n^\phi_i(k,x)$ represents the occupation number density
on phase space, which in thermal equilibrium  reduces to
$n^\phi_i(k,x)\rightarrow n^\phi_{\rm eq} = 1/({\rm e}^{\beta k_0} -1)$.
By making use of the sum rule~(\ref{sum_rule_scalars}) for the spectral function 
${\cal A}_\phi = (i/2)(\Delta^>-\Delta^<)$, one can show that 
the other Wigner function has to be of the form
\begin{equation}
 i\Delta_{\tt ii}^>(k,x) = 
  2\pi \delta\Big(k^2 - M_{\tt ii}^2\Big) {\rm sign}(k_0)
            \Big(1 + n^\phi_i(k,x)\Big)
\label{ce-diag-solution>}
\end{equation}
with the same density $n^\phi_i$ as in~(\ref{ce-diag-solution}).
These solutions are consistent provided the off-diagonals 
$i\Delta_{\tt 12}^<$ and $i\Delta_{\tt 21}^<$ are of order $\hbar$,
which, as we shall argue,
is a self-consistent assumption. The off-diagonal constraint
equation~(\ref{scalars-ce-12}) is solved by 
\beq
      i\Delta_{\tt 12}^<
  =   n^\phi_{12} \, \delta(k^2-\bar M^2) 
    - \frac 12 \Xi_{\tt 12} \cdot \del_k{\rm Tr}(i\Delta^<) 
    - \frac{2}{\delta(M^2)} \, k \cdot \Xi_{\tt 12} \, \delta(i\Delta^<)
\,,
\label{Delta12:solution}
\eeq
where $n^\phi_{12}=n^\phi_{12}(k,x)$ is a function that can be
determined from the boundary conditions. Since we are interested
in situations close to
equilibrium in which the off-diagonals are driven away from zero primarily
by the diagonals, we can set $n^\phi_{12}$ to zero for our purposes. Note that 
the solution~(\ref{Delta12:solution}) contains derivatives of the delta
function, and hence it does not strictly speaking represent an on-shell form.

\vskip 0.1in

Consider now the kinetic equations~(\ref{scalars-ke-d2}), which
in components read
\beqa
     \Big(  k \cdot \del
          + \frac 12 (\del M_{\tt ii}^2) \cdot \del_k
     \Big)
          i\Delta_{\tt ii}^<
 &=& \pm i \, k \cdot \Big(  \Xi_{\tt 12} i\Delta_{\tt 21}^< 
                           - \Xi_{\tt 21} i\Delta_{\tt 12}^<
                      \Big) 
\nonumber\\
 &-& \frac i4 \delta(M^2)\,
     \Big(  \Xi_{\tt 12} \cdot \del_k i\Delta_{\tt 21}^< 
          - \Xi_{\tt 21} \cdot \del_k i\Delta_{\tt 12}^<
     \Big) 
\label{scalars-ke-11-22}
\\
     \Big(  k \cdot \del
          + \frac 12(\del\bar M^2) \cdot \del_k \!
          + \! \frac i2 \delta(M^2) \!
          - \! i k\cdot\delta(\Xi)
     \Big)
          i\Delta_{\tt 12}^<  \!\!\!
 &=& \!\!\! - i k \cdot \Xi_{\tt 12} \delta(i\Delta^<)
     \!-\! \frac i4 \delta(M^2) \Xi_{\tt 12} \cdot \del_k {\rm Tr}(i\Delta^<)
\,,
\nonumber\\
\label{scalars-ke-12}
\eeqa
and the equation for $i\Delta_{\tt 21}^<$ is again
obtained by taking the complex conjugate of~(\ref{scalars-ke-12}).
To leading order in gradients the off-diagonal equation~(\ref{scalars-ke-12})
is solved by
\begin{equation}
      i\Delta_{\tt 12}^<
  = - \frac 12\,\Xi_{\tt 12}\cdot\partial_k{\rm Tr}(i\Delta^<)
 - \frac{2}{\delta(M^2)}\,  k\cdot \Xi_{\tt 12}\, \delta(i\Delta^<)
\,,
\label{scalars-ke-12-solution}
\end{equation}
and similarly for $i\Delta_{\tt 21}^<$. 
Remarkably, this corresponds precisely 
to the constraint equation solution~(\ref{Delta12:solution}), 
provided one sets $n^\phi_{12} =0$, representing a very nontrivial
consistency check of the kinetic theory for mixing particles.
By making use of the technique of Green functions, in 
Appendix~\ref{Gradient expansion in the off-diagonal scalar kinetic equation}
we show that the leading order result~(\ref{scalars-ke-12-solution}) 
represents a valid approximation, 
provided the condition for the gradient approximation
$k\cdot\partial \ll \delta(M^2)$ is satisfied, that is
one is not near the degenerate mass limit, $\delta(M^2) = 0$.

The example of mixing scalars represents indeed a nice illustration of
the workings of the constraint and kinetic equations. 
Unlike in the one field case, in which the kinetic flow term can be
obtained by acting with the bilinear $\diamond$ operator on the constraint
equation, in the mixing case the situation is more complex. 
All of the solutions of the kinetic equation~(\ref{scalars-ke-12}) are 
simultaneously solutions of the constraint equation~(\ref{scalars-ce-12}).
The converse is however {\it not} true: the constraint equation contains 
a larger set of solutions, which include the homogeneous 
solutions~(\ref{Delta12:solution})
that lie on a different energy shell, which are in fact excluded by
imposing the kinetic equation~(\ref{scalars-ke-12}).
This can be shown as follows:
the constraint equation~(\ref{scalars-ce-12}) can be obtained
from the kinetic equation~(\ref{scalars-ke-12}) by 
multiplying it by $\big(k^2 - \Tr(M^2)/2\big)/\delta(M^2)$,
followed by a partial integration. The constraint equation obtained this way
has of course a larger set of solutions. The additional solutions are of
the type: a function multiplying $\delta\big(k^2 - \Tr(M^2)/2\big)$.

Finally, upon inserting~(\ref{scalars-ke-12-solution}) into the diagonal 
equation~(\ref{scalars-ke-11-22}), we get
\beqa
   \Big(    k\cdot\partial
        +   \frac 12 (\partial M_{\tt ii}^2)\cdot \partial_k
        \pm i \, k\cdot
                 \big[\Xi_{\tt 12},\Xi_{\tt 21}\big]\cdot
                 \partial_k\, 
   \Big)
        i\Delta_{\tt ii}^<
 =  0
\,.
\label{scalars-ke-11-22b}
\eeqa
It is remarkable that not only the off-diagonals have disappeared 
from~(\ref{scalars-ke-11-22b}), but also the diagonals decouple.
The commutator
$k\cdot \Big[\Xi_{\tt 12},\Xi_{\tt 21}\Big]\cdot \partial_k$
does {\it not} vanish in general. However, for planar walls and 
in the wall frame,
\begin{equation}
     k\cdot \Big[\Xi_{\tt 12},\Xi_{\tt 21}\Big]
      \cdot \partial_k \;\rightarrow \;
   - k_z\Big(  \Xi_{z\tt 12}\,\Xi_{z\tt 21}
             - \Xi_{z\tt 21}\,\Xi_{z\tt 12}
        \Big)
             \partial_{k_z} 
 =   0
\label{vasnishing-commutator}
\end{equation}
vanishes identically, so that Eq.~(\ref{scalars-ke-11-22b}) reduces to
\begin{equation}
   \Big(  k \cdot \del 
        - \frac 12 (\del_z M_{\tt ii}^2)\del_{k_z}
   \Big)
        i\Delta_{\tt ii}^<
 = 0
\,.
\label{scalars-ke-11-22c}
\end{equation}
This proves that, although a naive analysis of 
Eqs.~(\ref{scalars-ce-d2}-\ref{scalars-ke-d2}) indicated that there might be
a CP-violating source in the kinetic equation at order $\hbar$, 
a more detailed look at the structure of these equations shows that 
for planar walls considered in the wall frame {\it no} such source appears.
Equation~(\ref{scalars-ke-11-22c}) thus represents a self-consistent kinetic
description of scalar fields accurate to order $\hbar$, such that the only
source in the kinetic equation is the classical force,
$\vec F_i = - \nabla \omega_{\phi i}
          = - \nabla M_{\tt ii}^2/ 2\omega_{\phi i}$.
In addition, we have shown that to order $\hbar$
the on-shell approximation for the diagonal occupation 
numbers~(\ref{ce-diag-solution}) with the classical dispersion relation
$k_0 = \pm \omega_{\phi i}$ still holds.
Even though the on-shell approximation fails for the off-diagonals, this
is only relevant for the dynamics at higher orders in gradient expansion.
Further, inclusion of collisions in the off-diagonal equations cannot 
change our conclusions concerning the source cancellation 
expressed in Eq.~(\ref{scalars-ke-11-22b}), even though it may introduce new 
collisional contributions. (For a detailed study of collisional sources
we refer to Paper~II.)
Needless to say, this analysis easily generalizes to
the case of N mixing scalars. This completes the proof that was originally 
presented in Ref.~\cite{{KainulainenProkopecSchmidtWeinstock:2001}},
which states that in the flow term of scalars there is no source
for baryogenesis at order $\hbar$ in gradient expansion.

\subsection{Boltzmann transport equation for CP-violating scalar densities}
\label{Boltzmann transport equation for CP-violating scalar densities}

For completeness, and to make a connection between the tree-level analysis
presented here and the analysis of the collision term in Paper~II,
we now show how -- starting with the
kinetic equation for the Wightman function~(\ref{scalars-ke-11-22c})
-- one obtains a Boltzmann equation for the CP-violating scalar particle
densities.
All flavor matrices are to be taken in the basis
where the mass is diagonal.
 
In the beginning we review some of the basics of the C (charge) and CP 
(charge and parity) transformations of quantum scalar fields. Under C and P,
a scalar field $\phi$ transforms as 
\begin{eqnarray}
      \phi^c(u) \equiv
      {\cal C}\, \phi(u)         \, {\cal C}^{-1} 
                                  = \xi_\phi^* \phi^*(u) 
 &,& \, {\cal C}\, \phi^\dagger(u) \, {\cal C}^{-1} \;
                                  = \xi_\phi \phi^T(u)            \,
\nonumber
\\
      \phi^p(u) \equiv
      {\cal P}\, \phi(u)         \, {\cal P}^{-1} 
                                  = \eta_\phi^* \phi(\bar u)
 &,&  {\cal P}\, \phi^\dagger(u) \, {\cal P}^{-1} 
                                  = \eta_\phi \phi^\dagger(\bar u)
\,,
\label{phi:C+CP-transform}
\end{eqnarray}
where $|\xi_\phi|=1$ and $|\eta_\phi|=1$ are global phases,
and $\bar u^\mu = (u_0,-\vec{u})$ denotes the inversion of the spatial part
of $u$. Note that our definition
of charge conjugation includes an additional transposition with respect
to the usual definition~\cite{ItzyksonZuber:1980}, which is required in 
the case of mixing scalar fields, when $\phi$ is a vector
in flavor space. From~(\ref{phi:C+CP-transform}) it 
follows that the scalar Wightman functions transform as 
\begin{eqnarray}
      i\Delta^<(u,v)
 &\stackrel{{\cal C}}{\longrightarrow}&
      i\Delta^>(v,u)^T
 \\
      i\Delta^<(u,v)
 &\stackrel{{\cal CP}  }{\longrightarrow}&
      i\Delta^>(\bar v,\bar u)^T
\,.
\label{phi_Wigner:C+CP}
\end{eqnarray}
When written in the Wigner representation, the equivalent transformations are 
\beqa
                                         i\Delta^<(k,x)
  &\stackrel{{\cal C}}{\longrightarrow}& i\Delta^>(-k,x)^T 
  \equiv                                 i{\Delta^c}^<(k,x)
\label{scalar_wigner_C}
\\
                                          i\Delta^<(k, x)
  &\stackrel{{\cal CP}}{\longrightarrow}& i\Delta^>(-\bar{k}, \bar{x})^T
  \equiv                                  i{\Delta^{cp}}^<(\bar{k}, \bar{x})
\,.
\label{scalar_wigner_CP}
\eeqa
The $CP$-conjugate of the Wigner function $i\Delta^<(k,x)$ is related
to physical objects at position $-\vec{x}$ with momentum $-\vec{k}$.
Therefore we introduced additional inversions of the spatial parts of
the position and momentum arguments in the definition of
$i{\Delta^{cp}}^<(k,x)$,
so that this object indeed describes particles at position $\vec{x}$ and with
momentum $\vec{k}$.
Note that this definition differs from what is usually found in textbooks.
In order to study the effects of the C and CP transformations on the distribution
functions, we proceed analogously to~(\ref{ce-diag-solution})
and define the spectral solutions
\begin{eqnarray}
      i{\Delta_{\tt ii}^{c/cp}}^<(k,x)
  &=& \frac{\pi}{\omega_{\phi i}}
       \left[ 
              \delta(k_0-\omega_{\phi i})
             -\delta(k_0+\omega_{\phi i})
       \right] n^{\phi c/cp}_i(k,x) \,
\label{scalar_C_CP_onshell<}
\\
      i{\Delta_{\tt ii}^{c/cp}}^>(k,x)
  &=& \frac{\pi}{\omega_{\phi i}}
       \left[
              \delta(k_0-\omega_{\phi i})
             -\delta(k_0+\omega_{\phi i})
       \right] \Big(1+n^{\phi c/cp}_i(k,x)\Big)
\,.
\label{scalar_C_CP_onshell>}
\end{eqnarray}
Now upon inserting this and~(\ref{ce-diag-solution}-\ref{ce-diag-solution>})
into (\ref{scalar_wigner_C}) and (\ref{scalar_wigner_CP}), we find
\beq
    n^{\phi c}_i  \big(\omega_{\phi i}, \vec{k}, x \big)
 =  n^{\phi cp}_i \big(\omega_{\phi i}, \vec{k}, x \big)
 = -\big[1 + n^\phi_i\big(-\omega_{\phi i}, -\vec{k} , x \big)
    \big] 
\,.
\label{scalar_neg-energiy_cp}
\eeq
Because of the additional inversions in the definition of ${i\Delta^{cp}}^<$,
the two densities are identical.

\vskip 0.1in

We shall now consider the kinetic equation for scalars. 
First we include the collision term into Eq.~(\ref{scalars-ke-11-22c}),
\begin{equation}
    \Big(  k\cdot\partial 
         - \frac 12 (\partial_z M_{\tt ii}^2)\partial_{k_z}
    \Big)
         i\Delta_{\tt ii}^<(k,x)
  = \frac 12 \big[  {\cal C}_{\phi\tt ii}(k,x)
                  + {\cal C}^\dagger_{\phi\tt ii}(k,x)
             \big]
\,,
\label{scalars-ke-11-22c+Col}
\end{equation}
where ${\cal C}_{\phi\tt ii}$ denotes the diagonal elements of
the collision term defined in Eq.~(\ref{Cphi}) after rotation into
the basis where the mass is diagonal. Next, we insert the spectral 
on-shell solution~(\ref{ce-diag-solution}) and 
integrate over positive frequencies, to obtain 
\begin{equation}
   \left( \del_t
          +\frac{\vec{k}\cdot{\nabla}}{\omega_{\phi i}}
          -\frac{(\partial_z{M_{\tt ii}^2}(z))}{2\omega_{\phi i}}\partial_{k_z}
   \right) f^\phi_{i+}(\vec k,x)
 = \frac{1}{2\pi}\int_0^\infty dk_0 \, \big[{\cal C}_{\phi\tt ii}(k,x) 
                                         +  {\cal C}^\dagger_{\phi\tt ii}(k,x)
                                       \big]
\,,
\label{scalar_pos_freq}
\end{equation}
which is the kinetic equation for the scalar distribution function
\begin{equation}
f^\phi_{i+}(\vec k,x) 
    \equiv  n^\phi_i\big(\omega_{\phi i}(\vec{k},x),\vec{k},x\big)
\,.
\label{scalar_distribution_function}
\end{equation}
Since $i\Delta^>(k,x)$ and  $i\Delta^<(k,x)$ satisfy identical equations,
we can immediately write 
\begin{equation}
\Big(-k \cdot \partial 
 + \frac 12 (\partial_z M_{\tt ii}^2(z))\partial_{k_z}\Big)
      i{\Delta_{\tt ii}^{cp}}^<(k,x)
  =  \frac 12
    \big[{\cal C}^*_{\phi\tt ii}(-k, x) 
      +  {\cal C}^T_{\phi\tt ii}(-k, x)
    \big]
\,,
\label{scalars-ke-11-22c+Col2}
\end{equation}
where we took account of~(\ref{scalar_wigner_CP}).
We can omit the transposition in the collision term, because we are
only interested in the diagonal elements.
Upon integrating over positive frequencies this becomes 
\begin{equation}
  - \left(\del_t
          +\frac{\vec{k}\cdot{\nabla}}{\omega_{\phi i}}
          -\frac{(\partial_{z}{M_{\tt ii}^2(z)})}
                {2\omega_{\phi i}}\partial_{k_z}
   \right) f^\phi_{i-}(\vec{k},x) = 
 \frac{1}{2\pi}\int_0^\infty dk_0 \, 
             \big[
                  {\cal C}_{\phi\tt ii}(- k, x)
               +  {\cal C}^\dagger_{\phi\tt ii}(- k, x)
             \big]
\,.
\label{scalar_KinEq_antiparticles}
\end{equation}
This transport equation is the CP-conjugate of~(\ref{scalar_pos_freq}) for  
antiparticles, where we defined the antiparticle distribution function as
\beq
          f^\phi_{i-}(\vec k,x)
  \equiv  n^{\phi cp}_i \big(\omega_{\phi i}, \vec k, x \big)
  =       - \big[ 1 +  n^\phi_i \big(-\omega_{\phi i}, -\vec{k}, x \big)
             \big] 
\,.
\label{particle-antiparticle}
\eeq
This then implies the following definition for the CP-violating particle 
density:
\begin{equation}
\delta f^\phi_i = f^\phi_{i+} - f^\phi_{i-}
\,.
\label{CP-violating density}
\end{equation}
The relevant kinetic equation for $\delta f^\phi_i$ 
is obtained simply by summing
(\ref{scalar_pos_freq}) and~(\ref{scalar_KinEq_antiparticles}),
\begin{eqnarray}
   \left(  \del_t
         + \frac{\vec{k}\cdot{\nabla}}{\omega_{\phi i}}
         - \frac{(\partial_z{M_{\tt ii}^2}(z))}{2\omega_{\phi i}}
           \partial_{k_z}
   \right)
   \delta f^\phi_i(\vec{k},x)
 &=& \frac{1}{2\pi}\int_0^\infty dk_0 \, 
                        \big[
                  {\cal C}_{\phi\tt ii}(k, x)
               +  {\cal C}^\dagger_{\phi\tt ii}(k, x)
\nonumber\\
&&\qquad\qquad\;\,\,
               +\,{\cal C}_{\phi\tt ii}(- k, x)
               +  {\cal C}^\dagger_{\phi\tt ii}(- k, x)
             \big]
\,.
\label{scalar_pos_freq_deltaCP}
\end{eqnarray}
It should be stressed that the absence of a CP-violating force in the
flow term at linear order in $\hbar$ is made explicit in this equation.
Further, this equation
makes it transparent how
to extract CP-violating contributions from the collision term.

\subsection{Applications to the stop sector of the MSSM}

This analysis is relevant for example for a calculation of the CP-violating
force in the  {\it stop} sector $\tilde q = (\tilde t_L,\tilde t_R)^T$
of the MSSM~\cite{HuetNelson:1995+1996,Riotto:1995,Riotto:1998,
ClineJoyceKainulainen:1998,CarenaQuirosRiottoViljaWagner:1997,
ClineJoyceKainulainen:2000+2001},
in which the mass matrix reads
\begin{equation}
       M_{\tilde q}^2 = \left(
             \begin{array}{cc} m_Q^2          & y( A^* H_2 + \mu H_1)   \\
                        y( A H_2 + \mu^* H_1) & m_U^2 \end{array} \right)
,
\label{mass-squarks}
\end{equation}
where $m_{Q}^2$ and $m_U^2$ denote the sum of the soft SUSY-breaking
masses, including D-terms and $m_t^2 = y^2 H_2^2$. Our analysis
immediately implies that, for squarks in the quasiparticle picture,
there is no CP-violating correction to the dispersion relation at first order
in gradients, and hence there is no CP-violating semiclassical force in
the flow term of the kinetic equation at order $\hbar$. This is in contrast
to what was found 
in Refs.~\cite{HuetNelson:1995+1996,Riotto:1995,Riotto:1998,
CarenaQuirosRiottoViljaWagner:1997}.

To investigate whether there is CP-violation in the off-diagonal sector
of the theory, we note that the mass matrix
is diagonalized by the unitary matrix
\begin{equation}
       U = \left(
            \begin{array}{cc} \cos\theta   & -\sin\theta\, {\rm e}^{-i\sigma}\\
            \sin\theta\, {\rm e}^{i\sigma} & \cos\theta \end{array} \right)
\,,
\label{U}
\end{equation}
where
\beq
       \tan 2\theta = \frac{2y|A^*H_2 + \mu H_1|}{m_Q^2-m_U^2}
\,,\quad
       \tan \sigma = \frac{|A| \tan\beta \sin\alpha + |\mu| \sin\zeta}
              {|A| \tan\beta \cos\alpha + |\mu| \cos\zeta}
\,,
\label{tan2theta}
\eeq
with $A^*H_2 + \mu H_1 = |A^*H_2 + \mu H_1| {\rm e}^{i\sigma}$,
$A^* =|A|{\rm e}^{i\alpha}$, $\mu = |\mu| {\rm e}^{i\zeta}$ 
and $\tan\beta = H_2/H_1$. From this we easily obtain
\begin{equation}
   \Xi_\mu \equiv iU\partial_\mu U^\dagger = \left(
          \begin{array}{cc} 0   & i{\rm e}^{-i\sigma}\\
          -i{\rm e}^{i\sigma} & 0 \end{array} \right) \partial_\mu\theta
 + \frac 12 \left(
 \begin{array}{cc} -(1 - \cos 2\theta) & \sin 2\theta\, {\rm e}^{-i\sigma}
     \\
   \sin 2\theta\, {\rm e}^{i\sigma} & 1 - \cos 2\theta \end{array} \right)
           \partial_\mu\sigma.
\label{Xi}
\end{equation}
%
%
Since the off-diagonal elements of $\Xi_\mu$
are precisely the
ones involved in the mixing of the diagonal and off-diagonal
equations, $\Xi_\mu$ is CP-violating whenever $\sigma \neq 0$ 
and $\tan\beta$ varies at the phase
interface, which is in general realized at the phase transition
in the MSSM by complex $A$ and $\mu$ parameters.


\cleardoublepage
\section{Kinetics of fermions: tree-level analysis}
\label{Kinetics of fermions: tree-level analysis}

In this section we consider the dynamics of fermions in the presence of 
scalar and pseudoscalar space-time dependent mass terms. Our analysis is
of relevance, for example, for electroweak baryogenesis calculations,
and establishes the nature of the propagating quasiparticle states 
in the electroweak plasma at the first order electroweak phase transition.
In supersymmetric models baryogenesis is typically mediated
by charginos and neutralinos~\cite{HuetNelson:1995+1996}. 
Since they both mix at the tree level, 
it is very important to consider the more general case of $N$ mixing
fermions~\cite{KainulainenProkopecSchmidtWeinstock:2001}. 
We derive the dispersion relation and 
kinetic equation for mixing fermions accurate to order $\hbar$. 
We also construct the equilibrium Wigner function to order $\hbar$,
which can then be used for baryogenesis source calculations.
This section is not just an overview of our original 
work~\cite{KainulainenProkopecSchmidtWeinstock:2001,
KainulainenProkopecSchmidtWeinstock:2002}, but it also contains 
new insights and results.

\subsection{Spin conservation and the fermionic Wigner function}

In section \ref{Wigner representation and gradient expansion}
we have already derived the equation of motion~(\ref{Wigner-space:fermionic_eom})
for the fermionic Wigner function.
For the moment we neglect the interactions induced by ${\cal L}_{\rm int}$
in the Lagrangean~(\ref{lagrangean}). 
%
%
%
The equations for $S^<$ and $S^>$ are identical, so we omit the label in the
following, indicating that the equations hold for both of them.
The equation of motion is
\begin{equation}
        {\cal D}S
 \equiv \left(  \kdag
              + \frac i2 \deldag
              - (m_h+i\gamma^5 m_a)
            \mbox{e}^{-\frac{i}{2}\stackrel{\leftarrow}{\del}\!\cdot\,\del_{k}}
        \right) S = 0 ,
\label{S_less_eom}
\end{equation}
where we introduced the symbol ${\cal D}$ for the kinetic
derivative operator.
For planar walls propagating in $z$-direction the mass 
terms in~(\ref{S_less_eom}) simplify to 
$m_{h,a}(z)
\mbox{e}^{\frac{i}{2}\stackrel{\leftarrow}{\del_z}\!\,\del_{k_z}}$,
when written in the wall frame.

Equation~(\ref{S_less_eom}) is the tree-level fermionic master equation,
which contains all the necessary information to study the nature of the 
propagating fermionic states in the presence of a space-time dependent mass
term. We shall now show that, for planar walls, Eq.~(\ref{S_less_eom})
contains a conserved quantity, 
corresponding to the spin pointing in the $z$-direction in 
the rest frame of the particle. This will then allow us to recast $S^{<,>}$
in a form which is block diagonal in spin, such that different spin blocks
completely decouple, provided, of course, they decouple at the boundaries.
A very nontrivial consequence of this observation is that 
a {\it quasiparticle picture}, when formulated in terms of the spin states, 
survives to order $\hbar$ in gradient expansion.

\subsubsection{The 1+1 dimensional ($\vec k_\| = 0$) frame}

In order to establish what precisely is the conserved quantity, we consider
a particle with arbitrary momentum $\vec{k}$ and perform a boost
into the frame where the particle momentum parallel to the wall vanishes,
while the momentum component perpendicular to the wall remains unaffected:
\beq
              k
 =           \left(k_0,         k_x, k_y, k_z \right)
 \rightarrow \tilde{k}
 =           \left(\tilde{k}_0, 0,   0,   k_z \right)
 \quad,\quad
  \tilde{k}_0 = \mbox{sign}(k_0) \sqrt{k_0^2-\vec{k}_\|^2}
\,.
\eeq
In this frame the particle moves only perpendicular to the wall, which
is why we call it ``1+1 frame''. The necessary boost is characterized by
\begin{equation}
    \vec v_\| = \frac{\vec k_\|}{k_0}, 
\qquad
    \gamma_\| = \frac{k_0}{\tilde k_0}
\,. 
\label{Lorentz-transform}
\end{equation}
In the spinor space it is represented by the operator
\begin{equation}
  L(k) = \frac{k_0 + \tilde{k}_0 
               - \gamma^0\vec{\gamma}\cdot\vec{k}_{\|}}
                {[{2\tilde{k}_0(k_0+\tilde{k}_0)}]^{1/2}}
\label{L-Lambda}
\end{equation}
and its inverse $L^{-1}(k_0,\vec{k}) = L(k_0,-\vec{k})$, which act like
\begin{equation}
  L(k)\kdag L^{-1}(k) \,=\,
     \gamma^0\tilde k_0 - \gamma^3 k_z \,\equiv \, \tilde{\kdag}
\,.
\label{transfercond}
\end{equation}
The mass terms in the equation of motion are transformed as follows:
\begin{equation}
   L(k) m_{h,a}(x)
        \mbox{e}^{-\frac i2 \stackrel{\leftarrow}{\del}
                            \!\cdot\,
                            \del_{k}}
   L^{-1}(k) 
 =      m_{h,a}(\tilde{x})
        \mbox{exp}\Big(-\frac i2 \stackrel{\leftarrow}{\tilde{\del}}
                            \!\cdot\,
                            \left( \del_{\tilde{k}}
                                  +\left[ L(k)\del_kL^{-1}(k), \,\cdot\,\right]
                            \right)
                  \Big)
\,.
\label{mass-terms:transform}
\end{equation}
Since $L(k)$ is a function of $k_0$ and $\vec k_\|$, the commutator
term
does in general not even vanish
for planar walls in the frames where $m=m(t,z)$.
The exception is the wall frame of a planar wall, where
$m=m(z)$, such that
$\left(\del m\right)\cdot\left(L(k)\partial_k L^{-1}(k)\right) = 0$,
and the commutator in~(\ref{mass-terms:transform}) vanishes. 
For this reason we work from now on in the {\it wall frame} of a planar wall.
In section~\ref{Plasma frame} below we remark on how to transform our results
to other frames, in particular to the plasma frame. 
Upon transforming~(\ref{S_less_eom}), we then get  
\begin{equation}
        \tilde {\cal D} \tilde S
 \equiv \left(  \tilde\kdag
              + \frac{i}{2}\tilde\deldag
              - (m_h + i\gamma^5 m_a)
            \mbox{e}^{\frac i2 \stackrel{\leftarrow}{\del_z}\!\,\del_{k_z}}
        \right) \tilde S  = 0 
\quad ({\tt 1+1\; dim.\;  frame}),
\label{S_less_eom-tilde}
\end{equation}
where the boosted Wigner function is 
\begin{equation}
       \tilde S(\tilde{k}) = L(k) S(k) L^{-1}(k)
\,.
\label{S_less-boost}
\end{equation}
Since the bubble wall, which is responsible for any dependence of the
Wigner function on the average coordinate $x^\mu$, is symmetric in the
$x$-$y$ plane, we can assume that the Wigner function has the same symmetry:
\begin{equation} 
   \tilde S = \tilde S(\tilde k,\tilde t,z)
\,.
\label{tildeS<:spin-conserving}
\end{equation} 
This form holds true for homogeneous boundary conditions, which is the case 
for most of the time during a first order phase transition, 
during which the bubbles are large, almost planar and far away from each other,
when measured in the units of a typical diffusion scale.
The derivative in~(\ref{S_less_eom-tilde}) then reduces like
$\tilde\deldag \rightarrow \gamma^0\partial_{\,\tilde t}+\gamma^3\partial_{z}$,
so that the spin operator in $z$-direction
\begin{equation}
  \tilde S_z = \gamma^0\gamma^3\gamma^5
\label{tildeSz}
\end{equation}
commutes with the kinetic operator
%
\begin{equation}
  [\tilde {\cal D}, \tilde S_z] = 0,
\label{tildeS_z-commutes}
\end{equation}
and hence in this frame the spin in $z$-direction is a good quantum
number. Assuming that the plasma is in thermal equilibrium before the
phase transition takes place, the fermions are described by the
equilibrium Wigner function~(\ref{Green_fermionic_eq_<}), which is
diagonal in spin in the 1+1 frame. The interaction with the bubble
wall introduces no spin mixing, so that we
can write $\tilde S$ in the spin-block diagonal form
\beq
    \tilde S = \sum_{s=\pm 1} \tilde S_s, 
\qquad 
    \tilde S_s \equiv \tilde P_s \tilde S \tilde P_s,
 \label{tildeS-block}
\eeq
where $\tilde P_s$ denotes the spin 
projector:
\begin{equation}
     \tilde P_s = \frac 12 (\mathbbm{1}+s\tilde S_z), 
\qquad
         \tilde P_s \tilde P_{s'} = \delta _{ss'} \tilde P_s,
\qquad   s = \pm 1
.
\label{tildePs}
\end{equation}
The spin diagonal Wigner function lives in the subalgebra of the 
the Clifford algebra of the Dirac matrices which is spanned
by the matrices commuting with $\tilde P_s$. We can
write down a decomposition of $\tilde S_s$ by using a suitable
hermitean basis of the subalgebra,
\beq
 \tilde S_s(\tilde{k}) = i\tilde P_s(k)
                   \left[  s\gamma^3\gamma^5 \tilde g^{s}_0(\tilde{k})
                         - s\gamma^3         \tilde g^{s}_3(\tilde{k})        
                         + {\mathbbm 1}      \tilde g^{s}_1(\tilde{k})
                         - i\gamma^5         \tilde g^{s}_2(\tilde{k})
                   \right]
\,,
\label{S<tilde_decomposition}
\eeq
with $\tilde{g}^s_a$, $a=0,1,2,3$ being scalar functions.
If we choose the following chiral representation of the Dirac matrices
\beqa
 \gamma^\mu = \left(\begin{array}{cc} 
               \mathtt{0}    & \sigma^\mu   \\
              \bar\sigma^\mu & \mathtt{0}
              \end{array}\right)
\,,
\label{chiral_basis_gamma}
\eeqa
where $\sigma^\mu = (\mathbbm{1},\sigma^i)$,
$\bar\sigma^\mu = (\mathbbm{1},-\sigma^i)$,
and $\sigma^i$, $i=1,2,3$ are the usual $2\times 2$ Pauli matrices,
then the spin operator is
\beq
  \tilde{S}_z = \gamma^0\gamma^3\gamma^5 = \mathbbm{1} \otimes \sigma^3
\eeq
and the spin block diagonality of the Wigner function becomes explicit:
\begin{equation}
        -  i \gamma^0 \tilde S_s 
        =  \frac 12 \rho^a \otimes (\mathbbm{1}+s\sigma^3) \tilde g_a^s
\,.
\label{tildeS<s}
\end{equation}

\subsubsection{The 3+1 dimensional (moving) frame}
\label{The 3+1 dimensional (moving) frame}

Knowing that in the 1+1 frame the spin in $z$-direction is
conserved by the interaction with the bubble wall, we can construct
the corresponding conserved quantity in the original frame by boosting
the spin operator $\tilde{S}_z$ back to the original frame:
\beq
        S_z(k)
 \equiv L^{-1}(k) \,\tilde{\!S}_z L(k)
 =      \frac{1}{\tilde k_0}
         \left(k_0 \gamma^0 - \vec k_\| \cdot \vec\gamma\right)\gamma^3\gamma^5
\,.
\label{Sz}
\eeq
In order to clarify the physical meaning of this operator, we recall that
the covariant form of the spin operator is given by the
Pauli-Lubanski tensor
\begin{equation}
  S_{PL}(k, n) \equiv - \frac{1}{e_0}\kdag\ndag\gamma^5
\,,
\qquad e_0 \equiv (k^2)^{1/2}
\,,
\label{Pauli-Lubanski}
\end{equation}
which measures the spin of a particle with momentum $k$ in the direction
$\vec n$ ($n^2 = -1$, $n\cdot k = 0$).
For simplicity we choose the normalization for $S_{PL}$ such that the
eigenvalues are $\pm 1$. 
In literature one often finds the normalization which 
corresponds to the eigenvalues $\pm\hbar/2$.
In the on-shell limit, $e_0$ reduces to the particle's mass. 
In the rest frame the Pauli-Lubanski tensor becomes 
$S_{PL}\stackrel{\vec k \rightarrow 0}{\longrightarrow}\gamma^0\npdag\gamma^5$,
where $n'$ is the spin vector in the rest frame, which is related to
$n$ by the corresponding Lorentz boost.
To measure the spin in $z$-direction in the 1+1-dimensional frame, we use
\begin{equation}
    \tilde k^\mu = (\tilde k_0,0,0,k_z), 
\qquad 
    \tilde n^\mu = \frac{1}{e_0}(k_z,0,0,\tilde k_0),
\label{tildek-tildes}
\end{equation}
and we find, of course,
$S_{PL}(\tilde k,\tilde n) = \gamma^0\gamma^3\gamma^5 = \tilde{S}_z$.
The same operator measures also the spin in $z$-direction in the rest frame,
because the spin operator~(\ref{Pauli-Lubanski}) is invariant under boosts
as long as the direction of the spin vector is parallel to the momentum,
$\vec n\, || \vec k$. 
Now, by setting $S_{PL}(k,n)$ equal to $S_z(k)$ in~(\ref{Sz}), we find that 
$S_z$ measures spin in the direction corresponding to 
\begin{equation}
  n^\mu(k) = \frac{1}{\tilde k_0 e_0}\left( 
        \begin{array}{c}  k_0 k_z \\   
                          k_x k_z \\   
                          k_y k_z \\   
                          \tilde k_0^2
        \end{array} 
      \right)
\,.
\label{spin-direction}
\end{equation}
The same result is of course obtained by boosting $\tilde n^\mu$ 
in~(\ref{tildek-tildes}) to the original (moving) frame $k^\mu$.
In the highly relativistic limit we have $\tilde k_0^2 \rightarrow k_z^2$, 
$k_0^2 \rightarrow {\vec k}^2$, 
and our special spin vector~(\ref{spin-direction}) becomes proportional
to $\vec k$, such that the spin operator $S_z$ approaches the helicity operator,
\begin{eqnarray}
  \hat H (\vec k) &=& -\frac{1}{e_0}\kdag\hdag\gamma^5 
\nonumber\\
         &=& \hat{\vec k}\cdot \gamma^0 \vec\gamma\gamma^5,
\qquad  
       h^\mu = \frac{1}{e_0}\left( 
        \begin{array}{c}  |\vec k| \\   
                           k_0 \hat{\vec k}
        \end{array} 
      \right)
\,,
\label{helicity-operator}
\end{eqnarray}
as one would expect. As usually, the helicity operator measures spin in the 
direction of a particle's motion, $\hat{\vec k} = \vec k/|\vec k|$.
As a consequence, 
for light particles with momenta of order the temperature, 
$k\sim T \gg m$, the spin states we consider here can be approximated
by the helicity states, which are often used in literature for baryogenesis
calculations.

The commutation relation~(\ref{tildeS_z-commutes}) in the 1+1 frame
now implies that by construction the spin operator $S_z(k)$ commutes
with the kinetic differential operator ${\cal D}$ in~(\ref{S_less_eom}),
provided the bubble wall is stationary and $x$-$y$-symmetric in the wall frame.
We can slightly relax this condition: $S_z$ and ${\cal D}$ commute even
for non-stationary and $x$-$y$-dependent bubble walls, as long as the
dependence is of the form
\beq
     S_s  =  S_s(k,t-\vec v_\|\cdot\vec x_\|, z)
\,,
\label{S<-tildeS<}
\eeq
such that $\nabla_\| = - \vec{v}_\|\partial_t$ in ${\cal D}$.
So we can treat time dependent problems, which can be used to study how the
system relaxes from some initial conditions to the stationary state,
admittedly only for quite special forms of the initial conditions.

As a consequence of the above discussion, 
even in the original moving frame, the problem splits 
into two non-mixing sectors labeled by the spin $s$:
\beq
      S = \sum_{s=\pm 1} S_s
\,, 
\qquad 
      S_s \equiv P_s S  P_s
\,,
 \label{S-block}
\eeq
where $P_s(k) = (\mathbbm{1}+sS_z(k))/2$, $s=\pm 1$ is the spin projector.
We can again write down a decomposition of $S_s$,
\begin{equation}
   S_s
 = iP_s \left[  s\gamma^3\gamma^5 g^{s}_0 
              - s\gamma^3         g^{s}_3 
              + {\mathbbm 1}      g^{s}_1 
              - i\gamma^5         g^{s}_2    \right]
\,,
\label{S<_decomposition2}
\end{equation}
using a set of matrices that commute with $P_s(k)$.
This is nothing else than the boosted 1+1 dimensional spin diagonal
Wigner function:
\beq
  \tilde{S}_s(\tilde{k}, \tilde{x}) = L(k) S_s(k, x) L^{-1}(k)
\,.
\eeq
As a consequence of the hermiticity property~(\ref{herm_ferm_Wigner}),
the scalar functions
$g^s_a(k, x) = \tilde{g}^s_a(\tilde{k}, \tilde{x})$ are all real.
Since the direction in which spin is measured now depends on the momentum,
it is not any more possible to find a representation of the Dirac algebra
in which the block diagonality of $S_s$ is explicitly displayed.

\vskip4truemm

\subsection{Constraint and kinetic equations}
\label{Constraint and kinetic equations}

We are now ready to study the tree-level fermion dynamics governed by
Eq.~(\ref{S_less_eom}).
In order to get the component equations, we take the block diagonal
form~(\ref{S<_decomposition2}) for the Wigner function $S_s$, 
multiply~(\ref{S_less_eom}) by
$P_s(k)\{1,s\gamma^3\gamma^5,-is\gamma^3,-\gamma^5\}$ and take the trace.
The resulting equations are
\begin{eqnarray}
   2i \hat{k}_0  g^s_0
  -2is \hat{k}_z g^s_3
  -2i \hat{m}_h  g^s_1
  -2i \hat{m}_a  g^s_2
  &=& \Tr P_s(k) {\mathbbm{1}} {\cal C_\psi}
\label{ga0_eom}\\
   2i \hat{k}_0  g^s_1
  -2s \hat{k}_z  g^s_2 \;
  -2i \hat{m}_h  g^s_0
 \,+2  \hat{m}_a g^s_3 \;
  &=& \Tr P_s(k)s\gamma^3\gamma^5 {\cal C_\psi}
\label{ga1_eom}\\
   2i \hat{k}_0  g^s_2
 \,+2s \hat{k}_z g^s_1 \;
  -2  \hat{m}_h  g^s_3 \;
  -2i \hat{m}_a  g^s_0
  &=& \Tr P_s(k)\left(-is\gamma^3\right) {\cal C_\psi}
\label{ga2_eom}\\
   2i \hat{k}_0  g^s_3
  -2is \hat{k}_z g^s_0
 \,+2 \hat{m}_h  g^s_2 \;
 \,-2\hat{m}_a   g^s_1 \;
  &=& \Tr P_s(k) \left(-\gamma^5\right) {\cal C_\psi}
\,,
\label{ga3_eom}
\end{eqnarray}
where we used the shorthand notations
\begin{equation}
  \hat{k}_0 = \tk + \frac i2 
         \frac{k_0\del_t+\vec k_\|\cdot\nabla_{\|}}{\tk},
  \quad\quad
  \hat{k}_z = k_z - \frac i2 \del_z
\,,
\label{shorthand_k0k3}
\end{equation}
and 
\begin{equation}
\hat{m}_{h,a} = 
 m_{h,a}(z) \mbox{e}^{\frac i2 \stackrel{\leftarrow}{\del_z}\!\,\del_{k_z}}
\,.
\label{hat-m-ha}
\end{equation}
For completeness, we have added the contributions from the fermionic
collision term. One should keep in mind that the dependence of the
Wigner function on the time and parallel coordinates is restricted
by~(\ref{S<-tildeS<}).

In the general case with fermionic mixing, each of the 
functions $g_a^s$ is a hermitean matrix in flavor space. 
The mixing is mediated through the off-diagonal elements of the 
mass terms~(\ref{hat-m-ha}) and thus appears already at the leading order
in gradient expansion; hence it would be inappropriate 
to work in the interaction basis~(\ref{ga0_eom}-\ref{ga3_eom}) without 
incorporating the flavor off-diagonal elements. Since
when integrated over the momenta
$\int d^4 k $ and $\int d^4 k (k^\mu/k_0)$, 
equations~(\ref{ga0_eom}-\ref{ga3_eom}) yield 
fluid equations, the same conclusions hold for fluid equations.
This is in discord with the strategy advocated in a recent 
work~\cite{CarenaQuirosSecoWagner:2002} on chargino-mediated 
baryogenesis in the Minimal Supersymmetric Standard Model,
where it was argued that, when considering the dynamics of CP-violating
sources, one should work in the weak interaction basis 
for charginos (without taking account of the off-diagonals). 

 Rather than considering the full dynamics of mixing fermions, we shall 
now argue that, in order to properly capture the dynamics of CP-violating
densities to order $\hbar$, it suffices to work in the spin and flavor 
diagonal bases, provided one transforms into the mass eigenbasis,
in which $m$ is diagonal. Here we generalize the analysis 
of Ref.~\cite{KainulainenProkopecSchmidtWeinstock:2001} to 
3+1 dimensions and include the space-time transients that accord
with~(\ref{S<-tildeS<}). More importantly, we fill in a gap in our
original derivation in~\cite{KainulainenProkopecSchmidtWeinstock:2001}.

\subsubsection{Flavor diagonalization}
\label{Flavor diagonalization}

In order to get to the mass eigenbasis, we diagonalize the fermionic 
mass matrix $m$. Since $m$ is in general nonhermitean, 
the diagonalization is exacted by the biunitary transformation
\begin{equation}
   m_d = U m V^\dagger
\,,
\label{mass_diagonalization}
\end{equation}
where $U$ and $V$ are the unitary matrices that diagonalize 
$mm^\dagger$ and $m^\dagger m$, respectively.
To make the analysis more transparent, we make 
an explicit separation of the spinor and flavor spaces by  
using a direct product notation $\otimes$.
The master equation~(\ref{S_less_eom}) then becomes 
\begin{equation}
  \left( \kdag \otimes {\mathbbm{1}}
         + \frac i2 \deldag \otimes {\mathbbm{1}} - \hat \bm
  \right) S = {\cal C}_{\psi},
\label{S_eom_mixing}
\end{equation}
where we have defined the mass term as
\begin{eqnarray}
\hat{\bm} &=& P_R \otimes \hat m + P_L \otimes \hat m^\dagger 
\nonumber\\
    &=& {\mathbbm 1}\otimes \hat m_h + i\gamma^5 \otimes \hat m_a
\,.
\label{bm}
\end{eqnarray}
It is now a simple matter to see that the mass matrix $\bm$ is diagonalized
by the unitary matrices
\begin{eqnarray}
  \bX &=& P_L \otimes V + P_R \otimes U 
       = {\mathbbm 1}\otimes \frac 12 (V+U) - \gamma^5 \otimes \frac 12 (V-U)
\nonumber\\
  \bY &=& P_L \otimes U + P_R \otimes V 
       = {\mathbbm 1}\otimes \frac 12 (V+U) + \gamma^5 \otimes \frac 12 (V-U)
\label{rotation_matrices}
\end{eqnarray}
as follows:
\begin{equation}
\bm_d   = \bX \bm \bY^\dagger
\,.
\label{diagonalization-full-m}
\end{equation}
The Wigner function then transforms as  
\begin{equation}
  S_d = \bY S \bX^\dagger 
\,,
\label{S_diagonalization}
\end{equation}
which is in general not diagonal.
From here on we work in the frame where the bubble wall is at rest,
so that we can make a spin-diagonal ansatz for the rotated Wigner
function $S_{sd}$ which is analogous to~(\ref{S<_decomposition2}).
Note that the spin projector $P_s$ commutes with the rotation matrices.
Since the mass term $\hat\bm$ mixes spinor and flavor, the rotation
matrices $\bY$ and $\bX$ do so as well. As a consequence, the component
functions $g^s_{ad}$ of the rotated Wigner function are not just the flavor
rotated components $g^s_a$ of $S_s$. Indeed,
inserting~(\ref{S<_decomposition2}) 
into~(\ref{S_diagonalization}) leads to the following relations
\begin{eqnarray}
  g^s_{0d} &=& \frac 12 \left[   V (g^s_{0}+g^s_{3}) V^\dagger
                            + U (g^s_{0}-g^s_{3}) U^\dagger \right]
\label{g0d_g0}\\
  g^s_{3d} &=& \frac 12 \left[ V (g^s_{3}+g^s_{0}) V^\dagger
                            - U (g^s_{0}-g^s_{3}) U^\dagger \right]
\label{g3d_g3}\\
  g^s_{1d} &=& \frac 12 \left[   V (g^s_{1}-ig^s_{2}) U^\dagger
                                +U (g^s_{1}+ig^s_{2}) V^\dagger
                        \right]
\label{g1d_g1}\\
  g^s_{2d} &=& \frac 12 \left[  V (g^s_{2}+ig^s_{1}) U^\dagger
                                 + U (g^s_{2}-ig^s_{1}) V^\dagger
                        \right] 
\,,
\label{g2d_g2}
\end{eqnarray}
that is $g_{0}^s$ mixes with $g_{3}^s$ and $g_{1}^s$ mixes with $g_{2}^s$.
We now transform~(\ref{S_eom_mixing}) 
by multiplying it from the left by $\bX$ and from the right by $\bX^\dagger$
to obtain
\begin{eqnarray}
&&
    \left( \kdag \otimes {\mathbbm 1} + \frac i2 \mDdag 
         - \bm_d\,
    {\rm e}^{\frac i2 \stackrel{\!\!\leftarrow\,}{{\mathbf D_z}}\partial_{k_z}}
    \right) \,\, S_d  
  = {\cal C}_{\psi d}
\,, 
\label{S_eom_mixing_diagonalization}
\end{eqnarray}
where ${\cal C}_{\psi d} = \bX {\cal C}_{\psi} \bX^\dagger$, 
and we defined a `covariant' derivative 
\begin{equation}
  {\mathbf D}_\mu 
   = {\mathbbm 1}\otimes{\mathbbm 1}\partial_\mu 
   - i [{\mathbbm 1}\otimes\Sigma_\mu,\cdot] 
   - i\{\gamma^5\otimes\Delta_\mu,\cdot\} 
\,,
\label{mD}
\end{equation}  
where $\Sigma_\mu$ and $\Delta_\mu$ are given in terms of the rotation 
matrices $U$ and $V$ as follows:
\begin{eqnarray}
 \Delta_\mu &=& \frac i2 \left( V \del_\mu V^\dagger 
                              - U \del_\mu U^\dagger \right)
\label{mixing_Delta}\\
 \Sigma_\mu &=& \frac i2 \left( V \del_\mu V^\dagger 
                              + U \del_\mu U^\dagger \right)
\,.
\label{mixing_Sigma}
\end{eqnarray}
Note that the covariant derivative~(\ref{mD}) can be obtained from
\begin{equation}
\partial_\mu\bm = \partial_\mu(\bX^\dagger\bm_d\bY) 
                 = \bX^\dagger({\mathbf D}_\mu\bm_d)\bY
\label{mD2}
\end{equation}  
%
The higher order derivatives are obtained simply by iteration. 
Already from~(\ref{S_eom_mixing_diagonalization}-\ref{mD}) it is clear 
that the kinetic and constraint equations, when written in the mass eigenbasis,
will contain extra commutator and anticommutator terms. 
The choice of the rotation matrices $U$ and $V$ is not unique. After an
$x$-dependent phase redefinition $U \rightarrow wU$ and $V \rightarrow wV$,
where $w$ is a diagonal matrix with eigenvalues of absolute value 1, $U$ and
$V$ still diagonalize $m$. This freedom to redefine the rotation matrices
was the source of some problems in finding the correct physical source in
the WKB approach. We will find, however, that only the diagonal elements of 
the matrix $\Delta_\mu$ are of relevance, and these are invariant under this
reparametrization.

We now project out the spinor structure in~(\ref{S_eom_mixing_diagonalization})
by multiplying by $P_s(k)\{1,s\gamma^3\gamma^5,-is\gamma^3,-\gamma^5\}$ and 
taking the spinorial traces. The procedure is identical to the derivation 
of (\ref{ga0_eom}-\ref{ga3_eom}) at the beginning of
section~\ref{Constraint and kinetic equations}, except for the subtlety 
related to the covariant derivative, because it contains spinor structure.
The resulting equations are
\begin{eqnarray}
&&\!\!\!\!\!\!\!\!
    \Big(2i\tilde k_0 - {\cal D}_t^-\Big) g^s_{0d}
  -s \Big(2ik_z + {\cal D}_z\Big) g^s_{3d}
  -2i {m}_{hd} \,
    {\rm e}^{\frac i2 {\stackrel{\!\!\leftarrow}{D_z}\partial_{k_z}}} g^s_{1d}
  -2i {m}_{ad}
    {\rm e}^{\frac i2 {\stackrel{\!\!\leftarrow}{D_z}\partial_{k_z}}} g^s_{2d}
\label{g0d_eom}
\\
&&\phantom{sssssssssssssssssssssssssssssssssssssssssssssssssssssssssssss} 
         = \Tr {\mathbbm 1}P_s{\cal C}_{\psi d}
\nonumber\\
&&\!\!\!\!\!\!\!\!
    \Big(2i\tilde k_0 - {\cal D}_t^+\Big) g^s_{1d}
  -s \Big(2k_z -  i {\cal D}_z\Big) g^s_{2d}
  -2i {m}_{hd}  
    {\rm e}^{\frac i2 {\stackrel{\!\!\leftarrow}{D_z}\partial_{k_z}}} g^s_{0d}
  +2  {m}_{ad}  
    {\rm e}^{-\frac i2 {\stackrel{\!\!\leftarrow}{D_z}\partial_{k_z}}} g^s_{3d}
\label{g1d_eom}
\\
&&\phantom{sssssssssssssssssssssssssssssssssssssssssssssssssssssssssssss} 
 = \Tr (s\gamma^3\gamma^5) P_s{\cal C}_{\psi d}
\nonumber
\\
&&\!\!\!\!\!\!\!\!
   \Big(2i\tilde k_0 - {\cal D}_t^+\Big)  g^s_{2d}
  +s \Big(2k_z - i {\cal D}_z\Big) g^s_{1d}
  -2  {m}_{hd}  
    {\rm e}^{\frac i2 {\stackrel{\!\!\leftarrow}{D_z}\partial_{k_z}}} g^s_{3d}
  -2i {m}_{ad}  
    {\rm e}^{\frac i2 {\stackrel{\!\!\leftarrow}{D_z}\partial_{k_z}}} g^s_{0d}
\label{g2d_eom}
\\
&&\phantom{sssssssssssssssssssssssssssssssssssssssssssssssssssssssssssss} 
 = \Tr \left(-is\gamma^3\right)P_s{\cal C}_{\psi d}
\quad
\nonumber\\
&&\!\!\!\!\!\!\!\!
   \Big(2i\tilde k_0 - {\cal D}_t^-\Big)  g^s_{3d}
  -s \Big(2ik_z + {\cal D}_z\Big) g^s_{0d}
  +2 {m}_{hd}  
    {\rm e}^{\frac i2 {\stackrel{\!\!\leftarrow}{D_z}\partial_{k_z}}} g^s_{2d}
  -2 {m}_{ad}
    {\rm e}^{\frac i2 {\stackrel{\!\!\leftarrow}{D_z}\partial_{k_z}}} g^s_{1d}
\label{g3d_eom}
\\
&&\phantom{sssssssssssssssssssssssssssssssssssssssssssssssssssssssssssss} 
 = \Tr \left(-\gamma^5\right)P_s{\cal C}_{\psi d}
\,,
\nonumber
\end{eqnarray}
where we have defined the derivatives 
\begin{eqnarray}
  D_z m_{hd}  &\equiv & \partial_z m_{hd} 
               - i[\Sigma_z,m_{hd}] - \{\Delta_z, m_{ad}\} 
\nonumber\\
  D_z m_{ad} &\equiv & \partial_z m_{ad}
               - i[\Sigma_z,m_{ad}] + \{\Delta_z, m_{hd}\} 
\label{Dm_ha}
\end{eqnarray}
and 
\begin{eqnarray}
{\cal D}_t^- &=& \gamma_\| \partial_t 
              + \gamma_\|\vec v_\|\cdot\nabla_\|
              - is[\Delta_z,\cdot]
\label{D_t-}
\\
{\cal D}_t^+ &=& \gamma_\| \partial_t 
              + \gamma_\|\vec v_\|\cdot\nabla_\|
              - is\{\Delta_z,\cdot\}
\label{D_t+}
\\
 {\cal D}_z  &=& \partial_z - i[\Sigma_z,\cdot] 
\label{D_z}
\end{eqnarray}
Note that due to the $\gamma^5$ the anticommutator in~(\ref{mD}) has become
the commutator in~(\ref{D_t-}). Moreover, the derivative
${\cal D}_t^+$ is not hermitean. Indeed, the
anticommutator term in~(\ref{D_t+}) is antihermitean.

\subsubsection{Constraint equations}
\label{Constraint equations}

The constraint equations correspond to the antihermitean
parts of the equations (\ref{g0d_eom})-(\ref{g3d_eom}).
Since we are interested in the dispersion relation accurate to first order
in $\hbar$, it suffices to consider these equations to first order in 
gradients:
\begin{eqnarray}
 \!  2\tilde k_0 g^s_{0d}
 - 2s k_z g^s_{3d}
 - \big\{m_{hd}, g^s_{1d}\big\}
 - \frac i2 \big[D_z m_{hd},\partial_{k_z} g^s_{1d}\big]
  - \big\{m_{ad}, g^s_{2d}\big\} 
  - \frac i2 \big[D_z m_{ad}, \partial_{k_z} g^s_{2d}\big]
\!\!  &=&\! {\cal C}^s_{0d}
\quad
\nonumber\\
\label{ce0d1}
\\
    2\tk g^s_{1d} + s\{\Delta_z, g^s_{1d}\}
  + s(\partial_z - i[\Sigma_z,\cdot])g^s_{2d}
 - \big\{m_{hd}, g^s_{0d}\big\}
 - \frac i2 \big[D_z m_{hd},\partial_{k_z} g^s_{0d}\big]
&&  
\label{ce1d1}
\\
  + \frac 12 \big\{D_z m_{ad}, \partial_{k_z} g^s_{3d}\big\} 
  -i \big[m_{ad}, g^s_{3d}\big]
 &=& {\cal C}^s_{1d}
\nonumber\\
    2\tilde k_0 g^s_{2d} 
  + s\{\Delta_z, g^s_{2d}\}
  - s(\partial_z - i[\Sigma_z,\cdot])g^s_{1d}
  - \frac 12 \big\{D_z m_{hd}, \partial_{k_z} g^s_{3d}\big\}
  + i\big[m_{hd}, g^s_{3d}\big]
&&
\label{ce2d1}
\\
  - \big\{m_{ad}, g^s_{0d}\big\} 
  - \frac i2 \big[D_z m_{ad}, \partial_{k_z} g^s_{0d}\big]
 &=& {\cal C}^s_{2d}
\nonumber\\
 \!  2 \tilde k_0 g^s_{3d}
 \!-\! 2s k_z g^s_{0d}
 \!+\! \frac 12 \big\{D_z m_{hd}, \partial_{k_z} g^s_{2d}\big\}
 \!-\! i\big[m_{hd}, g^s_{2d}\big]
  \!-\! \frac 12 \big\{D_z m_{ad},\partial_{k_z} g^s_{1d}\big\} 
  + i \big[m_{ad}, g^s_{1d}\big]
 \!\! &=& \!{\cal C}^s_{3d}
\,,
\quad
\nonumber\\
\label{ce3d1}
\end{eqnarray}
where 
\begin{eqnarray}
 {\cal C}^s_{0d} &=& \frac{1}{2i} \, {\rm Tr}{\mathbbm 1}
                     \Big( P_s(k){\cal C}_{\psi d}
                          -P^\dagger_s(k){\cal C}_{\psi d}^\dagger \Big)
\nonumber\\
 {\cal C}^s_{1d} &=& \frac{1}{2i} \, {\rm Tr}(s\gamma^3\gamma^5)
                     \Big( P_s(k){\cal C}_{\psi d}
                          -P^\dagger_s(k){\cal C}_{\psi d}^\dagger \Big)
\nonumber\\
 {\cal C}^s_{2d} &=& \frac{1}{2i} \, {\rm Tr}(-is\gamma^3)
                     \Big( P_s(k){\cal C}_{\psi d}
                          -P^\dagger_s(k){\cal C}_{\psi d}^\dagger \Big)
\nonumber\\
 {\cal C}^s_{3d} &=& \frac{1}{2i} \, {\rm Tr}(-\gamma^5)
                     \Big( P_s(k){\cal C}_{\psi d}
                          -P^\dagger_s(k){\cal C}_{\psi d}^\dagger \Big)
\,.
\label{ced:collision-terms:0-3}
\end{eqnarray}
The trace is only to be taken in spinor space, not in the fermionic
flavor space.
For simplicity we now perform the analysis for two mixing
fermions. Our findings are however valid for an arbitrary number of mixing 
fermions. Taking account of the fact that the off-diagonals are of the order
$\hbar$, equations~(\ref{ce0d1}-\ref{ce3d1}) now imply 
the following diagonal equations accurate to order $\hbar$,
\begin{eqnarray}
             2\tilde k_0\,{g_{0d}^s}_{{11}}
             - 2{m_{hd}}_{{1}}\, {g_{1d}^s}_{{11}}
             - 2{m_{ad}}_{{1}}\, {g_{2d}^s}_{{11}}
             - 2sk_z \, {g_{3d}^s}_{{11}}
             \phantom{s}
         &=& {{\cal C}^s_{0d}}_{{11}}
\label{ced:diag0}
\\
             - 2{m_{hd}}_{{1}}\, {g_{0d}^s}_{{11}}
             + 2\tilde k_0\,{g_{1d}^s}_{{11}}
             + 2s{\Delta_z}_{{11}} \, {g_{1d}^s}_{{11}}
             + s\partial_z \, {g_{2d}^s}_{{11}}
             + {(D_zm_{ad})}_{{11}}\partial_{k_z} {g_{3d}^s}_{{11}}
             \phantom{}
         &=& {{\cal C}^s_{1d}}_{{11}}
\label{ced:diag1}
\\
             - 2{m_{ad}}_{{1}}\, {g_{0d}^s}_{{11}}
             - s\partial_z \, {g_{1d}^s}_{{11}}
             + 2\tilde k_0\,{g_{2d}^s}_{{11}}
             + 2s{\Delta_z}_{{11}} \, {g_{2d}^s}_{{11}}
             - {(D_zm_{hd})}_{{11}}\partial_{k_z} {g_{3d}^s}_{{11}}
             \phantom{}
         &=& {{\cal C}^s_{1d}}_{{11}}
\label{ced:diag2}
\\
             - 2sk_z \, {g_{0d}^s}_{{11}}
             - {(D_zm_{ad})}_{{11}}\partial_{k_z} {g_{1d}^s}_{{11}}
             + {(D_zm_{hd})}_{{11}}\partial_{k_z} {g_{2d}^s}_{{11}}
             + 2\tilde k_0\,{g_{3d}^s}_{{11}}
             \phantom{s}
         &=& {{\cal C}^s_{3d}}_{{11}}
\,,
\label{ced:diag3}
\end{eqnarray}
and similarly for the $(22)$-components. 
Solving the latter three equations in gradient expansion in terms of 
${g_{0d}^s}_{{11}}$ we find
\begin{eqnarray}
{g_{1d}^s}_{{11}} &=& \frac{1}{\tilde k_0}
    \Big[
             {m_{hd}}_{{1}}\, {g_{0d}^s}_{{11}}
             - \frac{1}{\tilde k_0} s{\Delta_z}_{{11}} \, 
                  ({m_{hd}}_{{1}} \, {g_{0d}^s}_{{11}})
             - \frac{1}{2\tilde k_0} s\partial_z \, 
                  ({m_{ad}}_{{1}} \, {g_{0d}^s}_{{11}})
\nonumber\\
&& \phantom{ssss}
      - \frac{s}{2\tilde k_0}{(D_zm_{ad})}_{11}(1 + k_z\partial_{k_z})
          {g_{0d}^s}_{{11}}
      +  \frac 12{{\cal C}^s_{1d}}_{{11}}
    \Big]
\label{ced:diag1b}
\\
{g_{2d}^s}_{{11}} &=& \frac{1}{\tilde k_0}
    \Big[
              {m_{ad}}_{{1}}\, {g_{0d}^s}_{{11}}
             + \frac{1}{2\tilde k_0}s\partial_z \, 
                 ({m_{hd}}_{{1}} {g_{0d}^s}_{{11}})
             - \frac{1}{\tilde k_0} s{\Delta_z}_{{11}} 
                 ({m_{ad}}_{{1}} {g_{0d}^s}_{{11}})
\nonumber\\
&& \phantom{ssss}
             + \frac{s}{2\tilde k_0}{(D_zm_{hd})}_{{11}}
                 (1+k_z \partial_{k_z}){g_{0d}^s}_{11}
             + \frac 12 {{\cal C}^s_{2d}}_{{11}}
    \Big]
\label{ced:diag2b}
\\
{g_{3d}^s}_{{11}} &=& \frac{1}{\tilde k_0}
    \Big[
             sk_z \, {g_{0d}^s}_{{11}}
             + \frac{1}{2\tilde k_0}
                  {m_{hd}}_{{1}}{(D_zm_{ad})}_{{11}}\partial_{k_z}
                  {g_{0d}^s}_{{11}}
             - \frac{1}{2\tilde k_0}
                  {m_{ad}}_{{1}}{(D_zm_{hd})}_{{11}}\partial_{k_z}
               {g_{0d}^s}_{{11}}
             + \frac 12{{\cal C}^s_{3d}}_{{11}}
    \Big]
\,.\quad
\label{ced:diag3b}
\end{eqnarray}
Here ${m_{hd}}_{{1}}$ denotes the first diagonal element of $m_{hd}$,
${m_{ad}}_{{1}}$ is defined correspondingly.
Inserting these relations into~(\ref{ced:diag0}) we get the
constraint for the diagonal densities
\beqa
&&
 \frac{2}{\tilde{k}_0}
           \Big(  k^2
                 -|m_d|^2_{i}
                 + \frac{s}{\tilde k_0}
                    \big[  |m_d|^2_{i}(\partial_z {\theta_d}_{i} 
                          + 2{\Delta_z}_{\tt ii})
                    \big] 
           \Big)\,{g_{0d}^s}_{\tt ii}
\nonumber\\
&&\hphantom{XXXXX}
  =  {{\cal C}^s_{0d}}_{\tt ii}  
      + \frac{1}{\tilde k_0}
         \big(  {m_{hd}}_{i}{{\cal C}^s_{1d}}_{\tt ii}  
              + {m_{ad}}_{i}{{\cal C}^s_{2d}}_{\tt ii}  
              +  sk_z {{\cal C}^s_{3d}}_{\tt ii}
         \big)
\,, 
\label{ced:diag0b}
\eeqa
where we defined 
\begin{eqnarray}
  |m_d|^2_{{i}}
       &=&  {m_{hd}}_{i}^2 + {m_{ad}}_{i}^2
\nonumber\\
  |m_d|^2_{{i}}\partial_z {\theta_d}_{i} 
       &=&   {m_{hd}}_{{i}}\partial_z{m_{ad}}_{{i}}
        - {m_{ad}}_{{i}}\partial_z{m_{hd}}_{{i}}
\,,
\label{ced:diag0c}
\end{eqnarray}
where $i=1,2$.
Note that the energy shift in~(\ref{ced:diag0b}) can be written 
in terms of the rotation matrices $U$ as 
\begin{equation}
  |m_d|^2_{i}(\partial_z {\theta_d}_{i} + 2{\Delta_z}_{\tt ii})
   = - \Im[U(m\partial_z m^\dagger)U^\dagger]_{\tt ii}
\,.
\label{shift:U}
\end{equation}

\subsubsection{Quasiparticle picture and dispersion relation}
\label{Quasiparticle picture and dispersion relation}

Equation~(\ref{ced:diag0b})
specifies the spectral
properties of the plasma excitations. Since it
is an algebraic constraint, the {\it quasiparticle picture}
remains a valid description of the plasma to order $\hbar$,
as it was already pointed out 
in~\cite{KainulainenProkopecSchmidtWeinstock:2001,
KainulainenProkopecSchmidtWeinstock:2002}. 
Since the constraint equation 
contains no time dependence, it measures genuine 
spectral properties independent of the actual dynamical 
populations of the states.

The homogeneous solution of the constraint equation~(\ref{ced:diag0b})
has the following spectral form 
\beqa
     {g_{0d}^{<s}}_{\tt ii}(k,x)
 &=& 2\pi \delta\Big(  k^2
                     - |m_d|^2_{i}
                     + \frac{s}{\tilde k_0}
                       \big[  |m_d|^2_{i}(\partial_z {\theta_d}_{i} 
                            + 2{\Delta_z}_{\tt ii})
                       \big] 
                \Big)
     |\tilde{k}_0|\,n_{si}(k,x)
\nonumber\\
 &=& 2\pi \sum_{\pm} \frac{\delta(k_0\mp\omega_{\pm si})}
                          {2\omega_{\pm si}Z_{\pm si}} \,
     |\tilde{k}_0|\,n_{si}(k,x)
\,,
\label{spectral-solution:fermions}
\eeqa
with the following {\it dispersion relations} for fermions 
\begin{equation}
  \omega_{si} =  \omega_{0i}
             - \frac{s}{2\tilde \omega_{0i}\omega_{0i}}
   \big[  |m_d|^2_{i}(\partial_z {\theta_d}_{i} 
        + 2{\Delta_z}_{\tt ii})
   \big] 
,
\label{dispersion-relation:fermions}
\end{equation}
where 
\begin{eqnarray}
         \omega_{0i} &=& ({\vec k^2 + |m_d|^2_{i}})^\frac 12
\nonumber\\
   \tilde\omega_{0i} &=& ({k_z^2    + |m_d|^2_{i}})^\frac 12
\,.
\label{dispersion-relation:fermions2}
\end{eqnarray}
The normalization factors are
\beq
Z_{si} = 1 - \frac{s|m_d|^2_{i}(\partial_z {\theta_d}_{i} 
                        + 2{\Delta_z}_{\tt ii})}{2\tilde \omega_{0i}^3} 
\,.
\label{Zsipm}
\eeq
We have normalized the solution~(\ref{spectral-solution:fermions})
such that $n_{si}(k,x)$ represents the particle density in phase space
$\{k,x\}$.
We emphasize that the spectral solution~(\ref{spectral-solution:fermions})
and the dispersion relation~(\ref{dispersion-relation:fermions})
are valid in general for all plasma excitations:
equilibrium and stationary excitations,
as well as for non-equilibrium, time dependent transients that conserve spin,
that is for ${g^s_{0d}} = {g^s_{0d}}(k,t-\vec k_\|\cdot\vec x_\|,z)$,
assuming of course planar symmetry in the wall frame.

Since in the on-shell limit the Wigner functions $S^<$ and $S^>$ satisfy
identical Kadanoff-Baym equations~(\ref{Wigner-space:fermionic_eom}),
and in addition they are related by the fermionic 
sum rule~(\ref{sum_rule_fermions}), based on the above analysis
of $S^<$ we can easily reconstruct the spectral solution for the  
Wigner function $S^>$, which we write as
\begin{equation}
  S^>_d = \sum_{s=\pm 1} iP_s(k)
                   \left[  s\gamma^3\gamma^5 g^{>s}_{0d}
                         - s\gamma^3         g^{>s}_{3d}
                         + {\mathbbm 1}      g^{>s}_{1d}
                         - i\gamma^5         g^{>s}_{2d}  \right]
,
\label{S>_decomposition}
\end{equation}
where the spectral solution for $g^{>s}_{0d}$ reads
\begin{equation}
    g_{0d\tt ii}^{>s}(k,x)
  = - 2\pi \sum_{\pm} \frac{\delta(k_0\mp\omega_{\pm si})}
                          {2\omega_{\pm si}Z_{\pm si}} 
      \, |\tilde{k}_0| \, [1-n_{si}(k,x)]
\,.
\label{spectral-solution:fermions>}
\end{equation}
Note that $g_{0d\tt ii}^{>s}$
obeys the same equation~(\ref{ced:diag0b}) as $g_{0d\tt ii}^{<s}$. 
Similarly, $g_{1d\tt ii}^{>s}$, $g_{2d\tt ii}^{>s}$ 
and $g_{3d\tt ii}^{>s}$ are related to $g_{0d\tt ii}^{>s}$ 
the same way as indicated in 
Eqs.~(\ref{ced:diag1b}-\ref{ced:diag3b}).

\subsubsection{Plasma frame}
\label{Plasma frame}

 An interesting question is of course how to generalize the solution for
the Wigner function~(\ref{S<_decomposition2})
to other Lorentz frames, an important example being the plasma rest frame.
The natural solution is to apply a Lorentz boost 
on the Wigner function. The simplest boost
operator corresponding to a boost $v$ in $z$-direction 
can be easily constructed in analogy with~(\ref{L-Lambda}), and it reads
\begin{equation}
  L(\Lambda_z) = \frac{\gamma + 1 - \gamma v\cdot\gamma^0\gamma^3}
                       {[{2(\gamma+1)}]^{1/2}},
\label{L-Lambdaz}
\end{equation}
where $\gamma = (1-v^2)^{-1/2}$.
The Wigner function then transforms as
\begin{eqnarray}
\breve{S}_s &=& L^{-1}(\Lambda_z)S_sL(\Lambda_z)
\nonumber\\
                 &=& - \breve{P}_s(k) 
                  \left[
                      \gamma(1+v\gamma^0\gamma^3)  
                      \left(  s\gamma^3\gamma^5 \tilde g^{s}_0
                    - s\gamma^3 \tilde g^{s}_3
                  \right)
                    + {\mathbbm 1} \tilde g^{s}_1
                    - i\gamma^5 \tilde g^{s}_2           
                  \right]
\,,
\label{boost:v}
\end{eqnarray}
where
\begin{eqnarray}
    \breve{P}_s(\breve{k}) 
               = \frac{\gamma(\breve k_0 - v\breve{k}_z)}{\breve{\tilde k}_0}
                   \gamma^0\gamma^3\gamma^5
               - \gamma(1+v\gamma^0\gamma^3)
                  \frac{\breve{\vec k}_\|}{\breve{\tilde k}_0}
                   \cdot\vec \gamma_\|\gamma^3\gamma^5
\label{brevePs}
\end{eqnarray}
is the spin operator
with $\breve{\tilde k}_0^2 
 = \gamma^2(\breve k_0 - v\breve{k}_z)^2-\breve{\vec k}_\|^2$, and 
$g_a^s=g_a^s(\breve k,\breve x)$. We leave as an exercise
to the reader to find the direction of the spin vector $\breve s^\mu$ 
corresponding to the spin operator~(\ref{brevePs}). 
To get the final form for the Wigner function in the new frame,
one also needs to transform the coordinates and derivatives
in~(\ref{ced:diag1b}-\ref{ced:diag3b}) and~(\ref{spectral-solution:fermions}),
which makes the final expression for $i\breve S^<_s$ rather cumbersome.
To get the Wigner function 
$i\breve S_s$ in the {\it plasma frame}, one needs to choose the boost 
that corresponds to $v = v_w$ and $\gamma = \gamma_w = (1-v_w^2)^{-1/2}$.  
This exercise underlines clearly the advantages of working in the
wall frame, in which the Wigner function has a particularly simple form.
Some authors~\cite{CarenaMorenoQuirosSecoWagner:2000,
CarenaQuirosSecoWagner:2002} choose nevertheless to work in the plasma frame,
although in a simplistic disguise.

\subsubsection{Kinetic equations}
\label{Kinetic equations}

We now take  (minus) the hermitean part of 
Eqs.~(\ref{g0d_eom}-\ref{g3d_eom}) and get the following 
{\it kinetic equations}
%
\begin{eqnarray}
&&
{\cal D}_t^- g^s_{0d}
+ s {\cal D}_z g^s_{3d}
  - \{m_{hd}     
          \sin\big(\frac 12 {\stackrel{\leftarrow}{D_z}\!\partial_{k_z}}\big),
      g^s_{1d}\}
    + i[m_{hd}     
          \cos\big(\frac 12 {\stackrel{\leftarrow}{D_z}\!\partial_{k_z}}\big),
      g^s_{1d}]
\nonumber\\
&&
\phantom{sssssssssssssss}
  - \{m_{ad}     
          \sin\big(\frac 12 {\stackrel{\leftarrow}{D_z}\!\partial_{k_z}}\big),
      g^s_{2d}\}
    + i[m_{ad}     
          \cos\big(\frac 12 {\stackrel{\leftarrow}{D_z}\!\partial_{k_z}}\big),
      g^s_{2d}]
         = {\cal K}^s_{0d}
\quad
\label{ked0}
\\
&&
    ({\cal D}_t^+)_h g^s_{1d}
  +2sk_z g^s_{2d} 
  - \{m_{hd}     
          \sin\big(\frac 12 {\stackrel{\leftarrow}{D_z}\!\partial_{k_z}}\big),
      g^s_{0d}\}
    + i[m_{hd}     
          \cos\big(\frac 12 {\stackrel{\leftarrow}{D_z}\!\partial_{k_z}}\big),
      g^s_{0d}]
\phantom{sssssssss}
\nonumber\\
&&
\phantom{sssssssssssssss}
  - \{m_{ad}     
          \cos\big(\frac 12 {\stackrel{\leftarrow}{D_z}\!\partial_{k_z}}\big),
      g^s_{3d}\}
    - i[m_{ad}     
          \sin\big(\frac 12 {\stackrel{\leftarrow}{D_z}\!\partial_{k_z}}\big),
      g^s_{3d}]
 = {\cal K}^s_{1d}
\quad
\label{ked1}
\\
&&
    ({\cal D}_t^+)_h g^s_{2d}
  -2sk_z g^s_{1d} 
   + \{m_{hd}     
          \cos\big(\frac 12 {\stackrel{\leftarrow}{D_z}\!\partial_{k_z}}\big),
      g^s_{3d}\}
   + i[m_{hd}     
          \sin\big(\frac 12 {\stackrel{\leftarrow}{D_z}\!\partial_{k_z}}\big),
      g^s_{3d}]
\phantom{sssssssss}
\nonumber\\
&&
\phantom{sssssssssssssss}
  - \{m_{ad}     
          \sin\big(\frac 12 {\stackrel{\leftarrow}{D_z}\!\partial_{k_z}}\big),
      g^s_{0d}\}
    + i[m_{ad}     
          \cos\big(\frac 12 {\stackrel{\leftarrow}{D_z}\!\partial_{k_z}}\big),
      g^s_{0d}]
 = {\cal K}^s_{2d}
\quad
\label{ked2}
\\
&&
    {\cal D}_t^-  g^s_{3d}
  +s {\cal D}_z g^s_{0d}
  - \{m_{hd}     
          \cos\big(\frac 12 {\stackrel{\leftarrow}{D_z}\!\partial_{k_z}}\big),
      g^s_{2d}\}
  - i[m_{hd}     
          \sin\big(\frac 12 {\stackrel{\leftarrow}{D_z}\!\partial_{k_z}}\big),
      g^s_{2d}]
\nonumber\\
&&
\phantom{sssssssssssssss}
  + \{m_{ad}     
          \cos\big(\frac 12 {\stackrel{\leftarrow}{D_z}\!\partial_{k_z}}\big),
      g^s_{1d}\}
    + i[m_{ad}     
          \sin\big(\frac 12 {\stackrel{\leftarrow}{D_z}\!\partial_{k_z}}\big),
      g^s_{1d}]
 = {\cal K}^s_{3d}
\label{ked3}
\,,
\quad
\end{eqnarray}
where $({\cal D}_t^+)_h = \gamma_\|(\partial_t -i[\Sigma_t,\cdot])
 + \gamma_\|\vec v_\|\cdot \nabla_\| $
denotes the hermitean part of the derivative, and 
\begin{eqnarray}
 {\cal K}^s_{0d} &=& -\frac{1}{2} \, {\rm Tr}{\mathbbm 1}
                     \Big( P_s(k){\cal C}_{\psi d}
                          +P^\dagger_s(k){\cal C}_{\psi d}^\dagger \Big)
\nonumber\\
 {\cal K}^s_{1d} &=& -\frac{1}{2} \, {\rm Tr}(s\gamma^3\gamma^5)
                     \Big( P_s(k){\cal C}_{\psi d}
                          +P^\dagger_s(k){\cal C}_{\psi d}^\dagger \Big)
\nonumber\\
 {\cal K}^s_{2d} &=& -\frac{1}{2} \, {\rm Tr}(-is\gamma^3)
                     \Big( P_s(k){\cal C}_{\psi d}
                          +P^\dagger_s(k){\cal C}_{\psi d}^\dagger \Big)
\nonumber\\
 {\cal K}^s_{3d} &=& -\frac{1}{2} \, {\rm Tr}(-\gamma^5)
                     \Big( P_s(k){\cal C}_{\psi d}
                          +P^\dagger_s(k){\cal C}_{\psi d}^\dagger \Big)
\,.
\label{ked:collision-terms:0-3}
\end{eqnarray}
Again the trace must only be taken in spinor space.
Since we are interested in the order $\hbar$ effects, we can truncate
the kinetic equations~(\ref{ked0}-\ref{ked3}) to second order in gradients.
For example, the kinetic equation for particle density of spin 
$s$~(\ref{ked0}) reads 
\begin{eqnarray}
&&
   (\gamma_\|\partial_t + \gamma_\|\vec v_\|\cdot\nabla_\|
  - is[\Delta_z,\cdot]) g^s_{0d}
  + s(\partial_z -i[\Sigma_z,\cdot]) g^s_{3d}
\nonumber\\
&&
\phantom{sssssss}
  - \frac 12 \{D_z m_{hd}, \partial_{k_z} g^s_{1d}\}
    + i[m_{hd},  g^s_{1d}]
  - \frac 12 \{D_z m_{ad}, \partial_{k_z} g^s_{2d}\}
    + i[m_{ad}, g^s_{2d}]
         = {\cal K}^s_{0d}
\,,
\quad
\label{ked0-2}
\end{eqnarray}
where we dropped the second order terms in the commutator, which is legitimate
since they contribute only to the off-diagonal equations, 
and hence to order $\hbar^2$. 

 From~(\ref{ked0-2}) we can immediately write the kinetic equation 
for the diagonal densities 
\begin{eqnarray}
&&  \gamma_\|(\partial_t + \vec v_\|\cdot\nabla_\|)  {g^s_{0d}}_{11}
  +  s\partial_z {g^s_{3d}}_{11}
  -  (D_z m_{hd})_{11} \partial_{k_z} {g^s_{1d}}_{11}
  -  (D_z m_{ad})_{11} \partial_{k_z} {g^s_{2d}}_{11}
\nonumber\\
&\mp& is({\Sigma_{z}}_{12}{g_{3d}^s}_{21}-{\Sigma_{z}}_{21}{g_{3d}^s}_{12}
        + {\Delta_{z}}_{12}{g_{0d}^s}_{21}-{\Delta_{z}}_{21}{g_{0d}^s}_{12})
\label{ked0-diagB}
\\
&-& \frac 12\Big[
                 (D_z m_{hd})_{12} \partial_{k_z} {g^s_{1d}}_{21}
              +  (D_z m_{hd})_{21} \partial_{k_z} {g^s_{1d}}_{12}
              +  (D_z m_{ad})_{12} \partial_{k_z} {g^s_{2d}}_{21}
              +  (D_z m_{ad})_{21} \partial_{k_z} {g^s_{2d}}_{12}
            \Big]
         = {{\cal K}^s_{0d}}_{11}
.
\nonumber
\end{eqnarray}
In order to decouple the diagonal and off-diagonal equations we make use 
of the off-diagonal equations, which to leading order in gradients are
\begin{eqnarray}
&& \!\!\!\!\!\!\!\!   
        i\delta(m_{hd}){g_{1d}^s}_{12} 
     +  i\delta(m_{ad}){g_{2d}^s}_{12} 
\label{ke-off0}
\\
&&
      = - is{\Delta_{z}}_{12} \delta({g_{0d}^s})
        + \frac 12 (D_z m_{hd})_{12}\partial_{k_z} \Tr({g_{1d}^s})
        + \frac 12 (D_z m_{ad})_{12}\partial_{k_z} \Tr({g_{2d}^s})
        - is{\Sigma_{z}}_{12} \delta({g_{3d}^s})
        + {{\cal K}^s_{0d}}_{12}
\nonumber
\\
&& \!\!\!\!\!\!\!\!   
        i\delta(m_{hd}){g_{0d}^s}_{12} 
     +  2sk_z {g_{2d}^s}_{12}
     -  \Tr(m_{ad}){g_{3d}^s}_{12} 
\label{ke-off1}
\\
&&
      =   \frac 12 (D_z m_{hd})_{12}\partial_{k_z} \Tr({g_{0d}^s})
        - \frac i2 (D_z m_{ad})_{12}\partial_{k_z} \delta({g_{3d}^s})
        + {{\cal K}^s_{1d}}_{12}
\nonumber
\\
&& \!\!\!\!\!\!\!\!   
        i\delta(m_{ad}){g_{0d}^s}_{12} 
     -  2sk_z {g_{1d}^s}_{12}
     +  \Tr(m_{hd}){g_{3d}^s}_{12} 
\label{ke-off2}
\\
&&
      =   \frac 12 (D_z m_{ad})_{12}\partial_{k_z} \Tr({g_{0d}^s})
        + \frac i2 (D_z m_{hd})_{12}\partial_{k_z} \delta({g_{3d}^s})
        + {{\cal K}^s_{2d}}_{12}
\nonumber
\\
&& \!\!\!\!\!\!\!\!
        \Tr(m_{ad}){g_{1d}^s}_{12} 
     -  \Tr(m_{hd}){g_{2d}^s}_{12} 
\label{ke-off3}\\
&&    = - is{\Sigma_{z}}_{12} \delta({g_{0d}^s})
        + \frac i2 (D_z m_{ad})_{12}\partial_{k_z} \delta({g_{1d}^s})
        - \frac i2 (D_z m_{hd})_{12}\partial_{k_z} \delta({g_{2d}^s})
        - is{\Delta_{z}}_{12} \delta({g_{3d}^s})
        + {{\cal K}^s_{3d}}_{12}
\,.
\nonumber
\end{eqnarray}
Combining~(\ref{ke-off0}) and~(\ref{ke-off3}) we obtain
\begin{eqnarray}
  {g_{1d}^s}_{12} &=& \frac{1}{\delta(|m_d|)^2} 
  \Big\{
    - s(D_z m_{ad})_{12}\delta(g_{0d}^s)
    - is[{\Delta_z}_{12}\delta(m_{ad}) 
    - i{\Sigma_z}_{12}\Tr(m_{hd})]\delta(g_{3d}^s)
\nonumber\\
&&
\phantom{sssssssss}
- \frac i2  [ \Tr(m_{hd})(D_z m_{hd})_{12} \partial_{k_z}\Tr(g_{1d}^s)
            - \Tr(m_{ad})(D_z m_{ad})_{12} \partial_{k_z}\delta(g_{1d}^s)] 
\nonumber\\
&&
\phantom{sssssssss}
- \frac i2  [ \Tr(m_{hd})(D_z m_{ad})_{12} \partial_{k_z}\Tr(g_{2d}^s)
            + \Tr(m_{ad})(D_z m_{hd})_{12} \partial_{k_z}\delta(g_{2d}^s)] 
\nonumber\\
&&
\phantom{sssssssss}
           - i\Tr(m_{hd}) {{\cal K}^s_{0d}}_{12}
           + \delta(m_{ad}) {{\cal K}^s_{3d}}_{12}
  \Big\}
\label{ke-off1b}
\\
  {g_{2d}^s}_{12} &=& \frac{1}{\delta(|m_d|)^2} 
  \Big\{\phantom{ss}
    s(D_z m_{hd})_{12}\delta(g_{0d}^s)
    + is[{\Delta_z}_{12}\delta(m_{hd}) 
          + i{\Sigma_z}_{12}\Tr(m_{ad})]\delta(g_{3d}^s)
\nonumber\\
&&
\phantom{sssssssss}
- \frac i2  [ \Tr(m_{ad})(D_z m_{hd})_{12} \partial_{k_z}\Tr(g_{1d}^s)
            + \Tr(m_{hd})(D_z m_{ad})_{12} \partial_{k_z}\delta(g_{1d}^s)] 
\nonumber\\
&&
\phantom{sssssssss}
- \frac i2  [ \Tr(m_{ad})(D_z m_{ad})_{12} \partial_{k_z}\Tr(g_{2d}^s)
            - \Tr(m_{hd})(D_z m_{hd})_{12} \partial_{k_z}\delta(g_{2d}^s)] 
\nonumber\\
&&
\phantom{sssssssss}
           - i\Tr(m_{ad}) {{\cal K}^s_{0d}}_{12}
           - \delta(m_{hd}) {{\cal K}^s_{3d}}_{12}
  \Big\}
\,,
\label{ke-off2b}
\end{eqnarray}
where 
\begin{equation}
  \delta(|m_d|)^2 = \Tr(m_{hd})\delta(m_{hd})
                  + \Tr(m_{ad})\delta(m_{ad})
\,.
\label{deltamd2}
\end{equation}
From~(\ref{ke-off0}-\ref{ke-off3}) and~(\ref{ke-off1b}-\ref{ke-off2b})
we then get 
\begin{eqnarray}
  {g_{0d}^s}_{12} &=& \frac{1}{\delta(|m_d|)^2} 
  \Big\{
    -2k_z[ {\Sigma_z}_{12}\delta(g_{0d}^s) 
          +{\Delta_z}_{12}\delta(g_{3d}^s) ]
    + \frac 12 {\Sigma_z}_{12}\delta(|m_d|^2)\partial_{k_z}\Tr(g_{0d}^s)
\nonumber\\
&&
\phantom{sssssssss}\,
      + sk_z[ (D_z m_{ad})_{12}\partial_{k_z}\delta(g_{1d}^s)
            - (D_z m_{hd})_{12}\partial_{k_z}\delta(g_{2d}^s)]
\nonumber\\
&&
\phantom{sssssssss}\,
- \frac 12  [ \Tr(m_{hd})(D_z m_{ad})_{12} 
            - \Tr(m_{ad})(D_z m_{hd})_{12} ] 
                         \partial_{k_z}\delta(g_{3d}^s)
\nonumber\\
&&
\phantom{sssssssss}\,
           - i\Tr(m_{hd}) {{\cal K}^{s}_{1d}}_{12}
           - i\Tr(m_{ad}) {{\cal K}^{s}_{2d}}_{12}
           - 2isk_z {{\cal K}^{s}_{3d}}_{12}
  \Big\}
\label{ke-off0b}
\\
  {g_{3d}^s}_{12} &=& \frac{1}{\delta(|m_d|)^2} 
  \Big\{
    -2k_z[ {\Delta_z}_{12}\delta(g_{0d}^s) 
          +{\Sigma_z}_{12}\delta(g_{3d}^s) ]
    + \frac 12 {\Delta_z}_{12}\delta(|m_d|^2)\partial_{k_z}\Tr(g_{0d}^s)
\nonumber\\
&&
\phantom{sssssssss}\,
      - isk_z[ (D_z m_{hd})_{12}\partial_{k_z}\Tr(g_{1d}^s)
             + (D_z m_{ad})_{12}\partial_{k_z}\Tr(g_{2d}^s)]
\nonumber\\
&&
\phantom{sssssssss}\,
+ \frac i2  [ \delta(m_{ad})(D_z m_{ad})_{12} 
            + \delta(m_{hd})(D_z m_{hd})_{12} ] 
                         \partial_{k_z}\delta(g_{3d}^s)
\nonumber\\
&&
\phantom{sssssssss}\,
           - 2isk_z         {{\cal K}^s_{0d}}_{12}
           - \delta(m_{ad}) {{\cal K}^s_{1d}}_{12}
           - \delta(m_{hd}) {{\cal K}^s_{2d}}_{12}
  \Big\}
\,.
\label{ke-off3b}
\end{eqnarray}
Equations~(\ref{ke-off1b}-\ref{ke-off3b}) are the off-diagonal densities
correct to leading (first) order in gradients.
We are now ready to compute the contribution from the off-diagonal densities
in the kinetic equation~(\ref{ked0-diagB}). Making use of the hermiticity 
properties
${\Delta_z}_{12}^* = {\Delta_z}_{21}$, 
${\Sigma_z}_{12}^* = {\Sigma_z}_{21}$, 
${g_{ad}^s}_{12}^* = {g_{ad}^s}_{21}$, and~(\ref{ke-off1b}-\ref{ke-off3b}),
after some algebra one finds that the contribution from the off-diagonal
densities vanishes:
\begin{eqnarray}
{\Sigma_{z}}_{12}{g_{3d}^s}_{21}-{\Sigma_{z}}_{21}{g_{3d}^s}_{12}
        + {\Delta_{z}}_{12}{g_{0d}^s}_{21}-{\Delta_{z}}_{21}{g_{0d}^s}_{12}
    &=& 0 
\nonumber\\
                 (D_z m_{hd})_{12} \partial_{k_z} {g^s_{1d}}_{21}
              +  (D_z m_{hd})_{21} \partial_{k_z} {g^s_{1d}}_{12}
              +  (D_z m_{ad})_{12} \partial_{k_z} {g^s_{2d}}_{21}
              +  (D_z m_{ad})_{21} \partial_{k_z} {g^s_{2d}}_{12}
    &=& 0
\,, \qquad
\label{ked0-diagC}
\end{eqnarray}
where we neglected the collision terms that are suppressed by derivatives. 
With the help of this remarkable result and the constraint 
equations~(\ref{ced:diag1b}-\ref{ced:diag3b})  
the kinetic equation~(\ref{ked0-diagB}) finally reduces to the familiar 
form~\cite{KainulainenProkopecSchmidtWeinstock:2001,
KainulainenProkopecSchmidtWeinstock:2002}
\begin{equation}
      \frac{1}{\tilde{k}_0}\Big[ k\cdot\del
        -  \frac 12   \Big(\partial_z|m_d|^2_{i} 
        -  \frac{s}{\tilde k_0} 
           \partial_z\left[|m_d|^2_i(\partial_z\theta_{di}+2{\Delta_z}_{\tt ii})
           \right]\Big)\partial_{k_z} \Big]{g^s_{0d}}_{\tt ii}
  = {{\cal K}^s_{0d}}_{\tt ii} 
\,.
\label{kinetic_3+1}
\end{equation}
Note that this equation holds for both $g^{s<}_{0d}$ and $g^{s>}_{0d}$.
Recall furthermore that for spin conservation
the space-time transients are constrained to satisfy
${g_{0d}^s}_{\tt ii} = {g_{0d}^s}_{\tt ii}(k,t-\vec v_\|\cdot \vec x_\|,z)$.
The collisional contributions from the constraint equations are 
higher order in gradients and can be consistently neglected.

The functions $g_{ad}^s$ ($a=0,1,2,3$) satisfy four apparently
different equations~(\ref{ked0}--\ref{ked3}).
These functions are related
by the constraint equations~(\ref{ce0d1}--\ref{ce3d1}), which reduce the
number of independent functions to a single one, $g_{0d}^s$, projected on the
quasiparticle shell~(\ref{ced:diag0b}).
We will show in an appendix in Paper~II
that all four kinetic equations are actually mutually
dependent in a self-consistent manner, and thus equivalent to the kinetic
equation for $g_{0d}^s$, {\it plus} the on-shell 
equation~(\ref{ced:diag0b}). Our proof includes not only the flow term,
but also the collisional sources. 
The equivalence of these equations including the collision term is
a very nontrivial consistency check of our approach to the kinetics 
of fermions.

\subsection{Boltzmann transport equation for CP-violating fermionic densities}
\label{Boltzmann transport equation for CP-violating fermionic densities}

Since our primary interest is transport of CP-violating densities,
we show here how -- starting with the Kadanoff-Baym equations for 
fermions~(\ref{Wigner-space:fermionic_eom})
({\it cf.} also~(\ref{kinetic_3+1}))
-- one obtains the on-shell Boltzmann transport equations. The crucial
difference with respect to the scalar case discussed in 
section~\ref{Boltzmann transport equation for CP-violating scalar densities}
is the CP-violating {\it semiclassical force} in the flow term of the kinetic
equation for fermions. The force appears at second order in
gradients, or equivalently at first order in an expansion in $\hbar$.

\vskip 0.1in

We begin our analysis with recalling the C and CP transformations for 
the fermionic fields,
%
\begin{eqnarray}
          \psi^c(u)
 &\equiv& {\cal C} \psi(u) {\cal C}^\dagger
 =        i\gamma^0\gamma^2 {\bar\psi}^T(u) ,
\nonumber\\
          {\bar\psi}^c(u)
 &\equiv& {\cal C}\bar \psi(u) {\cal C}^\dagger
 =        \psi^T(u) i\gamma^0\gamma^2
\,,
\label{C-transformations-psi}
\end{eqnarray}
and for Dirac's $\gamma$-matrices,
\begin{eqnarray}
     {\cal C}{\gamma^\mu}^T  {\cal C}^\dagger
 &=& i\gamma^0\gamma^2 {\gamma^\mu}^T  (i\gamma^0\gamma^2)^\dagger
 =   - \gamma^\mu
\nonumber\\
     {\cal C} \gamma^5  {\cal C}^\dagger
 &=& i\gamma^0\gamma^2 \gamma^5  (i\gamma^0\gamma^2)^\dagger
  =  \gamma^5
\,.
\label{C-transformations-gamma}
\end{eqnarray}
In the relativistic limit 
the Standard Model fermions are chiral and couple differently to 
the weak interactions, so that both charge
{\cal C} and parity {\cal P} symmetries are strongly violated.
But the combined symmetry {\cal CP} is violated only very weakly, 
hence it is natural to consider antiparticles to be
related to particles by a CP transformation.
Therefore we also quote the parity transformations,
\begin{eqnarray}
          \psi^p(u) 
 &\equiv& {\cal P}\psi(u) {\cal P}^\dagger
 =        \gamma^0 \psi(\bar u),
\qquad\;\,
          {\cal P} \gamma^\mu   {\cal P}^\dagger
 =        \gamma^0 \gamma^\mu   \gamma^0
 =        {\gamma^\mu}^\dagger = \gamma_\mu
\nonumber\\
          {\bar\psi}^p(u)
 &\equiv& {\cal P}\bar\psi(u) {\cal P}^\dagger 
 =        {\bar \psi}(\bar u) \gamma^0
\,,\qquad
          {\cal P} \gamma^5  {\cal P}^\dagger
 =        \gamma^0 \gamma^5  \gamma^0 = - \gamma^5
\,.
\label{P-transformations-psi+gamma}
\end{eqnarray}
Combining these relations with the definitions of the fermionic
Wightman functions~(\ref{Green_fermionic_index}),
we easily find the relations for their C- and CP-transformations:
\begin{eqnarray}
 S^<(u,v)
 &\stackrel{{\cal C}}{\longrightarrow}&
 \gamma^0\gamma^2 {S^>}^T(v,u) \gamma^0\gamma^2
\label{Sc<}
\\
 S^<(u,v)
 &\stackrel{{\cal CP}}{\longrightarrow}&
 - \gamma^2 {S^>}^T(\bar v,\bar u) \gamma^2
\label{Scp<}
\,,
\end{eqnarray}
which in the Wigner representation become
\begin{eqnarray}
    S^<(k,x)
  &\stackrel{{\cal C}}{\longrightarrow}&
    \gamma^0\gamma^2 {S^>}^T(-k,x) \gamma^0\gamma^2
  \equiv
    {S^c}^<(k,x)
\label{C_transforms_fermions}
\\
    S^<(k, x)
  &\stackrel{{\cal CP}}{\longrightarrow}&
    \, - \gamma^2 {S^>}^T(-\bar{k}, \bar{x}) \gamma^2
    \hphantom{X}
  \equiv
    {S^{cp}}^<(\bar{k}, \bar{x}) 
\,;
\label{CP_transforms_fermions}
\end{eqnarray}
analogous relations hold for ${S^{c}}^>$ and ${S^{cp}}^>$.
As we did in the scalar case
in~(\ref{scalar_wigner_C}--\ref{scalar_wigner_CP}), 
we employ an additional inversion of
the spatial parts of position $\bar x^\mu = (x^0,-x^i)$
and momentum $\bar k^\mu = (k^0,-k^i)$ in the definition of $S^{cp}$.
In the case of several mixing fermions, the Wigner function is a matrix
in flavor space which is affected by the transposition, too.

\vskip 0.1in

We have argued in section~\ref{Reduction to the on-shell limit} 
that a weak coupling reduction to the on-shell limit of the equation of
motion~(\ref{Wigner-space:fermionic_eom})
for the fermionic Wigner function results in the equation
\begin{equation}
    \Big(  \kdag
         + \frac i2 \deldag_{\!x}
         - \left(m_h(x)+i\gamma^5m_a(x)\right)
           \mbox{e}^{-\frac i2 \overleftarrow{\del}_{\!x} \cdot \del_k}
    \Big) S^{>,<}(k,x)
  = {\cal C}_\psi(k,x) \,.
\label{ferm_eom_wigner}
\end{equation}
Taking account of the hermiticity property~(\ref{Wigner-space:fermionic_eom}),
hermitean conjugation and transposition lead to
\begin{equation}
      \Big(  \kdag^{\,T}
                   - \frac i2 \deldag_{\!x}^{\,T}
                   - \left(m_h^*(x)+i\gamma^5m_a^*(x)\right)
 \mbox{e}^{\frac i2 \overleftarrow{\del}_x \cdot \del_k}
              \Big)S^{>,<}(k,x)^T
  = -\gamma^0{\cal C}^*_{\psi}(k,x)\gamma^0 
\,.
\label{ferm_eom_wigner_T}
\end{equation}
By commuting $\gamma^0\gamma^2$ through and reversing the sign of the 
4-momentum, we get the following equation for the C-conjugate of the
Wigner function~(\ref{C_transforms_fermions}),
\begin{equation}
    \Big(  \kdag
         + \frac i2 \deldag_{\!x}
         - \left(m^*_h(x)+i\gamma^5m^*_a(x)\right)
           \mbox{e}^{-\frac i2 \overleftarrow{\del}_{\!x} \cdot \del_k}
    \Big) S^{c<,>}(k,x)
  = \gamma^2{\cal C}^*_\psi(-k,x)\gamma^2
\,.
\label{ferm_eom_wigner_C}
\end{equation}
Similarly, by commuting $\gamma^2$ through~(\ref{ferm_eom_wigner_T}),
we arrive at the equation for the CP-conjugate
Wigner function~(\ref{CP_transforms_fermions}):
\begin{equation}
    \Big(  \bar\kdag
         + \frac i2 \deldag_{\!\bar x}
         - \left(m^*_h(x)-i\gamma^5m^*_a(x)\right)
  {\rm e}^{-\frac i2 \overleftarrow{\del}_{\!\bar x}\cdot\partial_{\bar k}}
    \Big) S^{cp<,>}(k, x)
  = - \gamma^0\gamma^2{\cal C}^*_\psi(-k,x)\gamma^0\gamma^2
\,.
\label{ferm_eom_wigner_CPb}
\end{equation}

Several remarks are now in order. Note first that
the presence of imaginary elements in the mass matrices, 
$m_h \equiv (1/2)(m+m^\dagger)$ and $m_a \equiv (1/2i)(m-m^\dagger)$,
may violate both C and CP symmetry. This is true, of course, provided
the imaginary parts cannot be removed by field redefinitions. 
Second, while C is not in general violated 
by a (real) antihermitean mass term $m_a$,
CP is violated, provided $m_a=m_a(x)$ is space-time dependent, such that 
it cannot be removed by field redefinitions
\footnote{An example where 
these types of CP-violation can be of relevance for baryogenesis
has recently been considered in~\cite{GarbrechtProkopecSchmidt:2003},
where a model of {\it coherent baryogenesis} has been constructed, which
is typically operational on grand-unified scales.}. Finally, 
Eqs.~(\ref{ferm_eom_wigner_C}-\ref{ferm_eom_wigner_CPb}) illustrate where 
in the collision term are potential sources of C and CP-violation.
As expected, both C and CP may be violated in the presence of complex
Yukawa couplings, appearing in collision or mass terms.
This can be seen from the complex conjugation of the collision term,
${\cal C}_\psi^*$.
Further, the weak interaction vertices contain $P_{L,R}$, which is also 
a property of our model Lagrangean~(\ref{Yukawa}).  
Now since $\gamma^2 P_{L,R}\gamma^2 = P_{R,L}$ flips chirality, while 
$(\gamma^0\gamma^2) P_{L,R}(\gamma^0\gamma^2) = P_{L,R}$ leaves it invariant,
we conclude that the weak interaction vertices violate C,
but they are invariant under CP, as they should. 
Using the explicit expressions for the collision terms in Paper~II
and~(\ref{The 3+1 dimensional (moving) frame}),
one can show that the right hand side of
equation~(\ref{ferm_eom_wigner_CPb}) is equal to ${\cal C}_\psi$, with
all Wigner functions, both scalar and fermionic, replaced by their CP-conjugate
counterparts, and the Yukawa coupling matrices $y$ replaced by $y^*$.
Further details of the analysis of the collision term
are left for Paper~II.

\vskip 0.1in

A careful look at equation~(\ref{ferm_eom_wigner_CPb}) shows that the
kinetic operator for $S^{cp}(k)$ commutes with $P_s(\bar k)$
rather than with $P_s(k)$.
So the spin-diagonal ansatz for the CP-conjugate Wigner function, already
written in the mass diagonal basis, is
\begin{equation}
   S^{cp}_{d\tt ii}(k,x)
 = \sum_s S^{cp}_{sd\tt ii}(k,x)
 = \sum_s iP_{s}(\bar{k})
             \big[ s\gamma^3\gamma^5 g_{0d\tt ii}^{cps}
                  -s\gamma^3         g_{3d\tt ii}^{cps}
                  +\mathbbm{1}       g_{1d\tt ii}^{cps}
                  -i\gamma^5         g_{2d\tt ii}^{cps} \big](k,x)
\label{Sc<b}
\,.
\end{equation}
Now there are several ways to obtain the constraint and kinetic equation
for the component function $g_{0d\tt ii}^{cps}$. First, one could repeat
all the steps from the previous section. Since the mass term in the
CP-conjugate equation~(\ref{ferm_eom_wigner_CPb}) is the complex conjugate
of the mass term appearing in the equation for the original
Wigner function~(\ref{ferm_eom_wigner}),
we would begin by carrying out the flavor transformation, this time using the
rotation matrices $\bX^*$ and $\bY^*$, then taking the appropriate spinorial
traces. The antihermitean and hermitean parts would lead to constraint and
kinetic equations for the CP-conjugate component functions, respectively, and
finally to the desired equations for $g_{0d\tt ii}^{cps}$.

Another possibility is to note that equation~(\ref{ferm_eom_wigner_CPb}) for
$S^{cp}$ is obtained from equation~(\ref{ferm_eom_wigner}) for $S$ by
replacing the mass by its complex conjugate and replacing all
explicit occurrences of the momentum $k$ by $\bar{k}$, as well as
sending all spatial derivatives $\del_x$ to $\del_{\bar{x}}$, and finally
replacing the collision term on the right hand side, 
as indicated in Eq.~(\ref{ferm_eom_wigner_CPb}). When we apply exactly
the same changes to the constraint and the kinetic equation for $g_{0d}$,
which includes $\theta\rightarrow-\theta$ and $\Delta_z\rightarrow-\Delta_z$
because of the complex conjugated mass, we find the respective equations
for $g_{0d}^{cp}$.

The simplest method, however, is based on the observation that, because
of relation~(\ref{CP_transforms_fermions}), the Wigner function $S$ contains
complete information about its CP-conjugate $S^{cp}$, and the same then
holds for the component functions $g_0$ and $g_0^{cp}$ as well.
Relation~(\ref{CP_transforms_fermions}) is covariant under the flavor rotation,
\beq
   S_d^{cp<}(k, x)
 \equiv \bY^* S^{cp<}(k,x) \bX^T
 =      - \gamma^2 \left( \bY S^>(-k, x)  \bX^\dagger \right)^T \gamma^2
 =      - \gamma^2 {S_d^>}^T(-k, x) \gamma^2
\,,
\eeq
so we have to match the decomposition~(\ref{Sc<b}) for $S^{cp<}$ with
\beq
   - \gamma^2 [S^>_{d\tt ii}(-k, x)]^{\,T}\gamma^2 
 =  \sum_s iP_{s}(-\bar{k})
             \big[-s\gamma^3\gamma^5 g_{0d\tt ii}^{>-s}
                  +s\gamma^3         g_{3d\tt ii}^{>-s}
                  +\mathbbm{1}       g_{1d\tt ii}^{>-s}
                  +i\gamma^5         g_{2d\tt ii}^{>-s} \big](-k, x)
\label{eom:g2S>Tg2}
\,.
\eeq
This implies the identification
\beq
    g_{0d\tt ii}^{cps<}(k,x) = - g_{0d\tt ii}^{>-s}(-k, x)
\,,
\label{g0cp-g0>}
\eeq
where we used $P_s(-k) = P_s(k)$.
Looking at the constraint equation~(\ref{ced:diag0b}) or the spectral
ansatz~(\ref{spectral-solution:fermions>}) for $g_{0d\tt ii}^{>s}$,
we immediately see that the component functions of the CP-conjugate
Wigner function live on the same energy shells $\omega_{si}$ as the ones
of the original Wigner function.
In other words, the dispersion relation for antifermions is identical
to that for fermions, a conclusion that can also be reached by considering
the CP-conjugated constraint equations.
Therefore we can make the spectral ansatz
\begin{equation}
   g_{0d\tt ii}^{<cps}(k, x)
 = 2\pi\sum_{\pm} \frac{\delta(k_0\mp\omega_{\pm si})}
                          {2\omega_{\pm si}Z_{\pm si}} 
   \, |\tilde{k}_0| \, n^{cp}_{si}(k,x)
\,.
\label{spectral-solution:fermions>cp}
\end{equation}
Inserting this and the spectral ansatz~(\ref{spectral-solution:fermions>})
for $g^>_0$ into~(\ref{g0cp-g0>}) shows that the particle density at
negative energies is related to the density of antiparticles. For $k_0$
on the positive or negative energy shell we have
\begin{equation}
   n^{cp}_{si}(k_0, \vec k,x)
 = 1 - n_{-si}\big(-k_0,-\vec k,x\big)
\label{ncp_n}
\,.
\end{equation}
In order to find the kinetic equation for the CP-conjugate component function,
let us first recall the kinetic equation~(\ref{kinetic_3+1})
for $g_{0d\tt ii}^{s}$ from section~\ref{Kinetic equations}:
\begin{equation}
      \frac{1}{\tilde{k}_0}\Big[ k\cdot\del
        -  \frac 12   \Big(\partial_z|m_d|^2_{i} 
        -  \frac{s}{\tilde k_0} 
           \partial_z\left[|m_d|^2_i(\partial_z\theta_{di}+2{\Delta_z}_{\tt ii})
           \right]\Big)\partial_{k_z} \Big]  {g^{s}_{0d}}_{\tt ii}(k,x)
  = {{\cal K}^s_{0d}}_{\tt ii} (k,x)
\,.
\label{kinetic_3+1_repeat}
\end{equation}
Again we make use of relation~(\ref{g0cp-g0>}) and find
\beq
      \frac{1}{\tilde{k}_0}\Big[ k\cdot\del
        -  \frac 12   \Big(\partial_z|m_d|^2_{i} 
        -  \frac{s}{\tilde k_0} 
           \partial_z\left[|m_d|^2_i(\partial_z\theta_{di}+2{\Delta_z}_{\tt ii})
           \right]\Big)\partial_{k_z} \Big]
    g_{0d\tt ii}^{cps}(k, x)
  = - {\cal K}^{-s}_{0d\tt ii}(-k, x)
\,.
\label{kinetic_3+1_cp}
\eeq
Needless to say, all ways of obtaining these equations yield the same
results, although this might not be obvious for the collision term. Starting
with the right hand side of equation~(\ref{ferm_eom_wigner_CPb}),
one would expect to find
\beq
        {\cal K}^{cps}_{0d\tt ii}(k, x)
 \equiv  \frac 12 \mbox{Tr}
          \left(  \mathbbm{1} P_s(\bar{k})
                  \gamma^0\gamma^2 {{\cal C}^*_{\psi d}}_{\tt ii}(-k, x)
                  \gamma^0\gamma^2 
                + \text{h.c.}
          \right)
\label{coll_cp_a}
\eeq
as the collisional contribution to the kinetic equation for $g^{cp}_0$.
But since this is a trace in spinor space and we are looking at the
diagonal elements in flavor space, we can apply a transposition to find
\beqa
   {\cal K}^{cps}_{0d\tt ii}(k, x)
 &=& \frac 12 \text{Tr}
      \left(
              \left(\gamma^0\gamma^2 P_s(\bar{k}) \gamma^0\gamma^2\right)^T
              {{\cal C}^\dagger_{\psi d}}_{\tt ii}(-k, x)
            + \text{h.c.}
      \right)             
\nonumber\\
 &=& \frac 12 \text{Tr}
      \left(
              P_{-s}(k)
              {{\cal C}_{\psi d}}_{\tt ii}(-k, x)
            + P^\dagger_{-s}(k)
              {{\cal C}_{\psi d}}_{\tt ii}^\dagger(-k, x)
      \right)
  = - {\cal K}^{-s}_{0d\tt ii}(-k, x)
\,.
\label{coll_cp_b}
\eeqa

We define the distribution function for particles as the projection of the
phase space densities on the positive mass shell,
\beq
  f_{si+}(\vec{k}, x) \equiv n_{si}(\omega_{si}, \vec{k}, x)
\,,
\label{onshell-density}
\eeq
and then the kinetic equation for $g_0^{cp}$ suggests to define the
distribution function for antiparticles as
\beq
  f_{si-}(\vec{k}, x) \equiv n^{cp}_{si}(\omega_{si}, \vec{k}, x)
                      = 1 - n_{-si}(-\omega_{si}, -\vec{k}, x)
\,.
\label{onshell-density-cp}
\eeq
We will verify in Paper~II
that $f_{si\pm}(\vec{k})$
indeed measure the density of particles and antiparticles with momentum
$\vec{k}$ and spin $s$.
To obtain the Boltzmann equations for these densities
we have to integrate~(\ref{kinetic_3+1_repeat}) and~(\ref{kinetic_3+1_cp})
over positive frequencies:
\beqa
 \frac{1}{Z_{si}}
   \big( \del_t
        +\vec v_{si}\cdot \nabla_{\vec x}
        + F_{si}\partial_{k_z} \big) f_{si+}
  &=& \int _0^\infty \frac{dk_0}{\pi} {{\cal K}^s_{0d}}_{\tt ii}(k,x) 
\label{ferm_kin_pos}
\\
 \frac{1}{Z_{si}}
   \big( \del_t
        +\vec v_{si}\cdot \nabla_{\vec x}
        + F_{si}\partial_{k_z} \big) f_{si-}
  &=& \int _0^\infty \frac{dk_0}{\pi} {{\cal K}^{cps}_{0d}}_{\tt ii}(k, x) 
\,,
\label{ferm_kin_pos_cp}
\eeqa
with the quasiparticle velocity and semiclassical force
\begin{eqnarray}
\vec v_{si} &=& \frac{\vec{k}}{\omega_{si}}
\label{velocity_si+}
\\
F_{si} &=& -  \frac{1}{2\omega_{si}} 
              \Big(\partial_z|m_d|^2_{i} 
                   - \frac{s}{\tilde \omega_{si}} 
                     \partial_z\left[|m_d|^2_i(\partial_z\theta_{di}+2{\Delta_z}_{\tt ii})
                     \right]
              \Big)
\,.
\label{force_si+}
\end{eqnarray}
Note that, in addition to the classical terms,
both $\vec v_{si}$ and $F_{si}$ contain spin dependent 
contributions arising from derivatives of the pseudoscalar mass term, 
and thus a potential for inducing CP-violating effects. 

The Boltzmann equation~(\ref{ferm_kin_pos_cp}) for the antiparticle densities
can also be obtained by integrating the kinetic equation for $g_0$
over negative energies, and then using~(\ref{onshell-density-cp}).
This procedure has often been advocated as a way of identifying CP-violating
densities.
It works, because the complete information about $S^{<cp}$ is contained in
$S^>$, which in the on-shell limit is completely determined by $S^<$, as
can be shown with the help of the fermionic spectral sum 
rule~(\ref{sum_rule_fermions}).

\vskip 0.1in

For practical baryogenesis calculations one often deals with 
problems close to thermal equilibrium, in which case it is convenient to treat 
the Boltzmann equations~(\ref{ferm_kin_pos}) and~(\ref{ferm_kin_pos_cp})
in linear response approximation with respect to deviations from 
thermal equilibrium. We note that the thermal equilibrium distribution
function in the wall frame has the form 
\begin{equation}
f^{\rm eq}_{si}  = \frac{1}{e^{\beta \gamma_w(\omega_{si} + v_w k_z)}+1},
\label{distribution-fn-eq}
\end{equation}
with $\beta =1/T$, $\vec v_w = v_w \hat z$ is the wall velocity 
and $\gamma_w = (1-v_w^2)^{-1/2}$ is the corresponding boost factor. 
This is a generalization of the thermal equilibrium distribution 
function~(\ref{FermiDirac}) which includes both the effects of CP-violation
from a pseudoscalar mass and of a moving plasma on thermal equilibrium. 
We show in Paper~II
that this form for the equilibrium
distribution function leads to the correct energy conservation
in the collision term. Furthermore, the distribution 
function~(\ref{distribution-fn-eq}) satisfies
\begin{equation}
Z_{si}^{-1}\big(\partial_t + \vec v_{si}\cdot\nabla_{\vec x}
                              + F_{si}\partial_{k_z}\big) f^{\rm eq}_{si}
 = - v_w f^{\rm eq}_{si}(1-f^{\rm eq}_{si})Z_{si}^{-1}\beta F_{si}
\label{distribution-fn-eq_eom}
\end{equation}
such that, for a wall at rest, Eq.~(\ref{distribution-fn-eq}) satisfies
the flow equation exactly, as one would expect from the correct equilibrium
distribution function. Note that the degeneracy between the states
with opposite spin is broken by this equilibrium,
because particles with opposite spin satisfy different
energy-momentum dispersion relations.

Our main interest is the transport equation for a CP-violating 
distribution function that can eventually lead to the creation of
a Baryon asymmetry. We first define
\beq
f_{si\pm} = f^{\rm eq}_{si} 
          + \frac 12 \delta f_i^{\rm even} 
          + \delta f_{si\pm}
\,,
\label{distribution-fn_fsi_decompose}
\eeq
where $\delta f_{si\pm}$ are spin dependent correction functions.
Since spin dependence occurs only at order $\hbar$ in our Boltzmann equations
and all sources are wall velocity suppressed, as for example can be seen
in~(\ref{distribution-fn-eq_eom}), we find $\delta f_{si\pm} = O(v_w\hbar)$.
The distributions $f_{si\pm}$ can also contain a CP-even deviation from thermal
equilibrium, $\delta f_{i}^{\rm even}$, which is relevant for the dynamics
of phase
interfaces~\cite{MooreProkopec:1995,HuberJohnSchmidt:2001,JohnSchmidt:2000+2001}.
Note that, since $\delta f_i^{\rm even}$ is not suppressed by powers of
$\hbar$, it does not in general decouple from the equation for the 
densities $\delta f_{si\pm}$. In order to properly treat the dynamics
of CP-violating densities, one has to solve also for the dynamics of 
CP-conserving densities!
Together with $\delta f_{i}^{\rm even} = O(v_w\hbar^0)$,
Eq.~(\ref{distribution-fn-eq_eom}) implies that it suffices to 
solve the equation for $\delta f^{\rm even}_i$ to leading (classical)
order in gradients, which is obtained by
adding~(\ref{ferm_kin_pos}) and~(\ref{ferm_kin_pos_cp}),
\begin{equation}
   \big(\del_t  + \vec v_{0i}\cdot\nabla_{\vec x} 
              + F_{0i} \partial_{k_z}
   \big) \delta f_{i}^{\rm even} 
   + \beta f_0(1-f_0) v_w \frac{\del_z|m_d|^2_i}{\omega_{0i}}
 =  \int _0^\infty \frac{dk_0}{\pi}
     \left( {\cal K}^s_{0d\tt ii}(k) + {\cal K}^{cps}_{0d\tt ii}(k) \right)
\,,
\label{eom:delta_fi}
\end{equation}
where 
\begin{equation}
 \vec v_{0i} = \frac{\vec k}{\omega_{0i}}
\,,\qquad
   F_{0i}    = - \frac{\partial_z |m_d|^2_{i}}{2\omega_{0i}}
\label{v0i:F0i}
\end{equation}
are the classical particle velocity and the classical force, respectively,
and $f_{0i} \equiv 1/(e^{\beta\omega_{0i}}+1)$.

Based on the first order correction $\delta f_{si\pm}$, we can
form two CP-violating distribution functions:
\beqa
   \delta f^v_{si} &\equiv& \delta f_{si+} - \delta f_{si-}
\label{distribution-fv}
\\
   \delta f^a_{si} &\equiv& \delta f_{si+} - \delta f_{-si-}
\,.
\label{distribution-fa}
\eeqa
As we will explain in Paper~II,
$\delta f^v_{si}$ and $\delta f^a_{si}$ are related to the vector and axial vector
density in phase space, respectively. 
Working to second order in gradients and to linear
order in $v_w$, we obtain the equation for $\delta f^v_{si}$
by subtracting~(\ref{ferm_kin_pos_cp}) from~(\ref{ferm_kin_pos}):
\beq
   \big(  \del_t
        + \vec v_{0i} \cdot \nabla_{\vec x} 
        + F_{0i} \del_{k_z}
   \big) \delta f^v_{si} 
 = \int _0^\infty \frac{dk_0}{\pi}
    \left( {\cal K}^s_{0d\tt ii}(k) - {\cal K}^{cps}_{0d\tt ii}(k) \right)
\,.
\label{boltzmann:deltaf-v}
\eeq
Note that the flow term contains {\it no source} whatsoever.
This means that a CP-violating semiclassical force~(\ref{force_si+})
cannot create vector charge densities
(this remains no longer true, however, when one includes the possibility of
coherent particle production, which can be exacted by a dynamical tracing
of flavor mixing~\cite{GarbrechtProkopecSchmidt:2003}).

\vskip 0.05in

The transport equation for $\delta f^a_{si}$ is obtained
by changing the sign of $s$ in~(\ref{ferm_kin_pos_cp}) and then subtracting
it from~(\ref{ferm_kin_pos}). With the help
of~(\ref{distribution-fn-eq_eom}) we find
\beq
  \big(  \del_t
         + \vec v_{0i} \cdot \nabla_{\vec x} 
         + F_{0i} \partial_{k_z}
  \big)
       \delta f^a_{si} 
  + s{\cal S}_i^{\rm flow}
=  \int _0^\infty \frac{dk_0}{\pi}
      \left( {\cal K}^s_{0d\tt ii}(k) - {\cal K}^{cp-s}_{0d\tt ii}(k) \right)
\,.
\label{boltzmann:deltaf-a}
\eeq
This equation now has a source:
\beqa
{\cal S}_i^{\rm flow} 
 &=& - \frac 12 v_w f_{0i}(1-f_{0i})\beta
       \Big\{  \delta  F_i 
             + F_{0i}\frac{\delta\omega_i}{\omega_{0i}}
             + F_{0i}\delta\omega_i\beta(1-2f_{0i})
             + F_{0i}\delta Z_i
       \Big\}
\label{boltzmann:deltaf-a_source}
\\
&&
     + \frac 12 \Big\{
                \delta Z_i
                \big(  \del_t
                     + \vec v_{0i}\cdot\nabla_{\vec x} 
                     + F_{0i} \partial_{k_z}
                \big)
     +                 \frac{\delta\omega_i}{\omega_{0i}}
                       \vec v_{0i} \cdot \nabla_{\vec x}
                     + \Big[F_{0i}\frac{\delta\omega_i}{\omega_{0i}}
                     + \delta F_i \Big] \partial_{k_z}
               \Big\}
                     \delta f_{i}^{\rm even} 
\,,
\label{boltzmann:deltaf-a_even}
\quad\quad
\end{eqnarray}
where the definitions
\beqa
\delta F_i &\equiv& 
            \frac{\partial_z[|m_d|^2_i(\partial_z\theta_{di}+2{\Delta_z}_{\tt  ii})]}
                 {\omega_{0i}\tilde\omega_{0i}}
\nonumber\\
    \delta\omega_i 
           &\equiv& \frac{[|m_d|^2_i(\partial_z\theta_{di}+2{\Delta_z}_{\tt ii})]}
                           {\omega_{0i}\tilde\omega_{0i}}
\quad,\quad
    \delta Z_i 
           \equiv \frac{\omega_{0i}}{\tilde\omega_{0i}^2} \, \delta\omega_i 
\label{boltzmann:definitions}
\eeqa
have been used.
We now pause to comment on the physical meaning of the various terms appearing
in the Boltzmann transport 
equation~(\ref{boltzmann:deltaf-a}).
First, the CP-violating density, $\delta f_{si}^a$, evolves on phase space
according to the standard Boltzmann flow derivative,
$d/dt \equiv \del_t + \vec v_{0i}\cdot\nabla_{\vec x} + F_{0i} \partial_{k_z}$.
Second, various CP-violating sources are displayed 
in~(\ref{boltzmann:deltaf-a_source}).
The first term, $\delta F_i$, is the source
arising from the semiclassical force that has been usually taken
into account in the 
WKB-literature~\cite{JoyceProkopecTurok:1996,ClineJoyceKainulainen:2000+2001}.
Note that the second term has a similar origin.
Indeed, it arises from the CP-violating deviation in the energy-momentum
relation appearing in the force term, $-(\partial_z |m_d|_i)/2\omega_{si}$.
The third term originates from the CP-violating split in the dispersion 
relation in the equilibrium solution~(\ref{distribution-fn-eq}), 
and it thus vaguely resembles a local version of spontaneous baryogenesis.
And finally, the fourth term arises from the gradient ``renormalization''
of the Wigner function. All four source terms are of the same order
in gradients, and hence, {\it a priori}, they are all equally important. 
In Paper~II
we take moments of the
Boltzmann equation~(\ref{boltzmann:deltaf-a}) in order to obtain
fluid equations. Since the source term~(\ref{boltzmann:deltaf-a_source})
is symmetric in momentum, it will contribute to the zeroth moment equation.
The parametric dependence of this source becomes explicit by
rewriting it like
\beqa
     \int \frac{d^3k}{(2\pi)^3} {\cal S}_i^{\rm flow} 
  &=& v_w \Big( - \partial_z[|m_d|^2_i(\partial_z\theta_{di}+2{\Delta_z}_{\tt ii})]
                \frac{{\cal I}_a}{(2\pi)^2}
\nonumber\\
&&\hphantom{XX}
              + \beta^2 \left(\del_z|m_d|^2_i\right)
                [|m_d|^2_i(\partial_z\theta_{di}+2{\Delta_z}_{\tt ii})]
                \frac{{\cal I}_b}{(2\pi)^2}
        \Big)
\,.
\label{parametric-flow-source}
\eeqa 
In figure~\ref{figure:flow-source} we plot the dimensionless integrals
${\cal I}_a$ and ${\cal I}_b$ as a function of the mass.
When $|m_d|_i \ll T$ ($|m_d|_i \gg T$), the source ${\cal I}_a$ (${\cal I}_b$)
dominates. The sources are comparable in strength when $|m_d|_i \sim T$.
\begin{figure}[t]
  \unitlength=1in
  \begin{center}
    \psfrag{Ia}[r]{$\beta^2|m|^2 {\cal I}_a$}    
    \psfrag{Ib}[r]{$\beta^4|m|^4 {\cal I}_b$}    
    \psfrag{m}{$\hphantom{XXXXXXXXXXXX}\beta|m|\rightarrow$}
    \includegraphics[width=3.5in,angle=-90]{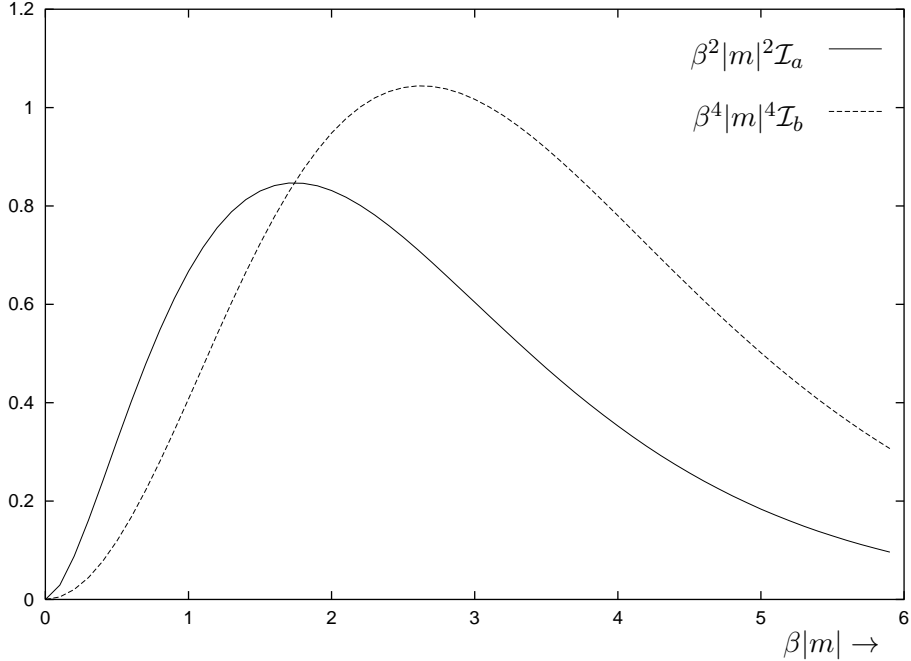}
  \end{center}
\lbfig{figure:flow-source}
\caption{
The integrals ${\cal I}_a$ and ${\cal I}_b$ in Eq.~(\ref{parametric-flow-source})
as a function of the mass.
We scaled ${\cal I}_a$ with $|m|^2$ because this is the way it appears
in the source
term, analogously ${\cal I}_b$ is scaled with $|m|^4$. Note that these
contributions enter the flow source with different signs.
}
\end{figure}
Third, the terms~(\ref{boltzmann:deltaf-a_even})
that couple the CP-even deviation form equilibrium have never been considered
in literature. Nevertheless, they are formally of the same order as 
the source terms~(\ref{boltzmann:deltaf-a_source}), and thus cannot be
neglected. While the first term in~(\ref{boltzmann:deltaf-a_even})
can be reexpressed in terms of the sum of the collision terms, as 
indicated by Eq.~(\ref{eom:delta_fi}), the latter terms cannot,
and represent genuine dynamical source terms, which have so far not
been included in baryogenesis calculations. Fourth, the 
right-hand-side of Eq.~(\ref{boltzmann:deltaf-a}) contains
the collision terms, to be considered in detail in Paper~II.
Finally, we remind the reader that all of the
source terms in Eq.~(\ref{boltzmann:deltaf-a}) 
are second order in derivatives (first order in 
$\hbar$), and are suppressed by the wall velocity.

\subsection{Applications}

 In this section we have developed a formalism for the treatment of
CP-violating effects of space-time dependent scalar and pseudoscalar 
mass terms in kinetic theory. We have shown how the effect of a CP-violating 
shift in the dispersion relation induces an order $\hbar$ semiclassical
force in the flow term of the Boltzmann transport equation. The force
acts on particles and antiparticles of opposite spin in opposite directions. 
Since we have included the possibility of fermionic mixing, 
this formalism is suitable for studies of baryogenesis in supersymmetric
theories at a strongly first order electroweak phase transition,
in which baryon production is typically biased by CP-violating
chargino or neutralino currents. Further, our formalism is also suitable for 
studying baryogenesis problems from coupling of fermions to the bubble wall 
in two Higgs doublet models. For completeness we shall now discuss
in some detail the case of charginos in the Minimal Supersymmetric 
Standard Model (MSSM) and its nonminimal extension, and finally remark 
on baryogenesis sources in two Higgs doublet models. 

\subsubsection{Minimal Supersymmetric Standard Model (MSSM)}
\label{Minimal Supersymmetric Standard Model (MSSM)}

Now that we have a general expression for the semiclassical source in the case of
mixing fermions, we want to study two explicit examples, which are of relevance
for baryogenesis. First we compute the source in the transport equations for the
chargino sector of the MSSM. The chargino mass term reads 
\begin{equation}
{\overline\psi_R}\, m \, \psi_L +{\rm h.c.} 
\,,
\label{chargino-mass}
\end{equation}
where $\psi_R = ({\tilde W}_R^+,{\tilde h}^+_{1,R})^T$
and $\psi_L = ({\tilde W}_L^+,{\tilde h}^+_{2,L})^T$ 
are the chiral fields in the basis of winos. 
The mass matrix is
\begin{equation}
     m = \left( \begin{array}{cc} m_2  & gH_2^* \\
                                    gH_1^* & \mu     \end{array} \right)
\,,
\label{CharginoMatrix_MSSM}
\end{equation}
where $H_1$ and $H_2$ are the Higgs field vacuum expectation values and 
$\mu$ and $m_2$ are the soft supersymmetry breaking parameters, which introduce
CP-violation~\footnote{We keep the possibility of complex Higgs vacuum
expectation values, because
we will reuse the formulas in the NMSSM case.}.
Since for a reasonable choice of parameters there is no transitional
CP-violation in the MSSM, we can take the Higgs expectation values
$H_1$ and $H_2$ to be real
\cite{HuberJohnSchmidt:2001,HuberJohnLaineSchmidt:2000}.
The matrix that diagonalizes $mm^\dagger$ can be parametrized 
as~\cite{ClineJoyceKainulainen:2000+2001}
\begin{equation}
     U = \frac{\sqrt{2}}{\sqrt{\Lambda(\Lambda+\Delta)}}
            \left( \begin{array}{cc} \frac 12 (\Lambda+\Delta) & a                \\
                                      -a^*                     & \frac 12 (\Lambda+\Delta)
                \end{array} \right) \,,
\label{Rotation_MSSM}
\end{equation}
where
\begin{eqnarray}
  a       &=& g(m_2H_1 + \mu^*H_2^*)                   \nonumber\\
  \Delta  &=& |m_2|^2  - |\mu|^2 +  g^2(h_2^2 - h_1^2) \nonumber\\
  \Lambda &=& \sqrt{\Delta^2 + 4|a|^2} \,,
\label{Rotation_MSSM_parameters}
\end{eqnarray}
and $h_i = |H_i|$.
The mass eigenvalues of the charginos are given by
\begin{equation}
  m_\pm^2 = \frac 12 \left( |m_2|^2 + |\mu|^2 + g^2(h_1^2+h_2^2) \right) 
            \pm \frac{\Lambda}{2}  \,.
\label{chargino_eigenstates}
\end{equation}
Upon inserting (\ref{CharginoMatrix_MSSM}-\ref{chargino_eigenstates})
into~(\ref{shift:U}), it is straightforward to show
that the chargino source term can be recast as 
\begin{equation}
 \left[|m_d|^2(\partial_z\theta+2\Delta_z)\right]_{\pm}  
   = \mp \frac{g^2}{\Lambda} \Im(\mu m_2)
  \partial_z\left(h_1h_2\right)
\,.
\label{chargino_source}
\end{equation}                                   
The sources figuring in the transport equation written for 
charginos in~(\ref{boltzmann:deltaf-a_source}) can be easily reconstructed 
from equations~(\ref{Rotation_MSSM_parameters}) and~(\ref{chargino_source}).
The result~(\ref{chargino_source}) agrees with the one found 
by WKB methods in~\cite{ClineJoyceKainulainen:2000+2001}. 
In~\cite{CarenaMorenoQuirosSecoWagner:2000,CarenaQuirosSecoWagner:2002}
however a different dependence on the Higgs fields was obtained.

\subsubsection{Nonminimal Supersymmetric Standard Model (NMSSM)}
\label{Nonminimal Supersymmetric Standard Model (NMSSM)}

We now consider an extension of the MSSM which contains a
singlet field $S$ in the Higgs sector, which can induce additional
CP-violation in the Higgs sector. In particular, 
the Higgs vacuum expectation values may become complex.
The singlet couples to higgsinos, and therefore we obtain the mass matrix
by generalizing the higgsino-higgsino component of the chargino mass
matrix~(\ref{CharginoMatrix_MSSM}) 
\begin{equation}
  \mu \rightarrow \tilde{\mu} = \mu + \lambda S \,,
\label{MSSM_to_NMSSM}
\end{equation}
where $\lambda$ is the coupling for the higgsino-higgsino-singlet interaction.
The field content we consider is the same as in the MSSM, so the mass matrix is
\begin{equation}
   m = \left( \begin{array}{cc} m_2    & gH^*_2      \\
                                gH^*_1 & \tilde{\mu}   \end{array} \right) \,.
\label{CharginoMatrix_NMSSM}
\end{equation}
This matrix is still diagonalized by $U$ in~(\ref{Rotation_MSSM}). 
We write the Higgs expectation values as
\begin{equation}
     H_i = h_i{\rm e}^{i\theta_i} \quad,\quad i=1,2 \,,
\end{equation}
where only one phase is physical. With the gauge 
constraint~\cite{HaberKane:1985}
\begin{equation}
       h^2_1\theta'_1 = h^2_2\theta'_2
\end{equation}
we can write
\begin{equation}
  \theta_1' = \frac{h_2^2}{h^2}\theta'
  \quad,\quad
  \theta_2' = \frac{h_1^2}{h^2}\theta' \,,
\end{equation}
where $\theta=\theta_1+\theta_2$ is the physical CP-violating phase,
and $h^2=h_1^2+h_2^2$.
Now everything is prepared to write the NMSSM-source term. It  can
be divided into three contributions, which have to be added. The first
one is a generalization of the chargino source~(\ref{chargino_source})
\begin{equation}
 \left(|m_d|^2(\partial_z\theta_d+2\Delta_z)\right)_{h_1h_2\pm}  
  =  \mp \frac{g^2}{\Lambda} \Im (\tilde{\mu}m_2{\rm e}^{i\theta}) (h_1h_2)'
\end{equation}
for the case involving a new scalar field $S$ and complex Higgs expectation values.
In addition to this there are two new types of sources. One of them is
proportional to a derivative of the CP-violating phase $\theta$ in the Higgs sector:
\begin{equation}
 \left(|m_d|^2(\partial_z\theta_d+2\Delta_z)\right)_{\theta\pm}  
 =  -\frac{g^2\theta'}{\Lambda}
    \bigg(
       \Big( \Lambda \pm (|m_2|^2+|\tilde{\mu}|^2)\Big)
           \frac{h_1^2h_2^2}{h^2}
       \mp \Re (\tilde{\mu}m_2{\rm e}^{i\theta})h_1h_2
    \bigg) \,.
\end{equation}
Finally, there is a source that can be written as a derivative of the 
singlet condensate:
\begin{eqnarray}
 \left(|m_d|^2(\partial_z\theta_d+2\Delta_z)\right)_{S\pm}  
&=&  \pm \frac{\lambda g^2}{\Lambda} \Im(m_2H_1H_2S')  \\
&&  + \frac{\lambda g^2}{2\Lambda}
     \Big( \Lambda \pm (|\tilde{\mu}|^2 + g^2h^2 - |m_2|^2) \Big)
      \Im(\tilde{\mu}^*S') \,.
\nonumber
\end{eqnarray}
The mass eigenvalues $m_\pm$, that is the diagonal elements of $|m_d|^2$,
can be obtained from the corresponding expression~(\ref{chargino_eigenstates})
in the MSSM part with the replacement $\mu\rightarrow\tilde{\mu}$.


\cleardoublepage
\section{Conclusion}

In this work we perform a controlled first principle derivation of
transport equations for a model Lagrangean of chiral fermions Yukawa-coupled
to a complex scalar. Our treatment is accurate  
to first order in an $\hbar$ expansion, or, more concretely, to
first order in a gradient expansion with respect to a slowly varying scalar
background field, which is formally valid when
$\hbar \partial \ll \hbar k$. Being valid to order $\hbar$, 
our treatment allows us to trace the propagation of CP-violating fluxes, 
which is of relevance, for example, for electroweak baryogenesis 
at a first order phase transition. We consistently include the collision term,
although in this first paper of our series only in a formal way.
The actual evaluation of these terms can be found in Paper~II.

\vskip 0.05in

We address an open question of electroweak scale 
baryogenesis mediated by mixing fermions, which couple in a CP-violating
manner to a propagating bubble wall of a first order phase transition:
which basis should be used to model the kinetics of mixing fermions?
We show that, if one is limited to a diagonal approximation,
the mass eigenbasis is singled out as the only basis in which 
the diagonal and off-diagonal elements of the distribution function
{\it decouple} at order $\hbar$ in a derivative expansion. 
Our derivation is valid when there are no nearly degenerate mass 
eigenvalues, that is when $\hbar k\cdot\partial\ll \delta(m_d^2)$,
where $\delta(m_d^2)$ denotes the (minimum) split in the mass eigenvalues.
No such claims can be made for the flavor (weak interaction) basis, 
in which flavor mixing is present already at the classical level $O(\hbar^0)$.
This indicates that the use of the flavor basis in transport equations
is at best problematic, unless flavor mixing is consistently included. 
Of course, the final resolution of this problem can only come from
a basis independent treatment, which would include the dynamics of 
both flavor off-diagonal and diagonal CP-violating densities. 
At the moment no such treatment is available, however.

\vskip 0.05in
 
 We also address various other issues, which comprise a proof that
the Kadanoff-Baym equation for a single scalar field can be reduced, in a 
weakly coupled regime, to an on-shell Boltzmann equation, which includes both
the self-energy and the collision term, approximated at the same order in a 
coupling constant expansion.
An inclusion of the hermitean self-energies would be desirable, but has not
been done yet.
These self-energies mix spin and moreover provide an additional source
of CP-violation, which we expect to be of a similar strength as the source 
in the collision term (see Paper~II).
Further, we demonstrate that no CP-violating source is present at order
$\hbar$ in the flow term of the scalar kinetic equation.
We then derive a Boltzmann
transport equation for the relevant fermionic quasiparticles with
a definite spin. 
We include all of the CP-violating sources arising from the flow term
present at the order $\hbar$, which, of course, include the semiclassical
force, but also a few, as-of-yet unencountered, sources. One of the new
sources is induced by a CP-even deviation from the equilibrium distribution
function, which itself is formally of order $\hbar^0$.

\vskip 0.05in

Our method represents a formalized and  controlled implementation
of the original heuristic semiclassical (WKB) treatment 
of the problem~\cite{JoyceProkopecTurok:1995,JoyceProkopecTurok:1996}, 
in which one calculates the relevant (fermionic) dipersion relation
accurate to order $\hbar$ by the means of a WKB analysis, and then inserts
the result in the appropriate kinetic equation. 
As we have shown here, when applied with due care, 
the semiclassical approach leads to a correct semiclassical force
in the flow term, but it does not permit the treatment of collisional
sources, nor does it allow a critical assessment of the quasiparticle picture 
of the plasma or the implementation of any effects beyond the quasiparticle
picture. None of these limitations apply for the approach presented here.

\vskip 0.05in

Our work is completed in Paper~II. There we
discuss the collision term, which in the present paper was
formally maintained in all equations, but never evaluated explicitly,
and we derive and study a set of fluid equations, which form
an approximation to the Boltzmann equations found here.

\vskip 0.1in

\section*{Acknowledgements}
We would like to acknowledge collaboration and engaged discussions
with Kimmo Kainulainen in earlier stages of this project.
In particular, the results in section~\ref{Reduction to the on-shell limit} 
have been largely developed based on work with Kimmo Kainulainen.
We would also like to thank Thomas Konstandin for constructive discussions.
The work of SW was supported by the U.S. Department of Energy
under Grant No. DE-AC02-98CH10886.
SW also thanks the Alexander von Humboldt Foundation
for support by a Feodor Lynen Fellowship.


\begin{appendix}

\cleardoublepage

\section{Gradient expansion in the off-diagonal scalar kinetic equation}
\label{Gradient expansion in the off-diagonal scalar kinetic equation}

We shall now prove that, provided
$k\cdot\partial, (\partial \bar M_d^2)\cdot \partial_k \ll \delta(M_d^2)$, 
gradient expansion applies for the off-diagonal equation. 
Equation~(\ref{scalars-ke-12}) is of the form
\begin{equation}
\Big(k\cdot\partial + \frac 12(\partial \bar M_d^2)\cdot \partial_k
     + \frac i2 \delta(M_d^2)  - ik\cdot \delta(\Xi)\Big)
         i\Delta_{12}^< = {\cal S}_{12},
\label{scalars-ke-12-2}
\end{equation}
where ${\cal S}_{12} = {\cal S}_{12}[\Delta_d]$ represents the source
composed of the diagonal elements. This equation can be solved by using
the method of Green functions as follows. Consider for simplicity
the following problem
\begin{equation}
   \big(k\cdot\partial + \frac i2 \delta(M_d^2)+\epsilon
   \big)G_\epsilon^r(k;x,x') 
 = \delta^4(x-x'),
\label{Green-function:eom}
\end{equation}
which is solved by the retarded Green function
\begin{equation}
 G_\epsilon^r(k;x,x') = \frac{1}{k_0}\theta(t-t')
               \delta^3\Big(\vec x-\vec x' - \frac{\vec k}{k_0}(t-t')\Big)
   \exp\Big({-\Big[\frac i2 \frac{\delta(M_d^2)}{k_0}+\epsilon\Big](t-t')}\Big)
,
\label{Green-function:eom2}
\end{equation}
where $\epsilon$ represents an (infinitesimal) positive dissipation term.
Formally, this solution contains rapid oscillations, since the exponent
varies at the scale $1/\hbar$, which characterizes the off-diagonal elements.
One may wonder what happened to these intrinsically quantum oscillations in
our solution~(\ref{Delta12:solution}) and~(\ref{scalars-ke-12-solution}). 
In order to answer this question we consider the solution 
to~(\ref{scalars-ke-12-2}) 
\begin{eqnarray}
 i\Delta_{12}(k,x) &=& \int d^4 x' G_\epsilon^r(k;x,x') {\cal S}_{12}(k,x')
\nonumber\\
   &=& \int ^t dt' {\cal S}_{12}\Big(k,\vec x-\frac{\vec k}{k_0}(t-t'),t'\Big)
 \frac{1}{k_0}
 \exp\Big({-\Big[\frac i2 \frac{\delta(M_d^2)}{k_0}+\epsilon\Big](t-t')}\Big)
.
\label{scalars-ke-12-solutionB}
\end{eqnarray}
The shift in the position, $\delta \vec x = \vec v(t-t')$, corresponds to 
the retardation for a particle moving with the velocity $\vec v = \vec k/k_0$. 
A similar retardation shift results when the $\partial_k$-derivative
is included in~(\ref{scalars-ke-12-2}-\ref{Green-function:eom}).
Since ${\cal S}_{12}$ is varying very slowly when compared with the
oscillatory term, we can expand around $t'=t$,
\begin{equation}
  {\cal S}_{12}\Big(k,\vec x-\frac{\vec k}{k_0}(t-t'),t'\Big) 
  = {\cal S}_{12}(k,x) 
  + (t'-t)\Big(\partial_t+\frac{\vec k}{k_0}\cdot\nabla\Big){\cal S}_{12}(k,x) 
  + O(\partial^2)
\end{equation}
and integrate~(\ref{scalars-ke-12-solutionB}) to obtain
\begin{eqnarray}
 i\Delta_{12}(k,x) &=& - \frac{2i}{\delta(M_d^2)}{\cal S}_{12}(k,x) 
 - \frac{4}{\delta(M_d^2)^2}\, k\cdot\partial\, {\cal S}_{12}(k,x) 
 + O(\partial^2).
\label{scalars-ke-12-solution2}
\end{eqnarray}
This converges provided 
\begin{equation} 
 k\cdot\partial \ll \delta(M_d^2),
\label{convergence:off-diagonal}
\end{equation}
which is precisely the criterion for the validity of the gradient expansion
in the off-diagonal equations. 

\vskip 0.1in

One may argue that the derivatives acting on the mass must be included as well.
To study the role of these derivatives, we assume that 
both the source and the masses depend on the time variable only. In this case 
the off-diagonal equation of motion reads
\begin{equation}
   \big[k_0\partial_t  +  \frac 12 (\partial_t \bar M_d^2)\partial_{k_0}
     +  \frac i2 \delta(M_d^2) - ik_0\delta(\Xi_t)
   \big]i\Delta_{12}(k_0,t) = {\cal S}_{12}(k_0,t) 
\label{Delta_12:eom1+1}
\end{equation}
for which the retarded Green function equation is 
\begin{equation}
   \big[k_0\partial_t  +  \frac 12 (\partial_t \bar M_d^2)\partial_{k_0}
     +  \frac i2 \delta(M_d^2) - ik_0\delta(\Xi_t) + \epsilon
   \big]G_\epsilon^r(k_0,k_0';t,t') 
 = \delta(k_0-k_0')\delta(t-t')
\,.
\label{Green-function:eom1+1}
\end{equation}
When written in the Fourier space
\begin{equation}
  G_\epsilon^r(k_0,k_0';t,t') = \int \frac{d\kappa}{2\pi} e^{i\kappa(t-t')}
                                g_\epsilon^r(k_0,k_0';\kappa)
\label{Fourier:Green-function:eom1+1}
\end{equation}
equation~(\ref{Green-function:eom1+1}) becomes 
\begin{equation}
   \big[ik_0\kappa  +  \frac 12 (\partial_t \bar M_d^2)\partial_{k_0}
     +  \frac i2 \delta(M_d^2)  - ik_0\delta(\Xi_t) + \epsilon
   \big]g_\epsilon^r(k_0,k_0';\kappa) 
 = \delta(k_0-k_0')
\label{Green-function:eom1+1:Fourier}
\end{equation}
where we have ignored the higher order gradients in time (that is 
we took $\partial_t \bar M_d^2$ and $\delta(M_d^2)$ to be time independent).
The retarded solution of~(\ref{Green-function:eom1+1:Fourier}) is given by (the solution
proportional to $-\theta(k_0'-k_0)$ corresponds to the advanced Green function),
\begin{equation}
 g_\epsilon^r(k_0,k_0';\kappa) = \frac{2}{\partial_t \bar M^2_d}\theta(k_0-k_0')
   \exp\Big(\frac{i}{\partial_t \bar M^2_d} \Big[
                 (\kappa-\delta(\Xi_t))
         ({k_0'}^2-k_0^2) + (\delta(M_d^2)-2i\epsilon)(k_0'-k_0)\Big]\Big)
.
\label{Green-function:solution:Fourier:1+1}
\end{equation}
This solution is not unique. Indeed, replacing ${k_0'}^2$ by 
${k_0'}[a {k_0'} + (1-a) k_0 ]$ is also a solution, which however differs
from~(\ref{Green-function:solution:Fourier:1+1}) at higher order in gradients,
and hence this ambiguity is irrelevant.
From~(\ref{Green-function:solution:Fourier:1+1}) we then have 
\begin{equation}
 G_\epsilon^r(k_0,k_0';t,t') 
= \frac{2}{\partial_t \bar M^2_d}\theta(k_0-k_0')
  \delta\Big(t-t'-\frac{k_0^2-{k_0'}^2}{\partial_t \bar M^2_d}\Big)
   \exp\Big(\frac{i(k_0'-k_0)}{\partial_t \bar M^2_d} 
    [\delta(M_d^2) - (k_0'+k_0) \delta(\Xi_t) - 2i\epsilon]\Big)
,
\label{Green-function:solution:1+1}
\end{equation}
such that Eq.~(\ref{Delta_12:eom1+1}) is solved by 
\begin{eqnarray}
 i\Delta_{12}(k,x) &=& \int dk_0'dt'
       G_\epsilon^r(k_0,k_0';t,t') {\cal S}_{12}(k_0',t')
\label{scalars-ke-12-solutionC}
\\
   &=& \int_{-\infty} ^{k_0} dk_0'
          {\cal S}_{12}\Big(k_0',t-\frac{k_0^2-{k_0'}^2}{\partial_t \bar M^2_d}\Big)
      \frac{2}{\partial_t \bar M^2_d}
      \exp\Big(\frac{i(k_0'-k_0)}{\partial_t \bar M^2_d} 
    [\delta(M_d^2) - (k_0'+k_0) \delta(\Xi_t) - 2i\epsilon]\Big)
.
\nonumber
\end{eqnarray}
The shift in time
$\delta t = -(k_0^2-{k_0'}^2)/(\partial_t \bar M^2_d)$ corresponds to 
the time retardation caused by the force term. 
Since in the limit of a very slowly varying field
($\partial_t \rightarrow 1/T \rightarrow 0$), 
${\cal S}_{12}$ is varying very slowly in comparison with the rapidly oscillating 
term, we can expand around $t'=t$ and $k_0'=k_0$,
\begin{equation}
  {\cal S}_{12}\Big(k_0',t-\frac{k_0^2-{k_0'}^2}{\partial_t \bar M^2_d}\Big)
  = {\cal S}_{12}(k_0,t) 
  + (k_0'-k_0)\Big(\partial_{k_0}
  + \frac{k_0'+k_0}{\partial_t \bar M^2_d}\partial_t\Big){\cal S}_{12}(k_0,t) 
  + {\cal O}(\partial_t^2,\partial_{k_0}^2)
\end{equation}
and integrate~(\ref{scalars-ke-12-solutionC}) to obtain
\begin{eqnarray}
 i\Delta_{12}(k_0,t) &=& - \frac{2i}{\delta(M_d^2) }
   \Big[1 
  + \frac{2i}{\delta(M_d^2)}\Big(k_0\partial_t 
                               + \frac 12 (\partial_t\bar{M}_d^2)\partial_{k_0}
                               - ik_0\delta(\Xi_t) \Big) 
\Big]{\cal S}_{12}(k_0,t) 
 + {\cal O}(\partial_t^2).\;\;
\label{scalars-ke-12-solution4}
\end{eqnarray}
One of the higher order terms we have dropped is, for example,
$4i[(\partial_t \bar M_d^2)/(\delta M_d^2)^3]\partial_t {\cal S}(k_0,t)$.
This converges provided 
\begin{equation} 
 k_0\partial_t, (\partial_t\bar{M}_d^2)\partial_{k_0} \ll \delta(M_d^2).
\end{equation}
From this and Eq.~(\ref{scalars-ke-12-solution2}-\ref{convergence:off-diagonal}) 
we conclude that the criterion for validity of the gradient expansion
in the off-diagonal equations reads
\begin{equation} 
 k\cdot\partial, (\partial\bar{M}_d^2)\cdot\partial_{k}, 
   k\cdot\delta(\Xi) \ll \delta(M_d^2).
\end{equation}

\end{appendix}


\cleardoublepage

\cleardoublepage

\end{document}